\documentclass{ar-1col}

\usepackage[comma]{natbib}
\usepackage{hyperref}
\usepackage{amssymb}
\usepackage{lipsum}
\usepackage{amsmath}
\usepackage{enumitem}

\newcommand{\hii}{\textsc{Hii}}
\newcommand{\ha}{$\rm H{\alpha}$}

\newcommand{\msun}{M_{\odot}}
\newcommand{\mstar}{M_{\star}}

\newcommand{\acounits}{\textup{M\ensuremath{_\odot}~pc\ensuremath{^{-2}} (K~km~s\ensuremath{^{-1}})\ensuremath{^{-1}}}}

\newcommand{\msunperpcsq}{\mbox{M$_\odot$\,pc$^{-2}$}}

\jname{Xxxx. Xxx. Xxx. Xxx.}
\jvol{AA}
\jyear{YYYY}
\doi{10.1146/((please add article doi))}

\begin{document}

\markboth{Schinnerer \& Leroy}{Cloud-Scale Gas \& Star Formation}

\title{Molecular Gas and the Star Formation Process on Cloud Scales in Nearby Galaxies}

\author{E. Schinnerer$^1$ and A.K. Leroy$^2$
\affil{$^1$Max Planck Institute for Astronomy, K\"onigstuhl 17, 69117 Heidelberg, Germany; email: schinner@mpia.de}
\affil{$^2$
Department of Astronomy, The Ohio State University, 140 West 18th Ave, Columbus, OH 43210, USA; email: leroy.42@osu.edu}}
\begin{abstract}

\begin{minipage}[l]{0.75\textwidth}

Observations that resolve nearby galaxies into individual regions across multiple phases of the gas-star formation-feedback ``matter cycle’’ have provided a sharp new view of molecular clouds, star formation efficiencies, timescales for region evolution, and stellar feedback. We synthesize these results, cover aspects relevant to the interpretation of observables, and conclude:

\begin{itemize}

\item The observed cloud-scale molecular gas surface density, line width, and internal pressure  all reflect the large-scale galactic environment while also appearing mostly consistent with properties of a turbulent medium strongly affected by self-gravity.

\item Cloud-scale data allows for statistical inference of both evolutionary and physical timescales. These suggest a period of cloud collapse of order the free-fall or turbulent crossing time ($\sim 10{-}30$~Myr) followed by forming massive stars and subsequent rapid ($\lesssim$ 5\,Myr) gas clearing after the onset of star formation. The star formation efficiency per free-fall time is well determined over thousands of individual regions at $\epsilon_{\rm ff}\approx 0.5_{-0.3}^{+0.7}\%$.

\item The role of stellar feedback is now measured using multiple observational approaches. The net yield is constrained by the requirement to support the vertical weight of the galaxy disk. Meanwhile the short gas clearing timescales suggest a large role for pre-supernova feedback in cloud disruption. This leaves the supernovae free to exert a large influence on the larger galaxy, including stirring turbulence, launching galactic-scale winds, and carving superbubbles.

\end{itemize}

\end{minipage}

\end{abstract}

\maketitle

\tableofcontents

\section{INTRODUCTION}
\label{sec:intro}

The conversion of interstellar gas into stars and the subsequent action of stellar feedback are central to the assembly and evolution of galaxies. These processes affect many other topics of current interest: the birth and evolution of star clusters, chemical evolution, the use of baryons to trace dark matter, the evolution of supermassive black holes, the coupling between galaxy disks and the circumgalactic medium, and planet formation.

Star formation and feedback represent crucial steps in a larger ``matter cycle'' that operates in galaxies. In this cycle, the neutral gas in the interstellar medium (ISM) cools from a warm to a cool atomic phase (\textsc{Hi}), then eventually transforms into cold, turbulent, molecular gas (H$_2$). A subset of this molecular gas becomes strongly gravitationally bound and dense, and stars can form from this cold, dense gas. These new stars exert feedback on their parent clouds and the surrounding galaxy via radiation, stellar winds, and supernovae. This feedback converts nearby gas to the ionized phase, stirs turbulence in the ISM, and reshapes the gas in the galaxy. Over time, this newly heated, lower density gas cools, once again becomes neutral and molecular, and this matter cycle begins again. It is now widely recognized that the cycle is not closed. Star formation over cosmic timescales is fueled by gas accretion onto galaxy disks from the circumgalactic medium (CGM), and ultimately from the intergalactic medium (IGM). Meanwhile, stellar feedback ejects matter and heavy elements from galaxy disks, and injects energy into the surrounding CGM and IGM.

\begin{figure}[t]
\begin{center}
\includegraphics[width=0.90\textwidth]{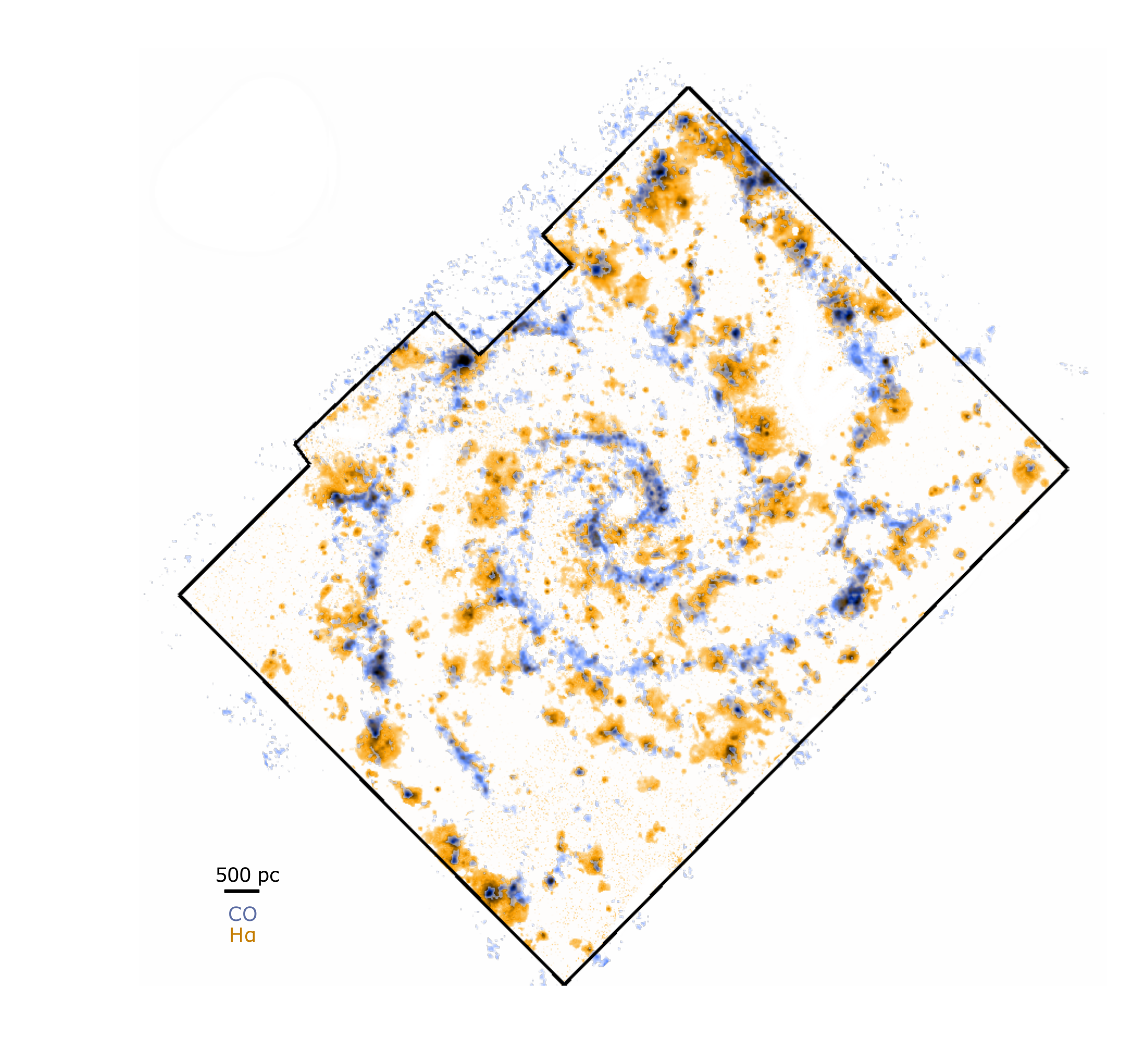} \\
\end{center}
\includegraphics[width=\textwidth]{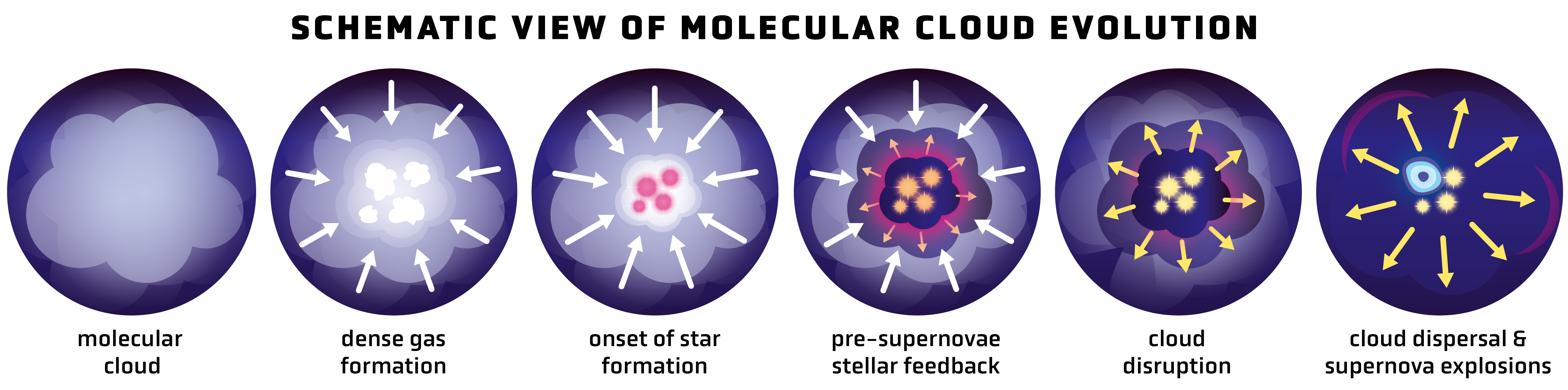}
\caption{
\textit{Top:} Visible separation of tracers of recent star formation (gold; VLT/MUSE H$\alpha$) and cold gas (blue; ALMA CO\,(2-1)) at ``cloud-scale'' resolution, here in NGC~628 from \citet{KRECKEL18SFR}. 
\textit{Bottom:} Schematic view of the evolution of a molecular cloud from formation to star cluster. The cloud begins as an over-density of cold, predominantly molecular gas. A subset of the gas achieves high column and volume densities. Stars form from this dense material. Newly formed massive stars rapidly impact their surrounding birth material via radiation and winds, reshaping or even disrupting the cloud. Over time, the continued energy and momentum input from these young massive stars disperses the gas cloud. As a result, many core-collapse supernovae explode in relatively low density, pre-cleared surroundings. In the sketch, gas density increases from blue to white, attenuation of stellar light decreases from red to yellow, light red colors indicate gas ionized by massive stars, and the cyan object in the right panel represents a supernova.}
\label{fig:sketch_cloud}
\end{figure}

Here, we review progress understanding the portion of this matter cycle that directly concerns the formation of stars (Fig. \ref{fig:sketch_cloud}). We focus on new observations of nearby galaxies, mostly at physical resolution $\theta \approx 100$~pc. Following emerging convention, we refer to these as \textbf{cloud-scale} observations. Such observations have resolution $\theta \sim 50{-}150$~pc, of order the size of a conventionally defined massive giant molecular cloud (GMC) or a giant \textsc{Hii} region \citep[e.g.,][]{MIVILLE17GMCS,OEY03HIIREGIONS}. Cloud-scale observations resolve emission \textit{into} populations of clouds or \textsc{Hii} regions, capturing their demographics, but they do not resolve individual regions. This $\theta \approx 100$~pc is also about the scale height of the molecular gas layer in massive disk galaxies \citep{YIM14EDGEON,HEYER15REVIEW}, and interstellar turbulence is driven at $\sim 50{-}500$ pc scales, with $\sim 100$\,pc often invoked as a typical outer scale \citep{MACLOW04REVIEW,ELMEGREEN04REVIEW}. 

In the last decade, multi-wavelength cloud-scale observations have seen immense growth, so that $\mu \lesssim 1''$ ALMA CO mapping (tracing the molecular gas), optical integral field unit (IFU) mapping (tracing ionized gas, recent star formation, and the underlying stellar populations), multiband HST imaging (tracing star clusters), and mid-IR JWST imaging (tracing embedded star formation and dust) are now available for $\sim 50{-}100$ nearby galaxies. This provides a systematic view of galaxies resolved into their constituent star-forming units (e.g., Fig. \ref{fig:sketch_cloud}). Because $\mu \lesssim 1''$ corresponds to $\theta \lesssim 100$~pc out to a distance of $d=20$~Mpc, this cloud-scale imaging spans a diverse set of galaxies, reflecting both the typical locations where stars form and rarer systems that probe extreme conditions.

We contrast cloud-scale observations with \textbf{resolved galaxy} observations, which have resolution $\theta \sim 500{-}2,000$~pc, of order a typical stellar disk scale length. These resolve the gas and stellar disks of galaxies, but do not access individual population constituents. We also distinguish cloud-scale observations from \textbf{resolved cloud} observations ($\theta \sim 5{-}20$~pc), which do resolve massive GMCs and \textsc{Hii} regions. The resolution here is about the size of a low-mass molecular cloud (like Taurus) or a typical Milky Way \textsc{Hii} region. This scale offers more direct access to the density structure, gravitational potential of clouds, and internal kinematics of the ionized and molecular gas. At even smaller scales, \textbf{cloud substructure} observations ($\theta \sim 0.1{-}1$~pc) isolate giant filaments and cores within molecular clouds. The resolution is of order the transverse size of a molecular filament or a core. Such studies are key to access the immediate sites of star formation and the local environments of star clusters. Neither resolved cloud studies nor cloud substructure studies are yet available covering representative samples of whole galaxies besides the Milky Way (MW).

\begin{marginnote}[]
\entry{Resolved galaxy observations}{observations that resolve the disks of galaxies ($\theta \approx 500{-}2,000$~pc) but do not isolate individual star-forming regions.}

\entry{Cloud-scale observations}{observations with resolution ($\theta \approx 50{-}150$~pc) matched to the size of individual star-forming regions (GMCs, \textsc{Hii} regions) to isolate but not resolve them.}

\entry{Resolved cloud observations}{observations that resolve massive molecular clouds or \textsc{Hii} regions ($\theta \approx 5{-}20$~pc).}

\entry{Cloud substructure observations}{observations that isolate and resolve individual dense cores and filaments ($\theta \approx 0.1{-}1$~pc).
}

\end{marginnote}

\begin{marginnote}[]
\entry{Stellar feedback}{Energy and momentum input to the ISM due to winds, radiation, and supernova explosions, mostly from massive ($M_{\star} \gtrsim 8\,M_{\odot}$) stars.}
\end{marginnote}

\subsection{Background and overview}
\label{sec:sketch}

This review follows the evolution of an idealized star-forming region shown in Fig.\ref{fig:sketch_cloud}. We first discuss the molecular clouds that host star formation, then focus on the dense gas where stars actually form, consider the tracers, timescales, and efficiencies of star formation, then review the effects of the stellar feedback exerted on the surrounding gas. Lastly, we discuss galaxy centers, the most accessible extreme environment for cloud-scale studies.

\subsubsection{Molecular clouds (\S \ref{sec:moleculargas})}

Star formation occurs inside cold, dense clouds of molecular gas. This gas represents the fuel and sets the initial conditions for star formation. Early studies of Galactic molecular clouds established a classical picture of molecular clouds often referred to as \textit{Larson's laws} \citep[][]{LARSON81GMCS,SCOVILLE87GMCS,SOLOMON87GMCS,BLITZ93GMCS}. In this picture molecular clouds have (i) mean column densities $N({\rm H}_2) \approx 10^{22}$~cm$^{-2}$, $\Sigma_{\rm mol} \approx 200$~M$_\odot$~pc$^{-2}$, or $A_V \approx 10$~mag; (ii) highly supersonic line widths and a dependence of line width on size scale that indicate turbulent motions to dominate the cloud kinematics; and (iii) a good correspondence between the dynamical mass inferred from the size and line width and the mass estimated from CO emission, implying gravitationally bound or even virialized clouds. Early CO studies of the nearest galaxies, especially the Local Group, largely confirmed this picture \citep[e.g.,][]{BOLATTO08GMCS,FUKUI10REVIEW}. These studies also often found relatively shallow cloud mass functions ($dN / d \log M_{\rm mol} \propto M_{\rm mol}^{-\alpha}$ with $\alpha \sim 0.5 {-} 0.7$) with the most massive clouds in the range $M_{\rm mol} \approx 10^{6}{-}10^{7}$~M$_\odot$. As a result of this shallow slope, most of the molecular mass resides in clouds with high masses, $M_{\rm mol} \approx 10^{5} {-} 10^{7}$~M$_\odot$, even though lower mass clouds remain more common by number. 

With new Milky Way surveys and extensive extragalactic cloud-scale CO mapping, this picture has evolved. The surface density and line width of molecular gas clearly vary within the Milky Way and across galaxies \citep[][]{HEYER09GMCS,HUGHES13GMCS,SUN18CLOUDS}. It is also clear that there is no single line width-size relation. Instead, the CO line width at a fixed size scale can vary dramatically from cloud to cloud \citep[][]{HEYER15REVIEW,SUN18CLOUDS}. The gas is hierarchically structured, perhaps fractal, consistent with a medium shaped by turbulence driven at multiple scales \citep{ELMEGREEN04REVIEW,MACLOW04REVIEW}. As a result, identification of individual clouds becomes subjective, with sizes and masses measured from CO emission closely reflecting the resolution and sensitivity of the observations \citep[][]{HUGHES13GMCS,LEROY16GMCS}. 

Considering this, a framework has emerged that de-emphasizes masses and sizes of specific objects in favor of more robust estimates of the physical state of the gas: surface density, line width at fixed size scale, internal pressure, and dynamical state. Central to this framework is a relation between line width-at-fixed size and surface density for a gravitationally bound object \citep[][]{HEYER09GMCS,KETO86GMCS}. This \textit{Heyer-Keto relation} (detailed in \S \ref{sec:comethods}) has become the point of reference for molecular cloud studies that span diverse environments, replacing the less general \textit{Larson's laws} framework. The relation can describe molecular gas across a wide range of environments, spatial scales, and surface densities \citep[][]{HEYER15REVIEW,SUN20GMCS}, reflecting that molecular gas often appears approximately bound by self-gravity. While the gas dynamical state varies weakly, its internal pressure, surface density, and line width all correlate with the environment where the gas is found, such that molecular clouds very much reflect their host galaxy properties.

\begin{textbox}[ht]
\section{Clouds}
\label{sec:clouds}
The idea of the ISM being organized into an ensemble of individual, self-contained clouds, which may each produce a population of stars, goes back many decades \citep[e.g., see early reviews by][]{HOYLE53REVIEW,DIETER66REVIEW,SPITZER68BOOK}. This concept now pervades the field and has been useful in many regards. We will invoke such a view throughout this article, as we consider, e.g., the dynamical state of molecular gas, the statistical inference of timescales, and even the idea of a cloud life cycle. However, we caution that while this concept remains useful, it has clear limitations. Observations suggest a fractal, turbulent medium with important features on multiple spatial scales and no well-defined boundaries \citep[e.g., see \citealt{ELMEGREEN04REVIEW}, \citealt{MACLOW04REVIEW}, or the recent reviews by][]{HACAR23FILAMENTS,chevance2023}. Thus, we will use the term and concept of ``cloud'' while also noting where a more modern view should apply. We will use it to refer to compact regions of enhanced gas density in the ISM that may or may not be bound by self-gravity, and also refer to each $\sim 100$~pc scale parcel of cold molecular gas as a ``cloud.''
\end{textbox}

\begin{marginnote}[]
\entry{$M_{\rm mol}$ and $\Sigma_{\rm mol}$}{molecular gas mass and surface density, including a factor of $1.36$ to account for the contribution of helium.}
\entry{Star formation rate, SFR}{Mass of new stars formed per unit time. Averaging over large spatial or time scales is required to obtain a meaningful SFR.}
\entry{Molecular gas depletion time, $\tau_{\rm dep}^{\rm mol}$}{Time over which the current SFR would exhaust all molecular gas, , $\tau_{\rm dep}^{\rm mol} \equiv M_{\rm mol}/{\rm SFR}$.
Typical values in local 
star-forming galaxies are $\tau_{\rm dep}^{\rm mol} \approx 1{-}2$~Gyr.}
\end{marginnote}

\subsubsection{Dense gas (\S \ref{sec:densegas})}

Within molecular clouds, stars form from a subset of dense, self-gravitating material. Milky Way disk studies find a direct link between the mass of high surface density gas and the rate at which a cloud forms stars \citep[][]{KAINULAINEN09DENSE,HEIDERMAN10DENSE,LADA10DENSE}. Meanwhile, extragalactic observations reveal a strong correlation between a galaxy's SFR (from IR emission) and the luminosity of ``dense gas tracers,'' especially HCN (1-0) and HCO$^{+}$ (1-0) \citep[][]{GAO04DENSE,GARCIABURILLO12DENSE}. Attempts to connect the Milky Way and extragalactic work \citep[e.g.,][]{WU10DENSE,LADA12DENSE} indicate that individual Galactic clumps or clouds obey approximately the same linear correlation between the mass of dense gas and the measured SFR as whole galaxies. We suggest to refer to the relationship between dense molecular material and the star formation rate as the \textit{Gao-Solomon relation} to honor the contributions to the field by Yu Gao (1963-2022) and Philip~M. Solomon (1938-2008). 

\begin{marginnote}[]
\entry{Dense molecular gas}{Typically refers to molecular gas with $\Sigma_{\rm mol} \gtrsim 100$~M$_\odot$~pc$^{-2}$, $A_V \gtrsim 8{-}10$~mag, or $n_{\rm H2} \gtrsim 10^4$~cm$^{-3}$. In Solar Neighborhood clouds, $\sim 10\%$ of the molecular mass is dense.}
\entry{Dense gas tracers}{Molecular transitions that have high Einstein A coefficients and critical densities well above the mean density of a Solar Neighborhood molecular cloud.}
\end{marginnote}

The existence of this relation and the popular view that the gravitational free-fall time ($\tau_{\rm ff} \propto \sqrt{\rho}$) is directly linked to the star formation rate \citep[e.g.,][]{KRUMHOLZ07EFF} raise key questions. How does the molecular gas density change as a function of environment across galaxies? How does the SFR per unit dense gas tracer change with environment? Observations of the Milky Way's Central Molecular Zone (CMZ) reveal a region rich in high density gas \citep[][]{BALLY87DENSE} but with little star formation compared to the amount of dense material \citep{LONGMORE13DENSE}, demonstrating that environment can affect both the gas density and star formation rate per unit dense gas.

Resolving dense cores at extragalactic distances remains out of reach, therefore addressing these questions in other galaxies requires observations of spectroscopic ``dense gas tracers'' like the aforementioned HCN and HCO$^{+}$ transitions. Though sensitive cloud-scale surveys of these lines are not yet widely available, new surveys have produced dozens of resolved maps of dense gas tracers across galaxies \citep[e.g.,][]{JIMENEZ19DENSE,NEUMANN23DENSE}. These studies reveal that physical gas density, traced by ratios like HCN/CO, changes systematically as a function of galactic environment. In the high stellar surface density, high pressure parts of gaalxies (e.g., galaxy centers) the gas density tends to increase while the SFR per unit dense gas tracer appears to decline. These results support a picture in which local over-densities, rather than a specific density threshold, represent the gas immediately available for star formation.

\begin{marginnote}[]
\entry{Free-fall time, $\tau_{\rm ff}$} {The time for a spherical cloud to collapse under its own self-gravity (Table \ref{tab:phystimes}).}
\entry{Turbulent crossing time, $\tau_{\rm cr}$} {The time it takes for gas moving at the rms turbulent velocity to cross a cloud (Table \ref{tab:phystimes}).}
\entry{Star formation efficiency ($\epsilon$)}{Fraction of molecular gas converted into stars over some fiducial timescale (Eq. \ref{eq:efficiency}).}
\entry{Star formation efficiency per free-fall time ($\epsilon_{\rm ff}$)}{Star formation efficiency adopting the gravitational free-fall time, $\tau_{\rm ff}$, as the fiducial timescale.}
\end{marginnote}

\subsubsection{Star formation, its efficiency and timescales (\S \ref{sec:timescales})}
\label{sec:intro_SF}

Molecular gas and star formation activity track one another when observations average together many individual regions. This results in a \textit{molecular Kennicutt-Schmidt relation} in which the surface densities of SFR, $\Sigma_{\rm SFR}$, and molecular gas mass, $\Sigma_{\rm mol}$, correlate well on the scale of integrated galaxies \citep[e.g.,][]{KENNICUTT98SFGAS} or large, $\sim$ kpc-sized regions within galaxies \citep[e.g.,][]{BIGIEL08SFGAS,LEROY13SFGAS}. Their ratio is often expressed as a molecular gas depletion time $\tau_{\rm dep}^{\rm mol}$, which varies weakly across galaxy disks or among massive galaxies on the star-forming main sequence \citep[e.g.,][]{YOUNG96SFGAS,LEROY08SFGAS,SAINTONGE17SFGAS} but shows stronger variations contrasting galaxy disks, centers, and merger-driven starbursts.

Physically, $\tau_{\rm dep}^{\rm mol}$ captures the SFR per unit gas mass, which should depend on the dynamical state of the gas (including magnetization), the free-fall and turbulent crossing times, and the impact of stellar feedback \citep[e.g.,][]{PADOAN11EFF,FEDERRATH12EFF}. Large-scale dynamical factors, including spiral arms and the rate of cloud collisions, likely matter as well \citep{FUKUI21COLLIDE}. Cloud-scale observations of the gas provide access to these theoretically important parameters via estimates of the mean gas density, turbulent velocity dispersion, and virial parameter. Comparing these properties to the measured $\tau_{\rm dep}^{\rm mol}$ offers both an avenue to test theories of star formation and a way to observationally establish the efficiency of star formation relative to a variety of physically relevant timescales, including the theoretically important star formation efficiency per free-fall time, $\epsilon_{\rm ff}$.

High resolution observations also isolate individual star-forming regions and catch each in a distinct evolutionary state, e.g., as a quiescent cloud or an \textsc{Hii} region. As a result, at cloud-scales the Kennicutt-Schmidt relation breaks down and $\tau_{\rm dep}^{\rm mol}$ shows increased scatter with a strong bias towards low or high values depending on whether one focuses specifically on molecular clouds or tracers of recent star formation \citep[][]{SCHRUBA10SFGAS,ONODERA10SFGAS}. This breakdown allows one to deploy statistical arguments to infer the timescales and evolutionary pathways associated with each stage of cloud evolution, star formation, and feedback \citep[e.g.,][]{KAWAMURA09TIMES,KRUIJSSEN14TIMES}. In its simplest form, this approach identifies a sample of regions likely to offer unbiased sampling of the full region evolution. This typically yields the relative timescales associated with each evolutionary step. Then, because the timescales associated with emission from young massive stellar populations are understood, these star formation tracers can be used as ``stellar clocks'' to help assign concrete values to statistically inferred timescales.

Now, $\epsilon_{\rm ff}$ has been measured for thousands of regions across $\sim 100$ galaxies \citep[e.g.,][]{SUN23SFGAS}. Star formation is inefficient, with $\epsilon_{\rm ff} \approx 0.5_{-0.3}^{+0.7}$
and $0.3$~dex scatter. Meanwhile, comparison of gas and star formation signatures at cloud-scale and better resolution suggests that gas is dispersed from star-forming regions rapidly after massive stars form, likely due to the effect of stellar feedback \citep[e.g.,][]{KIM22TIMES}. Furthermore, the earlier, quiescent ramp-up phase with strong CO emission and weaker signatures of massive star formation lasts of order a gravitational free-fall or turbulent crossing time. Although this first order picture appears well-supported by observations and in good agreement with current theoretical models, theoretically predicted variations in $\epsilon_{\rm ff}$ or $\tau_{\rm dep}^{\rm mol}$ as a function of cloud-scale gas properties have not been clearly observed and this may reflect some tension between observations and these models.

\begin{marginnote}[]
\entry{Dynamical equilibrium pressure, $P_{\rm DE}$}{Vertical pressure needed to support the weight of the ISM in the galaxy potential (see text box near Eq. \ref{eq:pde}).}
\end{marginnote}

\subsubsection{Stellar feedback (\S \ref{sec:feedback})}

The radiation, stellar winds, and supernovae (SNe) associated with young stars inject energy, momentum, and mass into the surrounding ISM, processes collectively referred to as \textit{stellar feedback}. This feedback is critical to set the overall low efficiency of star formation and shapes the evolution of star-forming regions \citep[e.g.,][]{HOPKINS12GALAXIES,chevance2023}. It influences the ISM across a wide range of spatial and time scales and involves complex coupling between radiation, shocks, and multiple gas phases. Given this complexity, constraining the efficiency, dominant mechanisms, and impact of stellar feedback requires multiple approaches and remains a fast-evolving topic. 

Averaged over large, $\sim$kpc, scales, stellar feedback likely often supplies the dynamical equilibrium pressure, $P_{\rm DE}$, needed to balance the vertical weight of the ISM in the galaxy potential \citep[][]{OSTRIKER10PRESS,OSTRIKER22PRESS}. SFR and $P_{\rm DE}$ can be estimated from observables and then contrasted to constrain the effective feedback yield, $\Upsilon_{\rm fb} \equiv P_{\rm DE} / \Sigma_{\rm SFR}$, placing a key constraint on the large-scale efficacy of stellar feedback.

Meanwhile, multi-wavelength cloud-scale observations imply fast, few Myr timescales for the cold gas to clear from an individual region after stars form. As stellar winds and radiation act rapidly compared to core-collapse SN explosions, this short clearing time suggests that pre-supernova (pre-SN) feedback plays a dominant role in molecular cloud disruption \citep[e.g.,][]{KRUIJSSEN19TIMES,CHEVANCE20TIMES}. During this pre-SN phase a variety of early feedback mechanisms act in concert, including radiation pressure, stellar winds, and expansion of the heated gas. To understand the time- and scale-dependence of each mechanism, direct estimates of the instantaneous strength of individual feedback terms acting within large samples of \hii\ regions have become a key tool \citep[e.g.,][]{LOPEZ11FEEDBACK,LOPEZ14FEEDBACK}.

SNe inject large amounts of energy and momentum into the ISM, and where they explode plays a major role in determining their impact \citep[e.g.,][]{WALCH15FEEDBACK}. Because SNe are delayed relative to wind and radiation feedback, some gas-clearing must happen before any SNe explode. Cloud disruption and kinematic drift may further separate a star from its birth site before the lowest mass core-collapse SNe explode at $\approx 30$~Myr. Fortunately, cloud-scale observations now yield both direct and indirect constraints on the gaseous environments of SNe. Observations also reveal the likely impact of these explosions, that is large-scale shells, bubbles, and turbulent motions seen in maps of gas and dust.

\subsubsection{Galactic centers, an extreme environment in galaxy disks (\S \ref{sec:centers})}

We contrast results between galaxy disks and galaxy centers, especially the bar-fed \textit{central molecular zones} (CMZs) common in disk galaxies. These regions are often covered as part of a single observation along with galaxy disks, and so represent the extreme environments most naturally accessed by cloud-scale surveys of normal star-forming galaxies. These central regions often exhibit the highest $\Sigma_{\rm mol}$ and $\Sigma_{\rm SFR}$, and shortest dynamical times in a galaxy, and these extreme conditions also manifest in the molecular gas and feedback properties of the regions. 

\begin{textbox}[ht]\section{Questions addressed}

\noindent \textit{1) What are the physical conditions in the molecular clouds where stars form?} What is the dynamical state, pressure, surface and volume density, and line width of the gas? How do these conditions change across galaxies? How do they depend on the galactic environment where the gas is found? (\S \ref{sec:moleculargas}, \ref{sec:densegas}, \ref{sec:centers})

\smallskip

\noindent \textit{2) How fast and how efficiently do molecular clouds make stars?} How efficient is star formation relative to key physical timescales, including the gravitational free-fall time? How quickly do star-forming regions evolve from quiescent clouds to star formation to exposed stellar populations? (\S \ref{sec:timescales}).

\smallskip

\noindent \textit{3) How does stellar feedback reshape the ISM?} What are the timescales and dominant mechanisms for dispersal of molecular clouds? What is the net efficiency of stellar feedback into the ISM? What impact do SN explosions have? (\S \ref{sec:feedback})

\smallskip
\noindent
\textbf{Using this review:} This review summarizes hard work by many researchers, and we urge anyone making use of this work \textit{to cite the primary sources} for key results. We adopt an observational focus, and have tried to include many useful equations, including scaling relations, translations between observables and physical quantities, and statistical measurements. Those looking for these numbers or equations \textit{will find them in the tables}. Given the limits set by the ARAA format, we have compiled a \textit{supplemental list of references} and we hope that readers interested in the field will take advantage of this \textbf{Supplemental Material}.

We also refer the reader to previous complementary reviews on the theory of star formation \citep{MCKEE07REVIEW}, gas and star formation in galaxies at larger scales \citep{KENNICUTT12REVIEW,TACCONI20REVIEW,SAINTONGE22REVIEW}, molecular gas \citep{BOLATTO13REVIEW,DOBBS14REVIEW,HEYER15REVIEW}, star clusters \citep{KRUMHOLZ19REVIEW}, and ionized gas in galaxies \citep{kewley2019,SANCHEZ20REVIEW}.

\end{textbox}

\subsection{Observational progress}
\label{sec:obsprogress}

Cloud-scale multi-wavelength imaging has now been obtained for $\approx 50{-}100$ galaxies. Here, we emphasize major advances from $\sim 1''$ resolution low-$J$ CO mapping by ALMA, high resolution, wide-field optical spectral mapping by VLT/MUSE and Keck, multiband near-UV through near-IR HST imaging sensitive to individual stars and clusters, and $\lesssim 1''$ near- and mid-IR imaging from JWST. We further highlight resolved galaxy mapping of dense gas tracers, which is closely physically related and also represents a major technical advance.

\smallskip

\textbf{High resolution CO surveys --} 
Emission from low-$J$ rotational transitions of the CO molecule remains the primary tool to survey the structure and motions of cool molecular gas in galaxies. Early extragalactic cloud-scale views of molecular gas came from single dish CO mapping of the Magellanic Clouds or interferometric CO studies of the Local Group \citep[e.g.,][]{BOLATTO08GMCS,FUKUI10REVIEW}, but this work was mainly restricted to nearby dwarf galaxies. The handful of $\mu \sim 1{-}3''$ wide area CO maps of more distant galaxies paved the way for cloud-scale studies \citep[e.g.,][]{PAWS13SURVEY,DONOVANMEYER13GMCS}, but required large time investments using previous-generation mm-interferometers.

ALMA and NOEMA have changed the landscape. Surveying CO~(2-1), ALMA can map a nearby galaxy at $\mu \approx 1''$ in a few hours. This has resulted in a boom of $\sim 1''$ resolution CO mapping of galaxies. Central to this review, PHANGS--ALMA \citep{PHANGSALMA21SURVEY} observed CO~(2-1) emission from 90 galaxies at $\sim 50{-}150$~pc physical resolution, sensitive to $M_{\rm mol} \sim$10$^5\,\msun$ clouds. PHANGS--ALMA focused on relatively massive, face-on star-forming galaxies. Meanwhile, WISDOM has surveyed $\approx 25$ early-type galaxies and galaxy centers \citep[][]{DAVIS22GMCS}, often targeting smaller areas in more distant systems with resolutions of $20{-}80$~pc. CO surveys of merging and starburst galaxies have also proceeded, with the PUMA project \citep{PUMA21SURVEYTWO} currently providing the most comprehensive high resolution view for ultra-luminous infrared galaxies (U/LIRGs).

\smallskip
\textbf{Resolved galaxy maps of dense gas tracers --}
While CO emission reveals the bulk structure of molecular gas, multi-line molecular spectroscopy can access physical gas densities. The key measurement is the ratio between emission from CO and a tracer sensitive to gas density. Because dense gas tracers like the low $J$ rotational lines of HCN, HCO$^+$, CS, or CN, tend to be $30{-}100$ times fainter than CO lines, high resolution, high completeness cloud-scale mapping of dense gas tracers is not widely available. Instead, current extragalactic dense gas tracer studies resemble CO-based studies about 20\,years ago. There have been extensive single-pointing surveys from merging starburst galaxies and galaxy centers \citep[e.g.,][]{GAO04SURVEY,GARCIABURILLO12DENSE,PRIVON15DENSE} and mapping surveys now capture dense gas tracer emission across the entire area of a few dozen galaxies \citep[e.g.,][]{JIMENEZ19DENSE,NEUMANN23DENSE,IMANISHI23DENSE}. Reflecting the faintness of the target lines, these surveys emphasize single dish telescopes or compact interferometer configurations in order to maximize surface brightness sensitivity. Despite large time investments, these observations still often reach only modest $S/N$ and many key results rely on binning or stacking to access mean trends. 

\smallskip
\textbf{Cloud-scale optical spectral mapping --}
Optical nebular lines and spectra of young stellar populations are key to tracing star formation activity and gauging the strength of stellar feedback. Over the last decade, optical integral field unit (IFU) technology has matured and new IFUs have come online that sample the point spread function while also covering a large field-of-view (FoV), including MUSE on the VLT. These capabilities allow one to isolate or even resolve individual \textsc{Hii} regions in galaxies out to $\sim 20$~Mpc\footnote{With MUSE, $\lesssim 1$\,h exposures for $\lesssim 20$~Mpc targets recover \ha\ luminosities $\rm log{(L_{H\alpha} [erg~s^{-1}])} \approx 36.5$, equivalent to ionization by a single late-type O star \citep[e.g.,][]{SANTORO22SFR}.} \citep[e.g.,][]{MAD19SURVEY,PHANGSMUSE22SURVEY}. The optical emission lines captured by these IFUs can be used to probe the metallicity, density, extinction, temperature, power source, illuminating radiation field, and ionizing photon production rate for individual nebulae \citep[e.g.,][]{kewley2019,SANCHEZ20REVIEW} and so infer both the strength and impact of stellar feedback \citep[e.g.,][]{mcleod2019}. Fitting of the stellar spectra provides insight into stellar population properties \citep[mass, age, metallicity distribution; e.g.,][]{walcher2011}.

\smallskip
\textbf{HST resolved cluster studies --} 
In the past decade, the quantitative study of populations of stellar clusters has taken a massive leap forward thanks to efforts using the \textit{Hubble} space telescope (HST). Clusters represent good candidates for simple stellar populations (SSPs), reasonably described by a single age and mass and able to serve as crucial ``clocks'' and key measurements of the output of the star formation process. However, they are small, with half-light radii of $0.5{-}10$~pc, and often $2{-}3$~pc \citep[e.g.,][]{KRUMHOLZ19REVIEW}. As a result, resolving clusters and separating their light from the rest of the galaxy requires high angular resolution and cluster studies have been a hallmark of HST since its earliest days. Over the last decade, the WFC3 upgrade and ambitious allocations have enabled multi-band surveys of star clusters spanning the ultraviolet to the near-IR and overlapping surveys at other wavelengths \citep[e.g., GOALS, PHAT, LEGUS, PHANGS--HST;][]{GOALS09SURVEY,PHAT12SURVEY,LEGUS15SURVEY,PHANGSHST22SURVEY}. These have helped bring cluster studies to maturity, enabling robust population studies and comparisons to other phases of the matter cycle \citep[see][]{KRUMHOLZ19REVIEW}.

\smallskip
\textbf{Cloud-scale mid-infrared imaging from JWST --}
In most star-forming regions in massive galaxies, most of the emission from massive stars is absorbed and re-emitted by dust. This makes mid-infrared (mid-IR) emission from hot dust a robust and powerful tracer of star formation activity \citep[e.g.,][]{CALZETTI05SFR,CALZETTI07SFR}. Unfortunately, beyond the nearest galaxies, the resolution of previous IR telescopes corresponded to a few $100$\,pc to $\sim$kpc. JWST is transforming this field. JWST maps mid-IR emission at $< 1''$ resolution, even at $\lambda \sim 20\mu$m wavelengths that probe thermal dust emission from star-forming regions, and opens the window to $\lesssim 100$~pc resolution mid-IR imaging out to $\sim 20$~Mpc. Already in its first year, this has yielded spectacular images that isolate thousands of individual star-forming regions, find embedded clusters, and even detect individual young stellar objects in the nearest galaxies \citep[e.g., Fig. \ref{fig:bubbles}][]{PHANGSJWST23SURVEY,LINDEN23CLUSTERS,LENKIC23YSOS}.

\smallskip
\textbf{Sampling the local galaxy population --}
New surveys align these multi-wavelength capabilities to study representative samples of galaxies. Here ``representative'' refers to sampling galaxies close to the \textit{star-forming main sequence} (SFMS), the narrow (rms $\sim 0.4$~dex) locus in SFR-M$_\star$ space where the majority of stars formation occurs \citep[e.g.,][]{TACCONI20REVIEW,SAINTONGE22REVIEW}. Galaxies vary along the SFMS in size, morphology, metallicity, dust-to-gas ratio, H$_2$-to-\textsc{Hi} ratio, and overall gas abundance. Lower mass galaxies are more gas-rich, have fewer heavy elements, less dust, and more atomic compared to molecular gas. 
The PHANGS surveys, the main datasets used in this review, aim to map out the matter cycle in a representative set of nearby SFMS galaxies \citep[][]{PHANGSALMA21SURVEY,PHANGSMUSE22SURVEY,PHANGSHST22SURVEY,PHANGSJWST23SURVEY}. Complementary efforts target less common, more extreme systems with the goal of understanding how extreme conditions affect the matter cycle or how this cycle operates during rare but important phases of galaxy evolution. These include surveys of early-type galaxies \citep[ATLAS-3D, WISDOM;][]{ATLAS3DSURVEY11,DAVIS22GMCS}, dwarf galaxies \citep[LITTLETHINGS, LEGUS;][]{LITTLETHINGS12SURVEY,LEGUS15SURVEY}, and merging galaxies \citep[GOALS, PUMA;][]{GOALS09SURVEY,armus2023,PUMA21SURVEYTWO}. (See also \textbf{Supplemental Material}.)

\section{MOLECULAR GAS ON CLOUD SCALES}
\label{sec:moleculargas}

The molecular phase of the ISM hosts star formation. This gas exhibits a wide range of physical densities, supersonic turbulent motions, and appears highly structured, organized on multiple scales into filaments, shells, and dense clumps. This structure reflects the competing effects of supersonic turbulence, gas self-gravity, and large-scale flows in the galactic potential. Because this gas presents the reservoir for star formation, measuring its properties and understanding how these properties link to the larger galactic context is a critical goal. In particular, the mean density, the magnitude of turbulent motions, and the balance between kinetic and gravitational potential energy are all thought to be key to the evolution of molecular gas clouds and their ability to form stars.

This molecular gas consists of mostly H$_2$. Because H$_2$ is hard to observe at the temperatures and extinctions found in molecular clouds, our knowledge of the molecular gas in other galaxies depends on the use of indirect tracers, especially spectroscopic mapping of the second most common molecule, CO. As discussed in \S \ref{sec:intro}, ALMA can efficiently map CO emission at cloud scales, $\theta \approx 50{-}150$~pc, from galaxies out to $\sim 20$~Mpc. This new capability has yielded the first complete, external picture of molecular gas at cloud scales across representative samples of relatively massive, star-forming galaxies enabling fundamental ``demographics'' of cloud-scale molecular gas surface density, line width, and more in star-forming, $z=0$ galaxies. Further, it reveals a close correspondence between the physical conditions in this molecular gas and the larger scale environments within the galaxy disk.

\begin{marginnote}[]
\entry{``Intensive'' cloud properties}{Molecular cloud properties normalized by mass, area, or line width. Such properties are more robust to resolution and algorithmic biases.}
\entry{``Extensive'' cloud properties}{Integrated properties such as mass, luminosity, or size. These tend to be highly sensitive to algorithmic and resolution biases.}
\end{marginnote}

\subsection{Measuring molecular gas properties at cloud scales}
\label{sec:comethods}

Advances in cloud-scale CO mapping have led to corresponding steps forward in analysis techniques. These methods aim at ensuring a fair comparative analysis between heterogeneous datasets targeting different galaxies, rigorous comparisons between observations and numerical simulation, and inferences of the physical and dynamical state of the molecular gas.

\subsubsection{Identification (or not) of clouds and need for data homogenization}

A first key point is that the ``clouds'' identified in extragalactic surveys almost always have sizes a few times the resolution \citep[e.g.,][]{PINEDA09GMCS,LEROY16GMCS}. This reflects that molecular gas traced by CO often does not show any clear preferred scale but instead appears clumpy, filamentary, and turbulent at multiple scales. When observing such a hierarchically structured medium, the local maxima picked out by cloud-finding algorithms are not unique objects but instead depend on the resolution and noise of the input image.

This coupling between resolution and the set and size of objects identified has crucial implications. First, any rigorous comparison between GMCs in different systems requires matching the resolution and sensitivity among the datasets compared \citep{HUGHES13GMCS,ROSOLOWSKY21GMCS}. Second, the distributions of ``extensive'' properties of clouds like mass, CO luminosity, and size are fundamentally linked to the resolution and sensitivity of the data. Therefore, we recommend that such extensive quantities be avoided entirely or their use restricted to comparative analysis among homogenized datasets.

Given these concerns, a simpler approach based on statistical image characterization has gained traction \citep{LEROY16GMCS,SUN18CLOUDS}. The distributions of surface density, line width, or related quantities are measured after convolving data to fixed spatial and spectral scales. No object identification is carried out, and the physical resolution of the data serves as the relevant length scale, with observations repeated across multiple resolutions to assess scale dependence. Such a structure-agnostic statistical approach is preferable to object identification for low to moderate inclination disks ($i \lesssim 70^\circ$). In such systems, there tends to be little line of sight shadowing, i.e. with $\lesssim 1$ distinct emission component per line of sight on average, and measurements can be carried out using standard image processing techniques. For systems with complex, multi-component spectra over individual lines of sight, like highly-inclined galaxies, merging galaxy systems or complex flows in, e.g., galaxy centers, then spectral decomposition \citep[e.g.,][]{HENSHAW16GMCS,MIVILLE17GMCS} can be critical. The Milky Way, with its combination of component crowding and large spread in distances step along each sight line remains a special case where object identification, or at least deprojection, is a complex but necessary step.

\subsubsection{Observables and formalism} 
\label{sec:cloudmethods}

Whether using object-identification techniques or fixed-spatial-scale statistics, emphasizing intensive quantities minimizes ``more is more'' effects and controls for biases related to the resolution of the data or the choice of segmentation algorithm. We emphasize four (mostly) intensive quantities that can be inferred from cloud-scale CO observations and together access the physical state of the gas.

\begin{marginnote}[]
\entry{Virial parameter, $\alpha_{vir}=2 K/U$}{Balance between kinetic, $K$, and potential, $U$, energy, with $\alpha_{\rm vir} = 1$ for a virialized cloud ($2 K=U$) and $\alpha_{\rm vir} = 2$ for a marginally bound cloud ($K = U$).}
\end{marginnote}

\smallskip
\textbf{Surface and volume density --} The gas mass surface density, $\Sigma_{\rm mol}$, descends directly from the observed CO intensity, $I_{\rm CO}$, or from the mass $M_{\rm mol}$ and radius $R_{\rm cl}$ of an object $\Sigma_{\rm mol} = M_{\rm mol} / ( \pi R_{\rm cl}^2 )$. Because it is normalized, intensive, and closely linked to observations, $\Sigma_{\rm mol}$ represents a key observable property of the molecular gas in a galaxy. At the scales discussed in this review, this quantity should be corrected for the effects of inclination $i$ by scaling by $\cos(i)$ \citep[see][]{SUN22CLOUDS}. The actual volume density, $\rho_{\rm mol} \sim \Sigma_{\rm mol} / H_{\rm mol}$, is also critical, especially to estimate the gravitational free-fall time, internal pressure, or the gravitational binding energy. This requires an estimate of the line of sight depth, $H$, and there are several approaches to this (see Eq. \ref{eq:hmol} and surrounding).

\smallskip
\textbf{Line width normalized by size scale --} Within individual clouds or among populations of clouds with similar $\Sigma_{\rm mol}$, the observed line width $\sigma_{\rm mol}$ scales with the size scale $R_{\rm cl}$ as $\sigma_{\rm mol} \propto R_{\rm cl}^{0.5}$ \citep[e.g.,][]{HEYER04GMCS}. The normalization of this relation varies systematically \citep[e.g.,][]{HEYER09GMCS,HUGHES13GMCS}. Contrasting the measured normalization with $\Sigma_{\rm mol}$ can reveal the dynamical state of the gas. As a result, the size-line width parameter, $\sigma_{\rm mol}/R_{\rm cl}^{0.5}$ or $\sigma_{\rm mol}^2/R_{\rm cl}$ (both are used), has become a key observable.

\smallskip
\textbf{Dynamical state and virial parameter --} The virial parameter, $\alpha_{\rm vir} = 2K/U$, is often used to express the dynamical state of a cloud. Here dynamical state refers to the balance of kinetic energy $K$ and potential energy $U$ due to self-gravity \citep[see][]{MCKEE92GMCS,BERTOLDI92GMCS}. Inferring the precise potential requires knowledge of the internal density structure of the gas, and the impact of external pressure, complex geometry, sources of gravitational potential other than gas self-gravity, and magnetic or other support not visible in the line width have all been considered. Most simply, 

\begin{equation}
\label{eq:avir}
\alpha_{\rm vir} \approx 5 f \sigma_{\rm mol}^2 R_{\rm cl} / G M_{\rm mol}
\end{equation}

\noindent with $f$ a factor set by the geometry of the cloud \citep[e.g.,][]{ROSOLOWSKY21GMCS,SUN22CLOUDS}. For a uniform density sphere with radius $R_{\rm cl}$ and mass $M_{\rm mol}$, $f=1$. More concentrated distributions yield lower $f$.\footnote{A common important case is that $R_{\rm cl}$ expresses the half-mass radius, which requires adjusting the formula for $\alpha_{\rm vir}$ accounting for only $1/2 M_{\rm mol}$; or taking $f=2$ \citep[e.g.,][]{ROSOLOWSKY21GMCS}.}

Formally, $\alpha_{\rm vir} = 1$ for a virialized cloud where $2 K=U$ and $\alpha_{\rm vir} = 2$ for a marginally bound cloud with $K = U$. However, because the true geometry, and so $f$, is generally imprecisely known, the main practical use of $\alpha_{\rm vir}$ is comparative, i.e., to capture variations in the strength of self-gravity in a homogeneously processed dataset.

\smallskip
\textbf{Internal pressure --} The mean internal kinetic pressure of the gas is $P_{\rm int} \sim \rho_{\rm mol} \sigma_{\rm mol}^2$ with $\rho_{\rm mol}$ estimated as above. This internal pressure is often loosely referred to as the ``turbulent'' pressure because CO line widths are essentially always too broad to be thermal. However, other non-turbulent sources may also contribute to the line width, including unresolved streaming motions, internal rotation, shearing motions, or collapse motions \citep[e.g.,][]{BALLESTEROS11GMCS,LIU21GMCS}. The quantity is often expressed as $P_{\rm int} / k_B$ in units of cm$^{-3}$~K for easy comparison to thermal ISM pressures, $n k_B T$. Note that this term will not account for any magnetic support, which does not contribute to the line width. In numerical simulations magnetic pressure can be dynamically important \citep[e.g.,][]{KIM21FEEDBACK}, but it remains difficulty to access via extragalactic observations.

\subsubsection{Heyer-Keto relation} 
\label{sec:heyerketo}

A single scaling relation relates $\Sigma_{\rm mol}$ to $\sigma_{\rm mol}/R_{\rm cl}^{0.5}$ for idealized spherical clouds with a fixed $\alpha_{\rm vir}$, i.e., a fixed ratio of kinetic energy to gravitational potential energy due to self-gravity. Following \citet{HEYER09GMCS} and \citet{KETO86GMCS}:

\begin{eqnarray}
\label{eq:heyerketo}
\frac{\sigma_{\rm mol}}{R_{\rm cl}^{0.5}} &=& \sqrt{\alpha_{\rm vir} \frac{G \pi \Sigma_{\rm mol}}{5~f} } 
\end{eqnarray}

\noindent so that for fixed $\alpha_{\rm vir}$ and $R_{\rm cl}$, $\sigma$ tracks $\Sigma_{\rm mol}^{0.5}$. This is the \textit{Heyer-Keto relation} referred to in \S \ref{sec:intro}. This relation has the advantage of leveraging intensive quantities, making it safer to compare across diverse datasets, and capturing the key physics in \textit{Larson's laws} \citep{LARSON81GMCS} in a single relation or a plot of $\sigma_{\rm mol} / R_{\rm cl}^{0.5}$ vs. $\Sigma_{\rm mol}$. Below we will use Eq. \ref{eq:heyerketo} for individual sight lines at fixed resolution, $\theta$, by setting $R_{\rm cl} \approx \theta/2$ and $f = \ln 2$ following \citet{SUN18CLOUDS,SUN20GMCS}\footnote{The factor $f=\ln 2$ for individual lines of sights reflects setting $\Sigma_{\rm mol} = M_{\rm mol} / A_{\rm beam}$ with $A_{\rm beam} = \pi / (4 \ln 2) \mu^2$ the area of a Gaussian beam with FWHM $\mu$.}.

Conversely, the internal pressure of a parcel of gas might be fixed. Holding $P_{\rm int}$ fixed predicts an orthogonal relationship (an isobar) in the same space:

\begin{eqnarray}
\label{eq:isobar}
\sigma_{\rm mol} &\approx& \left( 2~R_{\rm cl}~P_{\rm int} \right)^{0.5}~\Sigma_{\rm mol}^{-0.5}
\end{eqnarray}

\noindent here adopting a cylindrical geometry \citep[for consistency with][]{SUN18CLOUDS} so that $\rho_{\rm mol} \sim \Sigma_{\rm mol} / 2 R_{\rm cl}$. Using simple analytic cloud models, \citet{FIELD11GMCS} have considered how lines would connect for the case of an external confining medium of fixed pressure. Here we only note that one might expect a smooth transition between the isobar (Eq. \ref{eq:isobar}) and the self-gravitating line (Eq. \ref{eq:heyerketo} with $\alpha_{\rm vir} = 1{-}2$) as one considers higher and higher $\Sigma_{\rm mol}$ objects at fixed external pressure. We also note that while \citet{FIELD11GMCS} consider a general external pressure term, both \citet{LIU21GMCS} and \citet{MEIDT18GMCS} consider the broader galactic (mostly stellar) potential and model its impact on $\sigma_{\rm mol}^2/R_{\rm cl}$ and $\alpha_{\rm vir}$.

\begin{textbox}[ht]
\section{Line of sight depth in cloud-scale molecular gas studies}
\label{sec:thirddim}

The molecular gas volume density, $\rho_{\rm mol}$, is needed to calculate the gravitational potential and the gravitational free-fall time. Translating from the more observable $\Sigma_{\rm mol}$ to $\rho_{\rm mol}$ requires estimating the sight line depth, $H_{\rm mol}$. When using object-finding methods, this depth is often estimated by assuming isotropy, so that the radius, $R_{\rm cl}$, measured in the plane of the sky applies along the line of sight, $H_{\rm mol} \approx 2 R_{\rm cl}$. We discuss the close coupling between size measurements and data resolution in \S \ref{sec:comethods}. A similar simplifying assumption when considering pixel-wise measurements is to approximate $H_{\rm mol} \sim \theta$, i.e., that the third dimension is approximately equal to the resolution. Alternatively, one can make a dynamical estimate by leveraging some version of dynamical equilibrium formalism \citep[see discussion around Eq. \ref{eq:pde} and][]{KOYAMA09PRESS}. Depending on whether we consider the stellar-dominated regime or only gas self-gravity:

\begin{eqnarray}
\label{eq:hmol}
\textrm{stellar-dominated:\,\,} H_{\rm mol, z}^{\rm stellar} \approx \sigma_{\rm gas, z} \sqrt{\frac{2 H_\star}{G \Sigma_\star}}
\textrm{~; or~gas-dominated:\,\,}
H_{\rm mol, z}^{\rm gas} \approx \frac{2 \sigma_{\rm gas, z}^2}{\Sigma_{\rm mol} \pi G}
\end{eqnarray}

\noindent where $\Sigma_{\rm mol}$ and $\Sigma_\star$ are the surface densities of molecular gas and stars, $\sigma_{\rm gas, z}$ the vertical velocity dispersion of the gas in the disk, $H_\star$ is the stellar scale height, and $G$ the gravitational constant.

In the case where $\theta \gtrsim 100$~pc, we recommend using either $H_{\rm mol} = 100$~pc as a reasonable first guess or Eq. \ref{eq:hmol}. Then the effective three dimensional radius for a flattened object with radius $R_{\rm 2D}$ in the disk will be $R_{\rm 3D} = R_{\rm 2D}^{2/3}~\left(H_{\rm mol}/2\right)^{1/3}$ \citep[][]{ROSOLOWSKY21GMCS}. Given the importance of physical, local quantities, it is critical to leverage the clear perspectives from highly inclined galaxies to improve these prescriptions \citep[e.g., see][]{JEFFRESON22GMCS}.
\end{textbox}

\begin{table}[t!]
\tabcolsep7.5pt
\caption{Converting CO emission to molecular gas mass}
\label{tab:alphaco}
\begin{center}
\begin{tabular}{|l|c|l|}
\hline
Quantity $(Y)$ & Formula & Comments \\
\hline
\multicolumn{3}{c}{Basic conversions (using $\alpha_{\rm CO}$ or $X_{\rm CO}$ associated with specific CO transition)} \\
\hline
$M_{\rm mol}~\left[ {\rm M}_\odot \right]$ & $\alpha_{\rm CO}~L_{\rm CO}^\prime$
& luminosity $L_{\rm CO}^\prime$ in [K km s$^{-1}$ pc$^2$] \\
$\Sigma_{\rm mol}~\left[ {\rm M}_\odot~{\rm pc}^{-2}\right]$ & $\alpha_{\rm CO}~I_{\rm CO}~\cos i$ & line-integrated CO brightness $I_{\rm CO}$ in [K km s$^{-1}$]\\
$N ({\rm H}_2)~\left[ {\rm cm}^{-2} \right]$ & $X_{\rm CO}~I_{\rm CO}~\cos i$ & $I_{\rm CO}$ in [K km s$^{-1}$]\\
$\alpha_{\rm CO,MW}^{1-0}$ & $4.35$~\acounits & Milky Way CO~(1-0) conversion factor \\
\hline
\multicolumn{3}{c}{CO line ratios (following 
\citealt{LEROY22LINES}; see also text and Fig.\,\ref{fig:alphaco})} \\
\hline
$R_{21}$ & $I_{\rm CO}^{2-1} / I_{\rm CO}^{1-0}$, $\approx 0.65 \pm 0.18$ & $I_{\rm CO}$ in [K km s$^{-1}$]; mean value and scatter \\
$R_{32}$ & $I_{\rm CO}^{3-2} / I_{\rm CO}^{2-1}$, $\approx 0.50^{+0.09}_{-0.27}$ & $I_{\rm CO}$ in [K km s$^{-1}$]; mean value and scatter \\
$R_{31}$ & $I_{\rm CO}^{3-2} / I_{\rm CO}^{1-0}$, $\approx 0.31 \pm 0.11$ & $I_{\rm CO}$ in [K km s$^{-1}$]; mean value and scatter \\
$R_{21} (\Sigma_{\rm SFR})$ & $0.65\times \left(\frac{\Sigma_{\rm SFR}}{1.8 \times 10^{-2}~{\rm M}_\odot~{\rm yr}^{-1}~{\rm kpc}^{-2}} \right)^{0.125}$ with min 0.35, max 1.0
& CO\,(2-1) to CO\,(1-0) line ratio variations \\
$R_{31} (\Sigma_{\rm SFR})$ & $\langle R_{32} \rangle \times R_{21} (\Sigma_{\rm SFR}) \approx 0.5 \times R_{21} (\Sigma_{\rm SFR})$ & CO\,(3-2) to CO\,(1-0) line ratio variations \\
\hline
\multicolumn{3}{c}{Environment-dependent $\alpha_{\rm CO}$ in disk galaxies} \\
\hline
$\alpha_{\rm CO}^{1-0} (Z,\Sigma_\star)$ & $\alpha_{\rm CO,MW}^{1-0}~f(Z)~g(\Sigma_\star)$ & CO\,(1-0)-to-H$_2$ conversion factor \\
$\alpha_{\rm CO}^{2-1} (Z,\Sigma_\star,\Sigma_{\rm SFR})$ & $\alpha_{\rm CO,MW}^{1-0}~f(Z)~g(\Sigma_\star)~R_{21}(\Sigma_{\rm SFR})^{-1}$ & CO\,(2-1)-to-H$_2$ conversion factor \\
$\alpha_{\rm CO}^{3-2} (Z,\Sigma_\star,\Sigma_{\rm SFR})$ & $\alpha_{\rm CO,MW}^{1-0}~f(Z)~g(\Sigma_\star)~R_{31}(\Sigma_{\rm SFR})^{-1}$ & CO\,(3-2)-to-H$_2$ conversion factor \\
$f (Z)$ & $\left(\frac{Z}{Z_\odot} \right)^{-1.5}$ & ``CO-dark'' term, valid from $Z \sim 0.2{-}2.0 Z_\odot$ \\
$g (\Sigma_\star)$ & $\left( \frac{ {\rm max} \left( \Sigma_\star , 100 {\rm M}_\odot~{\rm pc}^{-2} \right) }{100~\msunperpcsq } \right)^{-0.25}$
& ``starburst'' emissivity term \\
\hline
\end{tabular}
\end{center}
\begin{tabnote}
\textsc{Notes} --- Expressions to estimate molecular gas mass or surface density from observables (see Fig. \ref{fig:alphaco}). All lines express $Y = \textrm{Formula}$ with notes and units in the last column. Here $i$ refers to the inclination of the disk; $\Sigma_\star$ and $\Sigma_{\rm SFR}$ are the stellar mass and SFR surface desnity measured at $\sim 1$~kpc resolution; $Z$ is the gas-phase metallicity, and $Z_\odot$ is the Solar value relevant to $\alpha_{\rm CO,MW}$; $R_{21}$, $R_{31}$, etc. are the line ratios among the CO~(1-0), CO~(2-1), and CO~(3-2) lines.
\end{tabnote}
\end{table}

\begin{figure}[t!]
\begin{center}
\includegraphics[width=1.0 \textwidth,angle=0]{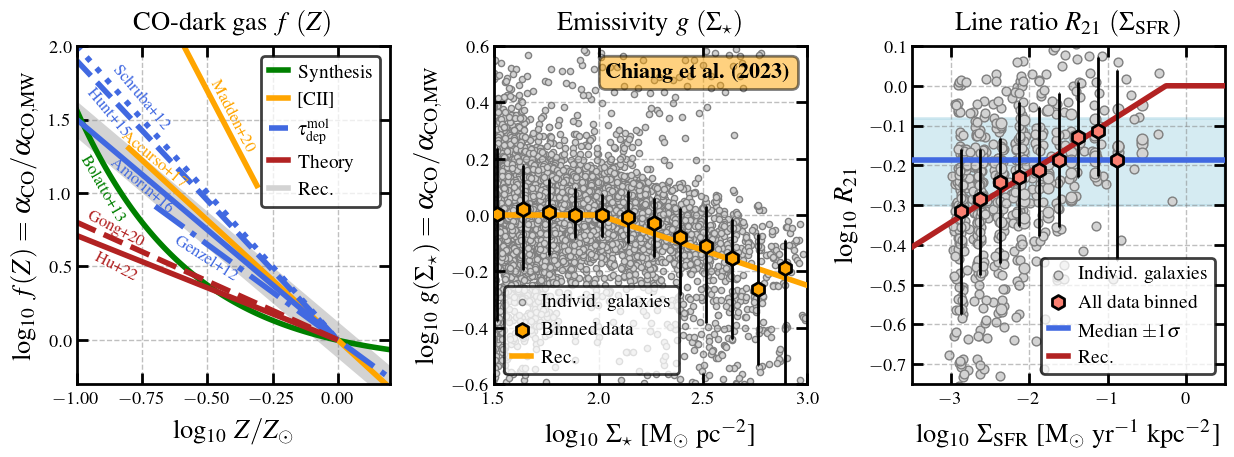}
\caption{Estimating molecular gas mass from CO emission. \textit{Left:} Estimates of the dependence of the conversion factor $\alpha_{\rm CO}$ on metallicity $Z$ \citep[based on Fig. 1 in][]{HU22XCO}. There is still a large spread among estimates of how $\alpha_{\rm CO}$ depends on $Z$, but many recent observational studies converge near $\alpha_{\rm CO} \propto Z^{-1.5}$ (gray). \textit{Middle:} Estimates of the impact of the ``starburst'' $\alpha_{\rm CO}$ term from \citet{CHIANG23XCO}. A clear decrease in $\alpha_{\rm CO}$ estimates based on dust emission with increasing stellar mass surface density $\Sigma_\star$ is evident in $\approx 40$ galaxies. Here, $\alpha_{\rm CO}$ for each galaxy is normalized to its mean value at $\Sigma_\star = 100$~M$_\odot$~pc$^{-2}$. \textit{Right:} $R_{21}$ line ratio as a function of SFR surface density $\Sigma_{\rm SFR}$ showing data from \citet[][]{DENBROK21LINES} and \citet[][]{LEROY22LINES}. Each gray point represents an average for one galaxy at that $\Sigma_{\rm SFR}$, and the red points show the median and 16{-}84\% range of the ratio over all data. The red line represents the fit reported in Table\,\ref{tab:lines}.}
\label{fig:alphaco}
\end{center}
\end{figure}

\subsection{Translating low-$J$ CO emission to molecular gas mass}
\label{sec:alphaco}

Much of our knowledge of cold gas requires translating low-$J$ CO emission into an estimate of molecular gas mass. Doing this relies on an overall calibration of the CO-to-H$_2$ conversion factor, $\alpha_{\rm CO}$, as well as accounting for variations in $\alpha_{\rm CO}$ due to metallicity, emissivity variations, and choice of CO line. Table \ref{tab:alphaco} presents a set of equations that reflect current best practice to account for each of these effects. These follow the framework of \citet{BOLATTO13REVIEW} in which $\alpha_{\rm CO}$ exhibits two major senses of variation. First, the fraction of ``CO-dark'' molecular gas will increase in regions with low metallicity and low dust-to-gas ratio, leading to high $\alpha_{\rm CO}$ in low-mass, low-metallicity galaxies \citep[e.g.,][]{MALONEY88XCO}. Second, the emissivity of the CO-emitting part of a cloud depends on the gas density, temperature, and optical depth, with the opacity also inversely dependent on the velocity gradient. This leads $\alpha_{\rm CO}$ to drop in the high-temperature, high-velocity dispersion conditions found in galaxy centers and merger-driven starbursts, a phenomenon sometimes referred to as the ``starburst'' conversion factor \citep[e.g.,][]{DOWNES98XCO}. Most analyses still adopt this basic physical picture, but since \citet{BOLATTO13REVIEW} there have been advances in observational work and major progress in numerical simulations, which now include realistic CO chemistry and cover regions large enough to capture the multi-phase structure of the ISM \citep[e.g.,][]{GONG20XCO,SEIFRIED20XCO,HU22XCO}.

\subsubsection{CO-dark gas and the metallicity dependence of $\alpha_{\rm CO}$} 

Thanks to hydrogen self-shielding, it is easier for H$_2$ to become the dominant form of hydrogen than for CO to become the dominant form of carbon. As a result, in the outer, weakly shielded parts of molecular clouds, there are regions where the gas is mostly H$_2$ but CO is mostly absent, with carbon being in $\rm C^0$ or $\rm C^+$. In the Solar Neighborhood, it appears likely that this CO-dark component makes up $f_{\rm CO-dark} \sim 20{-}60\%$ of the molecular gas mass \citep[e.g.,][]{PARADIS12XCO,PINEDA13XCO} in good agreement with numerical simulations \citep[e.g.,][]{SMITH14XCO,SEIFRIED20XCO,GONG20XCO,HU22XCO}. These works also suggest substantial small-scale (i.e., cloud-to-cloud) variations in $f_{\rm dark}$, due to variations in the column density distribution and illuminating radiation, and time evolution of cloud structure. 

This CO-dark component $f_{\rm dark}$ is included in the Milky Way $\alpha_{\rm CO}$ because most extragalactic observations average together the CO-dark and CO-bright parts of clouds. Therefore as metallicity decreases and the dust-to-gas ratio decreases, the mass in CO-dark regions increases, and $\alpha_{\rm CO}$ will increase. Observations of dust and $\rm C^+$, which should still be present in the CO-dark gas, demonstrate the need for high $\alpha_{\rm CO}$ in low-mass, low-metallicity galaxies \citep[e.g.,][]{LEROY11XCO,JAMESON16XCO}. Numerical simulations also reveal a strong dependence of $\alpha_{\rm CO}$ on metallicity, with significantly suppressed CO emission at low metallicity and low extinction \citep[e.g.,][]{GLOVER11XCO,HU22XCO}.

For simplicity, this behavior is often characterized as a power-law, with $\alpha_{\rm CO} \propto (Z/Z_\odot)^{-a}$ where $Z$ refers to the metallicity with $Z_\odot$ the Solar value. Current observational evidence suggests $a \approx 1{-}2$. A current commonly adopted value is $\alpha_{\rm CO} \propto (Z/Z_\odot)^{-1.5}$. This broadly agrees with \textsc{Cii}, dust, and depletion time based arguments \citep[e.g.,][]{GENZEL12XCO,ACCURSO17XCO,HUNT20XCO}. However, note that current simulation results often yield shallower dependence, with $a \sim 0.5{-}1$ being common \citep[e.g.,][]{GLOVER12XCO,GONG20XCO,HU22XCO}. Further, additional factors including the dust-to-metals ratio, interstellar radiation field, cosmic ray ionization rate, and the structure of the clouds themselves all play potentially important roles, effectively adding uncertainty or scatter to any purely metallicity-dependent prescription. 

If such a power-law dependence holds, it appears most likely to break down outside the range $Z \sim 0.2{-}2~Z_\odot$, i.e., below about the metallicity of the Small Magellanic Cloud and at super-solar metallicities ($12+\log_{10} {\rm O/H} \sim 8.0{-}9.0$). At high enough metallicity, CO-dark gas should make a negligible contribution to the total mass budget. Meanwhile, the CO emission in very low metallicity systems appears to be confined to very small, dense, and short-lived clumps, which should imply both a steeper increase and significant stochasticity in $\alpha_{\rm CO}$ at low $Z$ \citep[e.g.,][]{GLOVER16XCO}.

The metallicity dependence of $\alpha_{\rm CO}$ represents a major source of systematic uncertainty. Galaxies exhibit a mass-metallicity relation and have metallicity gradients. \textbf{The metallicity dependence of $\alpha_{\rm CO}$ will thus impact any correlations between molecular gas properties and any physical properties of interest that are covariant with galaxy mass or galactocentric radius}, which is most quantities of interest. Complicating the situation, only a fraction of local galaxies have high quality metallicity maps or gradients, and resolved estimates of the more physically relevant dust-to-gas ratio are even less common.

\subsubsection{``Starburst'' conversion factors, emissivity, and galactic centers} 

CO appears to become more emissive in galaxy centers and intensely star-forming regions. A wide variety of work targeting merging galaxies, especially local U/LIRGs, suggests low $\alpha_{\rm CO}$ in these systems \citep[e.g.,][]{DOWNES98XCO} and the same is true for galaxy centers \citep[e.g.,][]{SANDSTROM13XCO}. This results from a mixture of high temperatures and high cloud-scale velocity dispersions. These high dispersions reflect contributions beyond gas self-gravity to the turbulent line width, including the effect of the stellar potential, stellar feedback, and large-scale colliding flows. Whatever the cause, the broader dispersion leads to lower opacity, which in turn leads to high CO emission per unit mass. However, because of the difficulty in independently tracing the total gas, even the existence of $\alpha_{\rm CO}$ variations in merging galaxies remains controversial \citep[e.g.,][]{SCOVILLE14DUST,DUNNE22XCO}.

Because of these effects, both dust-based derivations of $\alpha_{\rm CO}$ \citep[e.g.,][]{SANDSTROM13XCO,CHIANG23XCO} and $^{12}$CO and $^{13}$CO spectral line modeling \citep[e.g.,][]{ISRAEL20XCO,TENG23XCO} suggest a significant decrease in $\alpha_{\rm CO}$ in the inner parts of galaxies compared to disks, a result that has the same sense but larger magnitude than theoretical work \citep[e.g.,][]{NARAYANAN12XCO,GONG20XCO}. The spectral line modeling work suggests that lower $^{12}$CO opacity plays a key role creating the low $\alpha_{\rm CO}$ in galaxy centers, and \citet{TENG23XCO} specifically showed an anti-correlation between $\alpha_{\rm CO}$ and both the $^{12}$CO opacity and the velocity dispersion measured from $^{12}$CO~(2-1) emission, supporting an important role for line broadening.

There are two main approaches to estimate the magnitude of these effects. The first approach identifies locations where the effect is likely to occur as indicated, e.g., by stellar or total mass surface density. In Table \ref{tab:alphaco} we recommend the first approach as an accessible, straightforward way to approximately account for the ``starburst'' $\alpha_{\rm CO}$ term in galaxy disks. We specifically recommend the prescription from \citet{CHIANG23XCO} in which $\alpha_{\rm CO}^{1-0}$ anti-correlates with $\Sigma_\star$ above $\Sigma_\star = 100$~M$_\odot$~pc$^{-2}$. This prescription, which is illustrated in Fig. \ref{fig:alphaco} resembles the \citet{BOLATTO13REVIEW} recommendation, which uses the total mass surface density. Both are calibrated to match dust-based, $\sim 2$~kpc resolution $\alpha_{\rm CO}$ estimates in star-forming disk galaxies, with \citet{CHIANG23XCO} using an updated dataset compared to the \citet{SANDSTROM13XCO} one on which \citet{BOLATTO13REVIEW} based their prescription.

A second approach leverages the properties of the CO emission itself to predict $\alpha_{\rm CO}$ variations. For example, \citet{GONG20XCO} suggest a prescription based on simulations in which the conversion factor depends on the observed scale and peak brightness temperature along the line of sight, $\alpha_{\rm CO}^{1-0} \propto T_{\rm peak}^{-0.64}~r_{\rm beam}^{-0.08}$, and/or the observed CO line ratios (see below). Meanwhile, \citet{TENG23XCO} find strong anti-correlations between $\alpha_{\rm CO}$ and the local cloud-scale velocity dispersion, as well as a close relationship between $\alpha_{\rm CO}$ and the observed $^{13}$CO/$^{12}$CO line ratios. These or similar cloud-scale prescriptions may prove highly valuable in the near future, but they do place more stringent requirements on the data needed to estimate $\alpha_{\rm CO}$.

\begin{marginnote}[]
\entry{$R_{21}$}{CO\,(2-1) to CO\,(1-0) line ratio, intensity $I$ expressed in [K km\,s$^-{1}$]. $R_{21} \equiv I_{2-1}/I_{1-0}$ (see Table \ref{tab:alphaco}).}
\entry{$R_{32}$ and $R_{31}$}{As $R_{21}$ but expressing the CO\,(3-2) to CO\,(2-1) ratio and CO\,(3-2)-to-CO\,(1-0) ratio (see Table \ref{tab:alphaco}).}
\end{marginnote}

\subsubsection{Ratios among low-$J$ CO lines} 

With improvements in submm-wave technology, observations of the CO~(2-1) and CO~(3-2) lines have become common. At $z=0$, ALMA maps CO~(2-1) emission faster than CO~(1-0) emission, while intermediate redshift studies often target CO~(2-1) or CO~(3-2) because these lines appear in favorable atmospheric windows \citep[e.g.,][]{TACCONI20REVIEW}. Translating results among these different lines is important to connect different surveys. The ratios among the lines also offer some access to the physical state of the cold gas. Tables \ref{tab:alphaco} and \ref{tab:lines} list the typical line ratios among CO~(1-0), CO~(2-1), and CO~(3-2), found in the disks of galaxies \citep[from a compilation by][]{LEROY22LINES}. These agree well with typical Milky Way disk ratios \citep{YODA10LINES,WEISS05LINES}. 

Line ratios vary in response to local conditions. They are generally higher at the centers of galaxies and, as shown in Fig. \ref{fig:alphaco}, correlate with $\Sigma_{\rm SFR}$ \citep[e.g.,][]{DENBROK21LINES,YAJIMA21LINES}.  The scatter in the measurements is large compared to the dynamic range in $R_{21}$. This reflects the challenge of accounting for pointing, calibration, and statistical uncertainties when combining multiple single dish telescope measurements \citep[e.g.,][]{DENBROK21LINES,YAJIMA21LINES}. Despite the scatter, a clear trend can be discerned, and in Table \ref{tab:alphaco} we recommended taking $R_{21} \propto \Sigma_{\rm SFR}^{0.125}$ at kpc-resolution \citep[updated from][]{LEROY22LINES}. This should capture large-scale variations in $R_{21}$. We also recommend a physically-informed bound on $R_{21}$: it should asymptote to $\lesssim 1$ in optically thick, thermally excited gas and should not drop to arbitrarily low values. Lacking a better prescription, we suggest to take $R_{31} = \left< R_{32} \right> \times R_{21} \left( \Sigma_{\rm SFR} \right)$, i.e., to use the $R_{21}$ prescription with a typical $\langle R_{32} \rangle \approx 0.5$. More observations of $R_{32}$ and $R_{31}$ are needed.

Physically, changes in the CO line ratios reflect the opacity, kinetic temperature, and density of the gas. These same factors affect $\alpha_{\rm CO}$ and in principle these line ratios represent a highly observable way to access these conditions while avoiding biases due to resolution and beam filling. \citet{GONG20XCO} have emphasized this potential, fitting functions to simulated observations that predict $\alpha_{\rm CO}$ based on $R_{21}$. The challenges of calibration and a small dynamic range in the ratio remain, but this direction has promise. To capture opacity in addition to excitation, approaches involving a $^{13}$CO line may have even more promise. 

The converse point also holds: the conditions that alter the line ratios also affect $\alpha_{\rm CO}^{1-0}$, though they will affect $\alpha_{\rm CO}^{2-1}$ and $\alpha_{\rm CO}^{3-2}$ even more strongly. Therefore, whenever variations of CO line ratios become significant and necessary to account for, so do variations of $\alpha_{\rm CO}^{1-0}$.

\subsection{Surface density, line width, pressure, and dynamical state}
\label{sec:cloudprops}

We now have cloud-scale CO mapping at $50{-}150$~pc resolution for $\sim 100$ galaxies that sample the star-forming main sequence over the stellar mass range $\rm log(M_{\star}/M_{\odot}) \approx 9.75{-}10.75$ \citep{PHANGSALMA21SURVEY}. The resulting view establishes a basic statistical picture of molecular gas in galaxies on these scales and shows how the gas varies as a function of environment.

\subsubsection{Surface density $\Sigma_{\rm mol}$, line width $\sigma_{\rm mol}$, and internal pressure $P_{\rm int}$}

\begin{table}[t!]
\tabcolsep7.5pt
\caption{Cloud-scale molecular gas properties}
\label{tab:cloudprops}
\begin{center}
\begin{tabular}{|l|c|c|c|c|l|}
\hline
Region & $\log_{10} \Sigma_{\rm mol}^{cloud}$ & $\log_{10} \sigma_{\rm mol}^{cloud}$ & $\log_{10} P_{\rm int}^{cloud}$ & $\log_{10} \alpha_{\rm vir}^{cloud}$ & Ref. \\
& (M$_\odot$~pc$^{-2}$) & (km~s$^{-1}$) & ($k_B$ cm$^{-3}$~K) & & \\
\hline
Disk Galaxies  & $1.3 \pm 0.9 \pm 2.9$ & $0.7 \pm 0.4 \pm 1.7$ & $4.3 \pm 1.6 \pm 6.0$ & $0.6 \pm 0.6 \pm 1.9$ & (1) \\
Spiral arms & $1.4 \pm 0.8 \pm 2.2$ & $0.7 \pm 0.4 \pm 1.3$ & $4.3 \pm 1.5 \pm 4.5$ & $0.5 \pm 0.6 \pm 1.6$ & (1) \\
Inter-arm & $1.2 \pm 0.7 \pm 2.2$ & $0.6 \pm 0.3 \pm 1.2$ & $4.0 \pm 1.3 \pm 4.2$ & $0.5 \pm 0.6 \pm 1.7$ & (1) \\
Centers & $2.3 \pm 1.4 \pm 3.4$ & $1.3 \pm 0.7 \pm 1.7$ & $6.3 \pm 2.6 \pm 6.5$ & $0.7 \pm 0.8 \pm 2.1$ & (1) \\
Merging galaxies & $2.3 \pm 0.9 \pm 2.5$ & $1.1 \pm 0.8 \pm 2.0$ & $6.4 \pm 2.4 \pm 5.8$ & $0.8 \pm 1.0 \pm 3.2$ & (2) \\
Early types & $1.7 \pm 1.6 \pm 3.1$ & $0.9 \pm 1.3 \pm 2.7$ & $5.1 \pm 3.9 \pm 7.8$ & $0.5 \pm 1.3 \pm 3.7$ & (3) \\
\hline
\end{tabular}
\end{center}
\begin{tabnote}
\textsc{Notes} --- Values report median $\log_{10}$ value. The first error bar encompasses $68\%$ of the data, i.e., the $1\sigma$ scatter. The second encompasses $\pm 99.7\%$ of the data, i.e., the $3\sigma$ scatter. References: (1) \citet{SUN20GMCS} based on PHANGS--ALMA at $\theta = 150$~pc; (2) \citet{BRUNETTI21GMCS} and Brunetti et al. (submitted) at $\theta = 80$~pc (Antennae) and $\theta=55$~pc (NGC 3256); (3) \citet{WILLIAMS23GMCS} at $\theta=120$~pc based on WISDOM. Compiled values use $\alpha_{\rm CO}$ as reported in the primary sources, which accounts for metallicity variations ($f (Z)$) in (1) but does not take the ``starburst'' term ($g (\Sigma_\star)$) into account. Further homogenization of these measurements remains an important next step.
\end{tabnote}
\end{table}

\begin{figure}[t!]
\begin{center}
\includegraphics[width=1.0 \textwidth,angle=0]{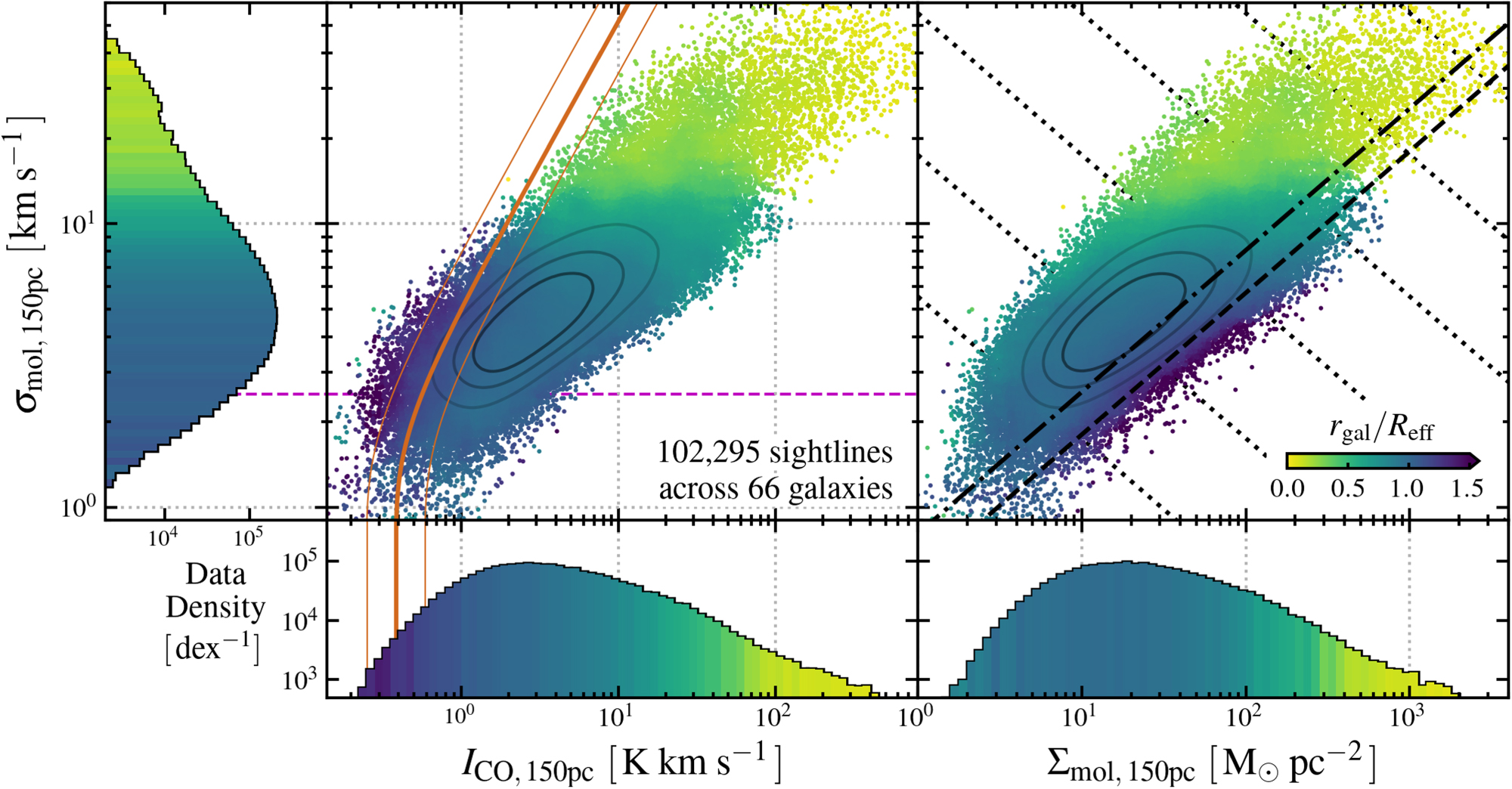}
\caption{
Cloud-scale surface density and line width across the local galaxy population, from \citet{SUN20GMCS}. \textit{Left:} Cloud-scale line width, $\sigma_{\rm mol, 150pc}$, as a function of CO~(2-1) intensity. \textit{Right:} $\sigma_{\rm mol, 150pc}$ as a function of cloud-scale $\Sigma_{\rm mol, 150pc}$. Both panels show $\sim 100,000$ independent cloud-scale sight lines across $66$ star-forming disk galaxies with measurements at fixed $150$~pc scale. Orange lines in the left panel indicate the sensitivity limits of the observations. Dashed and dashed-dotted lines show the \textit{Heyer-Keto relation} expected for clouds with fixed dynamical state (Eq. \ref{eq:heyerketo} with $\alpha_{\rm vir}=1$ and $2$). Dotted lines from top left to bottom right show isobars, indicating fixed internal pressure $P_{\rm int}$ at $P_{\rm int} / k_B = 10^3$, $10^4$, ..., $10^8$~cm$^{-3}$~K following Eq. \ref{eq:isobar}. Both panels illustrate a wide range of $\Sigma_{\rm mol}$, $\sigma_{\rm mol}$, and $P_{\rm int}$ with a narrower range of dynamical state, and the coloration by galactocentric radius illustrates that these are systematic variations.
\label{fig:heyerketo}
}
\end{center}
\end{figure}

Fig. \ref{fig:heyerketo} shows $\sigma_{\rm mol}$ as a function of $\Sigma_{\rm mol}$ at a fixed $\theta=150$~pc resolution for $66$ star-forming disk galaxies \citep[from][]{SUN20GMCS}. Table \ref{tab:cloudprops} reports the distribution of these properties. The data definitively support the \citet{HEYER09GMCS} revision to the original fixed-$\Sigma_{\rm mol}$ picture of GMCs \citep{LARSON81GMCS}. The surface density of molecular gas at fixed scale clearly varies, both within and among environments and from galaxy-to-galaxy. So does the line width. 

Fig. \ref{fig:heyerketo} also shows that $\Sigma_{\rm mol}$ and $\sigma_{\rm mol}$ correlate, lying approximately along the locus of $\sigma_{\rm mol} \propto \Sigma_{\rm mol}^{0.5}$ at fixed size scale as expected for clouds with fixed dynamical state, $\alpha_{\rm vir}$ (\S \ref{sec:heyerketo}, Eq. \ref{eq:heyerketo}). Most of the gas in local disk galaxies thus at least approximately obeys the \textit{Heyer-Keto relation} expected for gas bound by self-gravity with approximately fixed $\alpha_{\rm vir}$, i.e., an approximately fixed ratio between kinetic energy ($K$) and potential energy ($U$). Because both the geometry within the resolution element (and so the gravitational potential) and $\alpha_{\rm CO}$ have significant uncertainty, the absolute value of $\alpha_{\rm vir}$ also has significant uncertainty. As a result, the measurements in Fig. \ref{fig:heyerketo} and Table \ref{tab:cloudprops} do not strongly distinguish between freely collapsing clouds with $U \approx K$ and virialized clouds with $U \approx 2 K$ \citep[][]{BALLESTEROS11GMCS}. However, it would be surprising to see such a strong, widespread correlation between $\Sigma_{\rm mol}$ and $\sigma_{\rm mol}$ if self-gravity were not in approximate balance with kinetic energy, $U \sim K$, and our best estimate is that $\alpha_{\rm vir} \sim 1{-}2$ within a modest factor.

While the dynamical state, $\alpha_{\rm vir}$, appears approximately fixed, Fig. \ref{fig:heyerketo} and Table \ref{tab:cloudprops} show dramatic variations in the internal pressure of the molecular gas, $P_{\rm int} \sim \rho_{\rm mol} \sigma_{\rm mol}^2$. In \citet{SUN20GMCS}, weighting each 150~pc resolution element equally, the full sample shows median $P_{\rm int} / k_B = 1.8 \times 10^4$~K~cm$^{-3}$, a $1\sigma$ range of 1.6~dex and a $3\sigma$ range of 6~dex. In fact a reasonable description of Fig. \ref{fig:heyerketo} is that the gas shows a principal component of variation aligned with $P_{\rm int}$ and little variation along the orthogonal $\alpha_{\rm vir}$ axis.

This picture agrees well with the one that has emerged in the Milky Way, especially following \citet{HEYER09GMCS}, in which molecular cloud properties vary and reflect their environment \citep[e.g.,][]{FIELD11GMCS}. We have emphasized the results based on structure-agnostic statistics, but note that object-focused analyses yield results in close agreement with this picture \citep[e.g.,][]{HUGHES13GMCS,LEROY15GMCS,ROSOLOWSKY21GMCS}.

\begin{figure}[t!]
\begin{center}
\includegraphics[width=1.0\textwidth,angle=0]{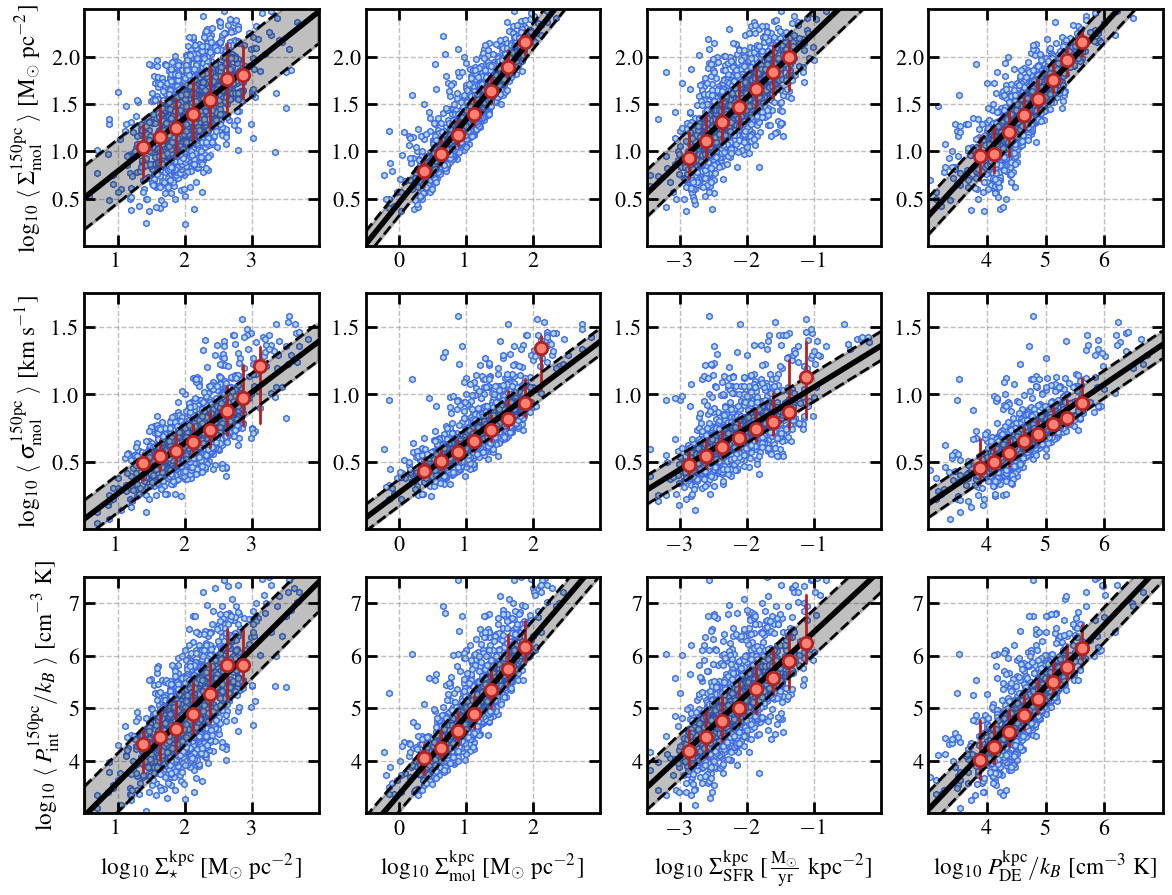}
\caption{Dependence of mean cloud-scale gas properties on environment. Each panel shows the correlation between mass-weighted mean cloud-scale gas properties measured in 1.5~kpc apertures from \citet{SUN22CLOUDS} and mean properties in the galaxy disk at $1.5$~kpc resolution. The mean cloud-scale $\langle \Sigma_{\rm mol}^{\rm 150pc} \rangle$, $\langle \sigma_{\rm mol}^{\rm 150pc} \rangle$, and $\langle P_{\rm int}^{\rm 150pc} \rangle$ (from top to bottom) show substantial systematic variation from region to region. These variations correlate with the structure of the galaxy disk (from left to right), traced by stellar surface density $\Sigma_{\star}^{\rm kpc}$, low resolution molecular gas surface density $\Sigma_{\rm mol}^{\rm kpc}$, the star formation rate surface density $\Sigma_{\rm SFR}^{\rm kpc}$, and the dynamical equilibrium pressure $P_{\rm DE}^{\rm kpc}$. See \citet{SUN22CLOUDS} for details on low-resolution $\Sigma_{\rm SFR}$ and $\Sigma_\star$ estimates, which follow conventional methods. Here blue points show results for individual 1.5~kpc apertures, red are binned averages. The black lines and shaded region show fits and robust scatter, which we provide in Table \ref{tab:cloudscaling}. 
}
\label{fig:cloudscaleenvironment}
\end{center}
\end{figure}

\begin{table}[t!]
\tabcolsep7.5pt
\caption{Cloud-scale gas properties as a function of local environment}
\label{tab:cloudscaling}
\begin{center}
\begin{tabular}{|c|c|c|c|c|c|l|}
\hline
$Y$ & $X$ & $m$ & $b$ & $\sigma$ & Range in $\log_{10} (X)$ \\
\hline
\multicolumn{6}{c}{Based on 1.5~kpc hexagonal apertures from \citet{SUN22CLOUDS}} \\
\multicolumn{6}{c}{Cloud-scale surface density $\langle \Sigma_{\rm mol}^{150pc} \rangle$ [M$_\odot$~pc$^{-2}$]} \\
\hline
$\langle \Sigma_{\rm mol}^{\rm 150pc} \rangle$ & $\Sigma_\star^{\rm kpc}$ & 0.56 & 0.23 & 0.34 & $1.25$ to $3.0$\\
$\langle \Sigma_{\rm mol}^{\rm 150pc} \rangle$ & $\Sigma_{\rm mol}^{\rm kpc}$ & 0.88 & 0.46 & 0.14 & $0.25$ to $2.0$ \\
$\langle \Sigma_{\rm mol}^{\rm 150pc} \rangle$ & $\Sigma_{\rm SFR}^{\rm kpc}$ & 0.68 & 2.94 & 0.25 & $-3.0$ to $-1.24$\\
$\langle \Sigma_{\rm mol}^{\rm 150pc} \rangle$ & $P_{\rm DE}^{\rm kpc}/k_B^{\rm kpc}$ & 0.67 & -1.71 & 0.20 & $3.75$ to $5.75$ \\
\hline
\multicolumn{6}{c}{Cloud-scale velocity dispersion $\langle \sigma_{\rm mol}^{\rm 150pc} \rangle$ [km~s$^{-1}$]} \\
\hline
$\langle \sigma_{\rm mol}^{\rm 150pc} \rangle$ & $\Sigma_\star^{\rm kpc}$ & 0.38 & $-0.11$ & $0.14$ & 1.25 to 3.25 \\
$\langle \sigma_{\rm mol}^{\rm 150pc} \rangle$ & $\Sigma_{\rm mol}^{\rm kpc}$ & $0.38$ & $0.27$ & 0.10 & 0.25 to 2.25 \\
$\langle \sigma_{\rm mol}^{\rm 150pc} \rangle$ & $\Sigma_{\rm SFR}^{\rm kpc}$ & 0.31 & 1.36 & 0.11 & -3.0 to -1.0 \\
$\langle \sigma_{\rm mol}^{\rm 150pc} \rangle$ & $P_{\rm DE}^{\rm kpc}/k_B$ & 0.30 & -0.70 & 0.10 & 3.75 to 5.75 \\
\hline
\multicolumn{6}{c}{Cloud internal pressure $\langle P_{\rm int}^{\rm 150pc}/k_B \rangle$ [cm$^{-3}$ K]} \\
\hline
$\langle P_{\rm int}^{\rm 150pc}/k_B \rangle$ & $\Sigma_\star^{\rm kpc}$ & 1.27 & 2.31 & 0.56 & 1.25 to 3.0 \\
$\langle P_{\rm int}^{\rm 150pc}/k_B \rangle$ & $\Sigma_{\rm mol}^{\rm kpc}$ & 1.50 & 3.36 & 0.32 & 0.25 to 2.0 \\
$\langle P_{\rm int}^{\rm 150pc}/k_B \rangle$ & $\Sigma_{\rm SFR}^{\rm kpc}$ & 1.19 & 7.64 & 0.43 & -3.0 to -1.0 \\
$\langle P_{\rm int}^{\rm 150pc}/k_B \rangle$ & $P_{\rm DE}^{\rm kpc}/k_B$ & 1.17 & -0.45 & 0.34 & 3.75 to 5.75 \\
\hline
\end{tabular}
\end{center}
\begin{tabnote}
\textsc{Notes} --- Fits consider all regions and span a range where the \citet{SUN22CLOUDS} database has at least 20 regions per 0.25~dex bin (see Fig. \ref{fig:cloudscaleenvironment}). All cloud-scale gas properties are measured at $150$\,pc resolution. Results are qualitatively similar but the sample sizes are smaller at $90$\,pc and $60$\,pc resolution \citep{SUN18CLOUDS,SUN22CLOUDS}. The format of the equations is $log_{10}(Y)=m \times log_{10}(X) + b$ with $\sigma$ as scatter from the fit.
\end{tabnote}
\end{table}

\subsubsection{Variations in cloud-scale gas properties as a function of galactic environment}

The wide range of $\Sigma_{\rm mol}^{\rm 150pc}$, $\sigma_{\rm mol}^{\rm 150pc}$, $P_{\rm int}^{\rm 150pc}$ seen across galaxies at 150~pc resolution (Fig.\,\ref{fig:heyerketo}) directly correlates with the larger scale environment in the galaxies (Fig.\,\ref{fig:cloudscaleenvironment}). Table \ref{tab:cloudscaling} reports scaling relations linking the mass-weighted mean cloud-scale properties to the larger scale galactic environment, as characterized by the $\sim 1.5$~kpc scale surface densities of stellar mass $\Sigma_{\star}^{\rm kpc}$, molecular gas mass $\Sigma_{\rm mol}^{\rm kpc}$, and star formation rate $\Sigma_{\rm SFR}^{\rm kpc}$, as well as the dynamical equilibrium pressure $P_{\rm DE}^{\rm kpc}$ (see Eq. \ref{eq:pde} and surrounding). These results, from \citet{SUN22CLOUDS} all show a clear link between the conditions in the molecular gas at cloud scales and the larger scale structure of galaxy disks. They agree well with earlier work \citep[e.g.,][]{HUGHES13GMCS,SCHRUBA19GMCS} and all have the sense that the molecular gas has, on average, higher $\Sigma_{\rm mol}^{\rm 150pc}$, $\sigma_{\rm mol}^{\rm 150pc}$, and $P_{\rm int}^{\rm 150pc}$ in environments that have high kpc-resolution $\Sigma_{\rm mol}^{\rm kpc}$, $\Sigma_{\star}^{\rm kpc}$, and high $P_{\rm DE}^{\rm kpc}$. These high surface density conditions also tend to occur in more massive, more actively star-forming galaxies. As a result, these cloud-scale gas properties also vary as a function of the integrated host galaxy properties.

These correlations do not have to exist \textit{a priori}. For example, variations in the $1.5$~kpc-resolution $\Sigma_{\rm mol}^{\rm kpc}$ could be entirely explained by changes in the filling factor of gas with fixed cloud-scale $\Sigma_{\rm  mol}^{\rm 150pc}$. \textbf{This cross-scale correlation argues that the cloud-scale $\Sigma_{\rm mol}^{\rm 150pc}$, $\sigma_{\rm mol}^{\rm 150pc}$, and $P_{\rm int}^{\rm 150pc}$ couple to, and perhaps descend from, the disk structure in important ways.} In this regard, we highlight the correlation between the cloud internal pressure $P_{\rm int}^{\rm 150pc} \sim \rho_{\rm mol} \sigma_{\rm mol}^2$ and the $1.5$~kpc scale dynamical equilibrium pressure required to support the disk, $P_{\rm DE}^{\rm kpc}$ (see panel containing Eq. \ref{eq:pde}). Large-scale variations in the distribution of gas and stars in galaxies lead $P_{\rm DE}^{\rm kpc}$ to vary by orders of magnitude. The cloud-scale $P_{\rm int}^{\rm 150pc}$ tracks these large-scale $P_{\rm DE}^{\rm kpc}$ variations but with the gas at cloud scales over-pressured relative to the large-scale mean by a factor of $\sim 2{-}3$, even before taking into account the potentially important contribution of magnetic fields. This might result naturally if the mean ISM self-regulates to $P_{\rm DE}^{\rm kpc}$ and then the dense molecular clouds represent modest over-pressurizations, e.g., due to gravitational collapse or dynamical mechanisms that collect and compress the gas. 

\textbf{Regardless of the physical reason, a key result is that we now can predict cloud-scale gas properties from large-scale conditions in disk galaxies.}

\begin{figure}[t!]
\begin{center}
\includegraphics[width=1.0 \textwidth,angle=0]{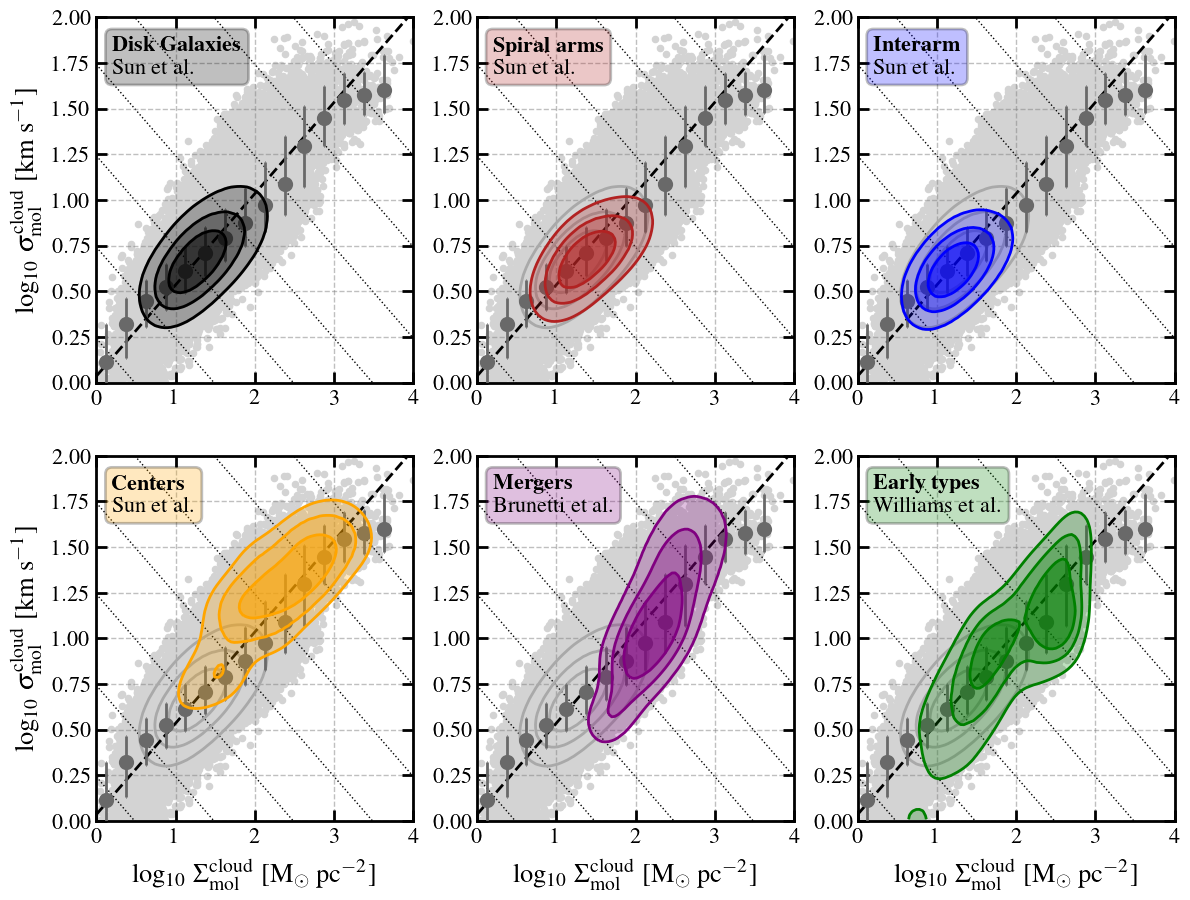}
\caption{Cloud-scale gas properties as a function of morphological environment. Each gray point represents an independent sight line at 150~pc resolution from \citet{SUN20GMCS}. The dashed black line corresponds to $U \approx K$ (i.e., $\alpha_{\rm vir} = 2$ with $f = \ln 2$ in Eq. \ref{eq:heyerketo}) under the assumption of spherical geometry with the size scale equal to the resolution element. The dotted gray lines show fixed $P_{\rm int}^{\rm cloud}$ as in Fig. \ref{fig:heyerketo}. Shaded colored contours show mass-weighted data density specifically for sight lines (from top left to bottom right) in disk galaxies, spiral arms, inter-arm regions, galaxy centers, merging galaxies, and early-type galaxies. Table \ref{tab:cloudprops} reports the corresponding quantitative distributions.
}
\label{fig:cloudmorph}
\end{center}
\end{figure}

\subsubsection{Spiral arms, centers, early-type galaxies, and interacting galaxies}
\label{sec:cloudmorph}

In addition to correlations with parameterized local conditions, cloud-scale gas properties vary between different dynamical environments (Fig.\,\ref{fig:cloudmorph}). Table \ref{tab:cloudprops} reports results for spiral arms, inter-arm regions, galaxy centers, mergers, and early-type galaxies. Each of these distinct environments shows different distributions of $\Sigma_{\rm mol}^{\rm 150pc}$, $\sigma_{\rm mol}^{\rm 150pc}$, $P_{\rm int}^{\rm 150pc}$, and sometimes $\alpha_{\rm vir}^{\rm 150pc}$. 

\smallskip
\textbf{Spiral arms --} Spiral arms are often striking in cloud-scale CO maps. In PHANGS--ALMA arms are associated with both concentration of gas, i.e., a higher filling factor, and higher cloud-scale $\Sigma_{\rm mol}^{\rm 150pc}$ and $P_{\rm int}^{\rm 150pc}$ \citep{SUN20GMCS,querejeta2021}. As Table \ref{tab:cloudprops} and Fig. \ref{fig:cloudmorph} show, gas in spiral arms has on average $\approx 1.5\times$ higher $\Sigma_{\rm mol}^{\rm 150pc}$ compared to that in inter-arm regions. Spiral arms also exhibit modestly larger line widths, $\sigma_{\rm mol}^{\rm 150pc}$ and $\sim2\times$ higher $P_{\rm int}^{\rm 150pc}$. \citet{MEIDT21ARMS} suggest that these arm/inter-arm contrasts in $\Sigma_{\rm mol}^{\rm 150pc}$ reflect spiral shocks that amplify the subtler arm/inter-arm contrasts seen in the underlying stellar potential; the contrasts observed in $\Sigma_{\rm mol}^{\rm cloud}$ correlate with those observed in $\Sigma_\star$ but typically have $\sim 3{-}4\times$ larger amplitude.

\smallskip
\textbf{Galaxy centers --} The population of high $\Sigma_{\rm mol}^{\rm 150pc}$, high $\sigma_{\rm mol}^{\rm 150pc}$ sight lines in Fig.\,\ref{fig:heyerketo}
arises mainly from the central regions of strongly barred galaxies (Fig.\,\ref{fig:cloudmorph}). The centers have high $\Sigma_{\rm mol}^{\rm 150pc}$, on average $\sim 10\times$ higher than in galaxy disks and often reaching $100{-}1000$~M$_\odot$~pc$^{-2}$, high $\sigma_{\rm mol}^{\rm 150pc} > 10$~km~s$^{-1}$, and correspondingly high $P_{\rm int}^{\rm 150pc}$, median $P_{\rm int}^{\rm 150pc}/k_B > 10^6{-}10^8$~cm$^{-3}$~K, almost $100\times$ that typical for galaxy disks. They also show high HCN/CO ratios (see \S\,\ref{sec:denseresults}), indicative of high physical densities. At $\sim 150$~pc resolution, the gas in these central regions also appears less bound by self-gravity compared to the gas in galaxy disks; $\langle \alpha_{\rm vir} \rangle \sim 5$ in centers and $\sim 3$ in disks. Taken at face value, this indicates that the gas in these regions may exist in a more extended phase bound by the combination of self-gravity and the stellar potential well. The existence of a diffuse but high column density phase in galactic centers is also supported by higher resolution observations, though we caution that the line profiles within these central regions can be complex, and the interpretation of line widths requires more subtleties than in extended galaxy disks (see discussion in \S \ref{sec:centers}). Above we discussed evidence that the broad line widths contribute to lower opacity in the CO and a likely lower $\alpha_{\rm CO}$ in these regions, which are not yet taken into account in the data that we compile. The effect of lower $\alpha_{\rm CO}$ will be a somewhat lower $\Sigma_{\rm mol}^{\rm 150 pc}$ while not affecting $\sigma_{\rm mol}^{\rm 150 pc}$, and so to further elevate $\alpha_{\rm vir}$ in these central regions.

\smallskip
\textbf{Merging galaxies --}
Merger-driven starbursts also produce high cloud-scale $\Sigma_{\rm mol}^{\rm cloud}$ and $\sigma_{\rm mol}^{\rm cloud}$, in some ways similar to galaxy centers. Data are more limited for these types of galaxies, but \citet[][\textcolor{red}{Brunetti et al. submitted}]{BRUNETTI21GMCS} measured cloud-scale gas properties for two of the closest gas-rich mergers, the Antennae (NGC\,4038/9) and NGC\,3256. Their results (see Fig.\,\ref{fig:cloudmorph} and Table\,\ref{tab:cloudprops}) show high $\Sigma_{\rm mol}^{\rm cloud}$, $\sigma_{\rm mol}^{\rm cloud}$, and $P_{\rm int}^{\rm cloud}$ in both systems, with most of the gas exhibiting conditions comparable to the extreme galaxy centers. The dynamical state of the gas appears more varied: $\alpha_{\rm vir}$ in the Antennae appears mostly similar to that in disk galaxies, while gas in NGC~3256 shows elevated $\alpha_{\rm vir}$. \citet[][Brunetti et al. submitted]{BRUNETTI21GMCS} suggest that this reflects the more advanced merger stage of NGC~3256, with stellar feedback causing the gas to be less bound by self-gravity.

\smallskip
\textbf{Early-type galaxies --}
\citet{WILLIAMS23GMCS} recently extended this picture to early-type galaxies using the WISDOM survey \citep{DAVIS22GMCS}. They observe gas with a wide range of $\Sigma_{\rm mol}$ and $\sigma_{\rm mol}$, including many regions with high values similar to gas in galaxy centers. The estimated internal pressures, $P_{\rm int}$, are also correspondingly high. Analyzing an overlapping set of systems, \citet{DAVIS22GMCS} show the gas to have a smoother and more diffuse morphology compared to star-forming disk galaxies, likely associated with the deeper stellar potential wells in the central regions of early-type galaxies (see also \S\,\ref{sec:centers_activity}).

\begin{textbox}[ht]
\section{Dynamical equilibrium pressure}
\label{sec:pde}

Galaxy disks exist in approximate vertical dynamical equilibrium, with the weight of the gas supported against collapse, likely often by pressure that is maintained by stellar feedback. From observables and assumptions about galactic structure, one can estimate the weight of gas in the galactic potential and infer the midplane pressure required to maintain equilibrium \citep[][]{ELMEGREEN89PRESS}. This ``dynamical equilibrium'' pressure ($P_{\rm DE}$) is a local quantity, like surface density. \citet{OSTRIKER10PRESS} developed a vertical self-regulation model for galaxy disks in which stellar feedback supplies the necessary $P_{\rm DE}$ (see \S \ref{sec:pdefeedback}). Both the approach to estimate $P_{\rm DE}$ and the validity of vertical force balance have been verified in simulations \citep[e.g.,][]{KOYAMA09PRESS,GURVICH20PRESS,OSTRIKER22PRESS}. A formula applicable inside massive disk galaxies at $z=0$ but neglecting dark matter and the self-gravity of clouds is

\begin{equation}
\label{eq:pde}
P_{\rm DE} = \frac{\pi G}{2} \Sigma_{\rm gas}^2 + \Sigma_{\rm gas} \sqrt{2 G \rho_\star} \sigma_{\rm gas,z}
\end{equation}

\noindent where $\Sigma_{\rm gas}$ is the surface density of gas, $\rho_\star$ the volume density of stars near the midplane, $\sigma_{\rm gas, z}$ the vertical velocity dispersion of the gas in the disk, and $G$ the gravitational constant.

The terms in Eq. \ref{eq:pde} can be estimated from observables for a face-on galaxy disk. Both $\Sigma_{\rm gas} = \Sigma_{\rm mol} + \Sigma_{\rm atom}$ and vertical velocity dispersion, $\sigma_{\rm gas,z}$, can be inferred from a mass-weighted combination of CO and \textsc{Hi} mapping. Because the scale dependence and weighting can be complex, a value of $\sigma_{\rm gas, z} \approx 10$~km~s$^{-1}$ is often assumed, in reasonable agreement with \textsc{Hi} velocity dispersion measurements. The midplane stellar density, $\rho_\star (0) = \Sigma_\star / 4 H_\star$ for an isothermal profile with $\rho_\star (z) \propto {\rm sech}^2 \left( z / 2 H_\star \right)$, depends on both the observed stellar mass surface density and an estimate of the stellar scale height, $H_\star$. Observations of edge-on disks suggest characteristic flattening ratios of $R_\star / H_\star \approx 7.3$ with $\approx 0.15$~dex scatter \citep[see][]{SUN20PRESS}. 

The close dependence of $P_{\rm DE}$ on $\Sigma_\star$, which varies dramatically across galaxies, means that correlations with $P_{\rm DE}$ may help explain the frequently noted close correspondence between $\Sigma_{\rm mol}$ and $\Sigma_\star$ \citep[e.g.,][]{YOUNG95SFGAS,REGAN01SFGAS,LEROY08SFGAS,SHI11SFGAS}. Note that similar to Toomre's $Q$ parameter \citep{TOOMRE64DISKS}, $P_{\rm DE}$ requires an assumed geometry, and treating a mixture of phases and observables. These details each merit discussion, but are beyond the scope of this review.
\end{textbox}

\section{DENSE GAS}
\label{sec:densegas}

Both observations and theory assign a central role to gas density in setting the star formation rate per unit gas. The CO imaging discussed in \S \ref{sec:moleculargas} accesses the mean density on cloud-scales, but each cloud harbors a wide range of physical densities. Observations and simulations of turbulent gas suggest a broad log-normal distribution of physical densities \citep[$\sigma \sim 0.9{-}1.4$~dex for Mach numbers $\mathcal{M} \sim 5{-}10$ following][]{PADOAN02TURB} with a power-law distribution at high densities, likely due to the influence of self-gravity. The mean physical density, width of the density distribution, and fraction of gas at high physical densities are all key parameters to understand the star formation process.

Milky Way observations isolate high-column, high-density gas and find a clear connection between recent star formation activity and gas with $A_V \gtrsim 8$~mag and/or $n_{\rm H2} \gtrsim 10^4$~cm$^{-3}$ \citep[e.g.,][]{LADA10DENSE,LADA12DENSE}. These dense substructures tend to be physically small, $\lesssim 1$~pc \citep[e.g.,][]{EVANS20DENSE}, and cannot be resolved with current radio telescopes at the distance to representative sets of nearby galaxies, $d \gtrsim 10{-}20$~Mpc. Instead, extragalactic observations almost always integrate over whole clouds or even whole regions or whole galaxies. As a result, they average together emission from a wide range of physical densities within each resolution element. Given this constraint, multi-species, multi-transition \textbf{density-sensitive spectroscopy} has emerged as the key method to access the gas density distribution in other galaxies. In this approach, one observes lines sensitive to a variety of different excitation conditions, particularly lines with different effective critical densities, and contrasts them to constrain the physical density distribution in the resolution element. The challenge has been that even the brightest high critical density transitions are faint compared to current observational capabilities (see Table\,\ref{tab:lines}). Although the state-of-the-art for these spectroscopic studies lags behind CO mapping, there is now resolved (sub-kpc) galaxy mapping of the brightest high critical density, low-$J$ rotational lines, HCN\,(1-0) and HCO$^{+}$\,(1-0), across several dozen nearby galaxies. Cloud-scale density-sensitive spectroscopy is beginning to emerge, particularly in targeted studies of galaxy centers, but mostly remains in the regime of case studies.

\subsection{Density-sensitive spectroscopy}
\label{sec:densegasmeaning}

\begin{figure}[t!]
\begin{center}
\includegraphics[width=1.0 \textwidth,angle=0]{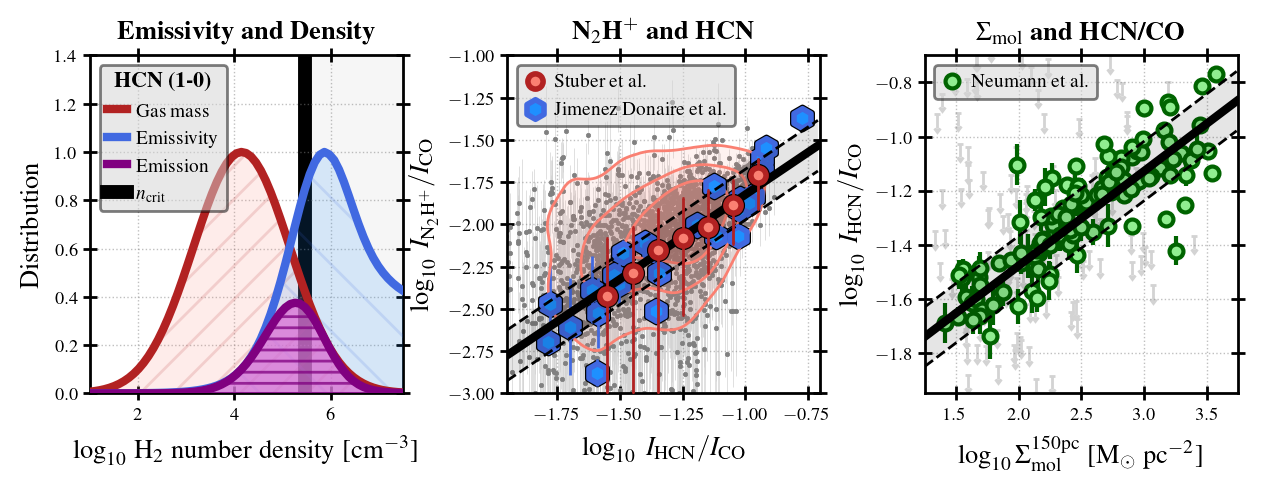}
\caption{Dense gas tracers and gas density. \textit{Left:} Illustration of how the density-dependent emissivity of a transition, here HCN\,(1-0) in blue, convolves with the physical gas density distribution, in red, to produce emission from a range of densities, in purple \citep[following][]{LEROY17DENSE}. Because HCN and similar lines often sample a steeply falling part of the density distribution, a sizeable fraction of  HCN emission often arises from gas below the critical density (black vertical line). A bulk gas tracer like CO would be sensitive to almost the full range of physical densities here. \textit{Middle}: Correlation between HCN~(1-0)/CO~(1-0) and N$_2$H$^+$~(1-0)/CO~(1-0) for regions within galaxies. The blue points from \citet{JIMENEZ23DENSE} show regions in NGC 6946 and a literature compilation. The red points show the binned trend from mapping of M51 by \citet{STUBER23DENSE}, with gray dots showing individual M51 sight lines and red contours indicating data density. N$_2$H$^+$ emerges primarily from cold, dense regions where CO freezes out, but is too faint to be surveyed in a wide range of extragalactic systems. The good correspondence shows that the brightest, most accessible extragalactic dense gas tracer, HCN, yields consistent results with N$_2$H$^+$. \textit{Right}: Correlation between HCN/CO, a spectroscopic tracer of gas density, and the mean molecular gas surface density at 150~pc scales inferred from high resolution CO imaging \citep[from][]{NEUMANN23DENSE}. The excellent correspondence between these two extragalactic tracers of density supports the interpretation of HCN/CO as a density-sensitive line ratio and demonstrates a close link between cloud-scale mean density and physical density.
}
\label{fig:densetracing}
\end{center}
\end{figure}

Density-sensitive spectroscopy involves contrasting dense gas tracers, lines that emit preferentially at high physical density, e.g., transitions of HCN, HCO$^+$, or CS (see Table \ref{tab:lines}) with bulk gas tracers, lines that emit effectively across most densities found in molecular clouds (see Fig.\,\ref{fig:densetracing}). The emissivity of a transition, i.e., the luminosity per unit total gas mass, increases with increasing collider density below $n_{\rm crit}$ and then plateaus well above $n_{\rm crit}$. Rotational transitions with high Einstein $A$ coefficients and modest optical depth, like those in Table \ref{tab:lines}, have high $n_{\rm crit}$ whereas bulk gas tracers like the low-$J$ CO lines\footnote{Radiative trapping due to the high optical depth of the CO lines lowers the effective $n_{\rm crit}$ to below the typical physical density of most molecular gas \citep[e.g.,][]{BOLATTO13REVIEW}.} have low effective $n_{\rm crit}$, and most molecular gas emits effectively in these transitions. 
Observing both a dense gas tracer and a bulk gas tracer and then measuring their ratio offers one of our most direct observational probes of the physical density distribution.

A key point when interpreting these ratios is that dense gas tracers do emit in gas below their effective critical densities, they simply emit less effectively. As a result, a significant fraction of the integrated emission from dense gas tracers can arise from subthermal excitation in lower density gas (see left panel of Fig.\,\ref{fig:densetracing}). Though the emissivity of HCN is decreased in low-density gas, this same lower density gas tends to also be much more abundant than high-density gas. These two effects compete, and a wide range of gas densities can produce dense gas tracer emission. The importance of this widespread subthermal emission has been emphasized in discussions of model calculations \citep[e.g.,][]{LEROY17DENSE}, numerical simulations \citep[e.g.,][]{ONUS18DENSE,JONES23DENSE}, and Milky Way studies that resolve individual clouds \citep[e.g.,][]{KAUFFMANN17DENSE,PETY17DENSE}.

The presence of widespread and variable subthermal excitation means that most dense gas tracers do not uniquely trace gas above a specific density threshold. Instead, the density of the gas that produces the emission from any particular line will always reflect the convolution of the density-dependent emissivity with the distribution of gas densities within the beam. Despite this caveat, it is important to emphasize, that \textbf{because the emissivity of dense gas tracers depends on density below $n_{\rm crit}$, line ratios like HCN/CO are highly sensitive to the physical density distribution within the beam}. That is, the line ratios contrasting dense gas tracers with bulk gas tracers are density-sensitive, even if a single luminosity-to-dense gas mass conversion factor does not apply. Given the difficulty of directly accessing gas density in external galactic disks, the need for a more nuanced interpretation of such measurements seems like a modest concern compared to the promise offered by density-sensitive spectroscopy.

Variables beyond density also control dense gas tracer emission. Both theoretically and in observations of local molecular clouds, temperature influences the emission of these lines 
\citep[e.g.,][]{TAFALLA21LINES,TAFALLA23LINES,PETY17DENSE}. The optical depth of each transition also affects the effective critical density, with higher optical depth lowering the effective density needed for excitation \citep[e.g.,][]{JIMENEZ17DENSE}. Moreover, though the collision partner for dense gas tracers is typically taken to be H$_2$, collisions with electrons do efficiently excite high dipole moment molecules \citep[][]{FAURE07LINES}. Electrons could dominate the collisional excitation of HCN at ionization fractions above $x_e \gtrsim 10^{-6}$ \citep[][]{WALTER17DENSE,GOLDSMITH17DENSE}. Finally, the abundance of each molecule relative to H$_2$ is critical to quantitative modeling, and abundance effects tied to AGN feedback and photodissociation region structure have been invoked to explain line ratio variations among dense gas tracers \citep[e.g.,][]{GRACIACARPIO08DENSE}. Both the absolute abundances and environmental variations of these molecules remain weakly constrained, but projects like the ALMA line survey ALCHEMI \citep{ALCHEMI21SURVEY} promise to improve the situation.

\subsubsection{Evidence that density-sensitive line ratios do trace physical density} 

Despite the caveats above, there is good empirical evidence that line ratios like HCN/CO and HCO$^+$/CO do track the physical density distribution. First, these line ratios correlate with the mean surface density, $\Sigma_{\rm mol}$, measured from cloud-scale imaging of CO. The cloud-scale CO imaging captures the mean density of the molecular gas at $\sim 10{-}200$~pc scales, which offers an observational tracer of the mean volume density (see \S \ref{sec:cloudmethods} and text box near Eq. \ref{eq:hmol}). \citet{GALLAGHER18HCNCLOUD} and \citet{NEUMANN23DENSE} showed that in the disks of $29$ star-forming disk galaxies, the HCN/CO ratio correlates well with the mass-weighted mean $\Sigma_{\rm mol}^{\rm cloud}$ derived from cloud-scale CO imaging (Fig.\,\ref{fig:densetracing} right panel). \citet{TAFALLA23LINES} show that almost the same relation holds within local molecular clouds. Thus, these two methods of tracing density in extragalactic molecular gas yield consistent results.

Second, the density sensitive line ratios accessible to extragalactic observations, especially HCN/CO, show clear correlations and nearly fixed ratios with the N$_2$H$^+$/CO ratio. Following \citet{BERGIN07REVIEW}, N$_2$H$^+$ has been viewed in Galactic work as a gold-standard to identify dense, cold molecular material because it only becomes abundant once CO is heavily depleted onto grains and because the emission correlates well with high extinction regions in local clouds \citep[e.g.,][]{FERNANDEZLOPEZ14DENSE}. On larger scales, the typical ratio between N$_2$H$^+$~(1-0) and CO~(1-0) is $\sim 1:275$, making extensive surveys of N$_2$H$^+$ prohibitive with current instruments. However, recently both \citet{JIMENEZ23DENSE} and \citet{STUBER23DENSE} conducted pioneering case studies in nearby spiral galaxies. They show that, where N$_2$H$^+$ has been detected in the disk, the intensity of N$_2$H$^+$ and HCN trace one another well with a ratio of $\sim 0.15$. They also show that the density-sensitive ratios HCN/CO and N$_2$H$^+$/CO correlate approximately linearly over a $\gtrsim 1$ dex range (Fig. \ref{fig:densetracing} middle panel). Thus, on moderate ($\gtrsim 100$\,pc) scales in other galaxies, the brighter dense gas tracers do appear to track the fainter, more selective ``chemically-motivated'' tracers used in Milky Way work, though spatial separation may arise in resolved cloud observations.

Thus, for physical reasons we expect density variations to drive line ratio variations and diverse tracers of gas density correlate well. Of course, many additional factors can affect excitation and bedevil the inference of molecular column densities \citep[see][]{MANGUM15LINES}. These may prevent these lines from ever being truly precise indicators of the gas at a specific density, but current evidence strongly suggests that HCN/CO, HCO$^+$/CO, and CN/CO are useful first-order tools to assess density variations in the gas of other galaxies.

\subsubsection{Dense gas conversion factors} 
\label{sec:denseconversion}

Dense gas tracers capture emission from a wide range of densities and their luminosity will depend on factors including temperature and abundance. Conversion factors that translate from line luminosity to mass of gas above a specific density should therefore be treated with caution, and we re-emphasize our recommendation to primarily utilize line ratios, like HCN/CO, as first-order density tracers. Nonetheless, the ratio of dense gas mass to luminosity (and its scatter from location to location) can be useful to build physical intuition and make order of magnitude estimates and we therefore note some current estimates.

Most dense gas conversion factors frame their results around HCN~(1-0), defining $\alpha_{\rm HCN} \equiv M_{\rm dense} / L_{\rm HCN}^\prime$ with $L_{\rm HCN}^\prime$ the luminosity of HCN~(1-0) in standard K~km~s$^{-1}$~pc$^2$ units and $M_{\rm dense}$ the mass of molecular gas in regions where the H$_2$ volume density exceeds some threshold $n_{\rm thresh}$. Many recent studies adopt $n_{\rm thresh} = 10^4$~cm$^{-3}$, which corresponds to $A_V \sim 8$~mag, $N_H \sim 8 \times 10^{21}$~cm$^{-2}$, and $\Sigma_{\rm mol} \sim 160$~M$_\odot$~pc$^{-2}$ in local clouds. The formalism mirrors that used for the CO-to-H$_2$ conversion factor, $\alpha_{\rm CO}$ \citep[e.g.,][]{BOLATTO13REVIEW} and the resulting $\alpha_{\rm HCN}$ has units of \acounits . (See also \textbf{Supplemental Material}.)

If one simply wants a number, our assessment is that the current literature supports a median $\alpha_{\rm HCN} \approx 15$~\acounits\ for $n_{\rm thresh} \approx 10^4$~cm$^{-3}$, but with significant region-to-region variations spanning $\alpha_{\rm HCN} \approx 3{-}30$~\acounits\ and with even higher values in some cases \citep[e.g.,][and see especially the compilation of results in \citealt{BARNES20DENSE}]{GARCIABURILLO12DENSE,ONUS18DENSE,JONES23DENSE}. Lower values of $\alpha_{\rm HCN}$ are more common in dense environments, starburst environments and high-mass star-forming clouds, whereas higher values appear more common when considering small scales and specific, dense sub-regions.

\begin{table}[t!]
\tabcolsep7.5pt
\caption{Selected rotational lines tracing molecular gas}
\label{tab:lines}
\begin{center}
\begin{tabular}{|l|c|c|c|}
\hline
Species & Transition & CO~(1-0) ratio$^{\rm a}$ & $n_{\rm crit}$$^{\rm b}$ \\
& & (K units) & (cm$^{-3}$) \\
\hline
$^{12}$CO$^{\rm c}$ & (1-0) & $1.0$ & $5.7 \times 10^2$ \\
$^{12}$CO$^{\rm c}$ & (2-1) & $0.65$ & $4.4 \times 10^3$ \\
$^{12}$CO$^{\rm c}$ & (3-2) & $0.31$ & $1.5 \times 10^4$ \\
$^{13}$CO & (1-0) & $0.091$ & $4.8 \times 10^2$ \\
C$^{18}$O & (1-0) & $0.015$ & $4.8 \times 10^2$ \\
HCN & (1-0) & $0.024$ & $3.0 \times 10^5$ \\
HCO$^+$ & (1-0) & $0.0081$ & $4.5 \times 10^4$ \\
CS & (2-1) & $0.023$ & $8.5 \times 10^4$ \\
HNC & (1-0) & $0.011$ & $1.1 \times 10^5$ \\
CN & ($1_{3/2}$-$0_{1/2}$)$^{\rm d}$ & $0.022$ & $2.4 \times 10^5$ \\
N$_2$H$^+$ & (1-0) & $0.0036$ & $4.1 \times 10^4$ \\
\hline
\end{tabular}
\end{center}
\begin{tabnote}
$^{\rm a}$ Typical ratio of the line in question to $^{12}$CO\,(1-0) in recent local disk galaxy surveys. This offers a first-order typical line ratio relevant for $z=0$ massive ($M_\star \gtrsim 10^{10}$~M$_\odot$) main sequence star-forming galaxies, but these ratios all vary significantly with environment within and among galaxies (see Figs. \ref{fig:alphaco}, \ref{fig:densetracing}, \ref{fig:densehcnco}). \\
$^{\rm b}$ Critical density of colliding $n_{\rm H2}$ for the optically thin case at $T_{\rm kin} = 20$~K with no background radiation, following \citet{SHIRLEY15LINES}. Because of the competing impacts of the density distribution and the density-sensitive emissivity, much emission arises from lower densities. See text and Fig. \ref{fig:densetracing}. \\
$^{\rm c}$ In practice, the effective critical density will be reduced by a factor $\sim 1/ (1 + \tau )$ relative to the nominal $n_{\rm crit}$. This effect will be strongest for $^{12}$CO and weaker for its optically thinner isotopologues $^{13}$CO and C$^{18}$O. \\
$^{\rm d}$ Values for the stronger hyperfine grouping near $\nu_0 = 113.5$~GHz.\\
\textsc{References}: For line ratios \textbf{CO\,(2-1)} and \textbf{CO\,(3-2)} --- \citet{LEROY22LINES}, and see \citet{DENBROK21LINES}, \citet{YAJIMA21LINES}, \citet{LAMPERTI20LINES}, and Fig. \ref{fig:alphaco} ; 
\textbf{$^{13}$CO} --- \citet{cormier2018}, and see \citet{morokuma-matsui2020} and \citet{cao2023}; 
\textbf{C$^{18}$O} --- C$^{18}$O/$^{13}$CO from \citet{JIMENEZ17COLINES} scaled by typical $^{12}$CO/$^{13}$CO quoted here;
\textbf{HCN, HCO$^+$, CS~(2-1)} --- calculated weighting all $\theta = 2.1$~kpc lines of sight in ALMOND \citep{NEUMANN23DENSE} by CO luminosity. See also \citet{JIMENEZ19DENSE}, \citet{USERO15DENSE}, \citet{GALLAGHER18DENSE}, and Fig. \ref{fig:densehcnco}; 
\textbf{HNC} --- \citet{JIMENEZ19DENSE};
\textbf{CN} --- CN/HCN from \citet{WILSON18LINES,WILSON23LINES} scaled by the typical HCN/CO ratio; 
\textbf{N$_2$H$^+$} --- N$_2$H$^+$/HCN ratio from \citet{JIMENEZ23DENSE} scaled by the typical HCN/CO quoted here, see also \citet{STUBER23DENSE} and Fig. \ref{fig:densetracing}.  References for collisional rate coefficients used for critical density calculations: \textbf{CO} --- \citet{YANG10COLLIDE}; \textbf{HCN, HNC} --- \citep{DUMOUCHEL10COLLIDE}; \textbf{HCO$^{+}$} --- \citep{DENISALPIZAR20COLLIDE}; \textbf{CS} --- \citep{DENISAPLIZAR18COLLIDE}; \textbf{CN} --- \citep{LIQUE10COLLIDE}; \textbf{N$_2$H$^+$} --- \citep{FLOWER99COLLIDE} extrapolated from HCO$^+$. \\
\end{tabnote}
\end{table}

\begin{table}[t!]
\tabcolsep7.5pt
\caption{Dense gas scaling relations}
\label{tab:densescaling}
\begin{center}
\begin{tabular}{|c|c|c|c|c|c|l|}
\hline
$Y$ & $X$ & $m$ & $b$ & $\sigma$ & Range in $\log_{10} (X)$\\
\hline
\multicolumn{6}{c}{Gao-Solomon relation} \\
\hline
$L_{\rm TIR}$ & $L_{\rm HCN}$ & $1$$^{\rm a}$ & $2.84$ & $0.47$ & 1.0 to 9.0 \\
${\rm SFR}$~$^{\rm b}$ & $L_{\rm HCN}$ & $1$ & $-7.00$ & $0.47$ & 1.0 to 9.0 \\
$\tau_{\rm dep}^{\rm dense}$~$^{\rm c}$ & --- & --- & $8.17$ & --- & --- \\
$\epsilon_{\rm ff}^{\rm dense}$~$^{\rm c}$ & --- & --- & $-2.53$ & --- & --- \\
\hline
\multicolumn{6}{c}{HCN/CO and environment} \\
\multicolumn{6}{c}{(combining \citet{JIMENEZ19DENSE} and \citet{NEUMANN23DENSE}; see Fig. \ref{fig:densehcnco})} \\
\hline
$L_{\rm HCN}/L_{\rm CO 1-0}$ & $\Sigma_{\star}^{\rm kpc}$ & $0.52$ & $-2.92$ & $0.14$ & 1.7 to 3.3 \\
$L_{\rm HCN}/L_{\rm CO 1-0}$ & $\Sigma_{\rm mol}^{\rm kpc}$ & $0.45$ & $-2.40$ & $0.14$ & 0.5 to 2.6 \\
$L_{\rm HCN}/L_{\rm CO 1-0}$ & $P_{\rm DE}^{\rm kpc}/k_B$ & $0.33$ & $-3.45$ & $0.16$ & 4.2 to 6.6 \\
\hline
\multicolumn{6}{c}{HCN/CO and cloud-scale mean properties} \\
\multicolumn{6}{c}{(\citet{NEUMANN23DENSE}; see Fig. \ref{fig:densetracing})} \\
\hline
$L_{\rm HCN}/L_{\rm CO (2-1)}$ & $\left< \Sigma_{\rm mol}^{\rm 150pc} \right>$ & $0.41$ & $-2.51$ & $\approx 0.15$ & $1.5{-}3.5$\\
$L_{\rm HCN}/L_{\rm CO (2-1)}$ & $\left< P_{\rm turb}^{\rm 150pc} \right>$ & $0.19$ & $-1.5$ & $\approx 0.15$ & $5{-}8.5$ \\
\hline
\multicolumn{6}{c}{SFR/HCN and environment} \\
\multicolumn{6}{c}{(combining \citet{JIMENEZ19DENSE} and \citet{NEUMANN23DENSE}; see Fig. \ref{fig:densehcnco})} \\
\hline
${\rm SFR}/L_{\rm HCN}$ & $\Sigma_{\star}^{\rm kpc}$ & $-0.36$ & $-6.01$ & $0.16$ & 1.7 to 3.3 \\
${\rm SFR}/L_{\rm HCN}$ & $\Sigma_{\rm mol}^{\rm kpc}$ & $-0.31$ & $-6.37$ & $0.17$ & 0.5 to 2.6 \\
${\rm SFR}/L_{\rm HCN}$ & $P_{\rm DE}^{\rm kpc}/k_B$ & $-0.23$ & $-5.65$ & $0.16$ & 4.2 to 6.6 \\
\hline
\end{tabular}
\end{center}
\begin{tabnote}
\textsc{Notes} --- All equations have the format $\rm log_{10}(Y)=m\times log_{10}(X)+b$ with $\sigma$ as scatter from the fit. All line luminosities $L$ are reported in [K~km~s$^{-1}$~pc$^2$]; $L_{\rm TIR}$ in [$L_\odot$]; SFR in [M$_\odot$~yr$^{-1}$]; depletion times in [yr]; $\Sigma_\star$ and $\Sigma_{\rm mol}$ in [M$_\odot$~pc$^{-2}$]; and $P_{\rm DE}/k_B$ in [cm$^{-3}$~K]. \\
$^{\rm a}$ For the \textit{Gao-Solomon relation} we report ratios and assume the slope to be one (\S \ref{sec:denseresults}, Fig. \ref{fig:gaosolomon}). \\
$^{\rm b}$ Converting from IR luminosity to equivalent SFR and vice versa following \citet{MURPHY11SFR}. For resolved galaxy mapping and integrated galaxy measurements, continuous star formation and fully sampled IMF should be good assumptions, but these do break down on a local level (\S \ref{sec:SF}). \\
$^{\rm c}$ Assuming $\alpha_{\rm HCN} = 15$~\acounits\ and $n_{\rm H2} \sim 10^4$~cm$^{-3}$ and converting the ratio reported in the previous line (but see \S \ref{sec:denseconversion} for discussion and variations).
\end{tabnote}
\end{table}

\subsection{Gas density, star formation, and environment}
\label{sec:denseresults}

\begin{figure}
\begin{center}
\includegraphics[width=1.0 \textwidth,angle=0]{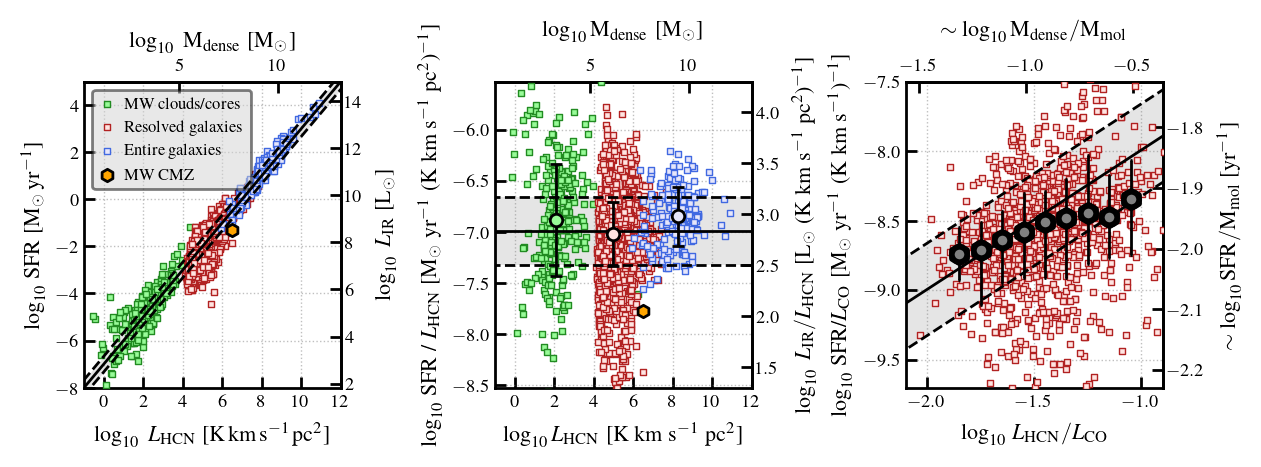}
\caption{Star formation and dense gas tracers. In all panels, blue points show whole galaxies, red points regions at kpc to cloud-scale regions within galaxies, and green points individual Milky Way clouds or clumps. Also in all panels the black line shows a linear scaling with a fixed SFR-to-HCN ratio with the shaded region and dashed lines indicating the $1\sigma$ scatter about that value found considering all data (see Table \ref{tab:densescaling}). \textit{Left:} The \textit{Gao-Solomon relation} showing star formation rate (SFR) or equivalent IR luminosity ($L_{\rm IR}$) as a function of the luminosity of the dense gas tracer HCN\,(1-0). \textit{Middle:} IR-to-HCN or SFR-to-HCN ratio as a function of integrated HCN luminosity for the data in the left panel. The approximately constant ratio between SFR tracers and tracers of high-density molecular gas across a wide range of scales and systems have motivated the popular \textit{Gao-Solomon hypothesis} that star formation proceeds in a universal way in dense gas. \textit{Right:} SFR/CO, tracing the star formation rate per unit molecular gas as a function of HCN-to-CO, tracing the density of molecular material. The gray circles and black lines show the binned trend and scatter for the resolved galaxy data. SFR/CO correlates with HCN/CO, indicating that denser gas appears to form stars more efficiently. However the spread in the data reflects significant systematic changes in SFR/HCN (Fig. \ref{fig:densehcnco}). Plots based on \citet{JIMENEZ19DENSE} using an expanded literature sample including that compiled by \citet{NEUMANN23DENSE}. See \textbf{Supplemental Material} for full reference list.
}
\label{fig:gaosolomon}
\end{center}
\end{figure}

As shown by \citet{GAO04DENSE} and \citet{LADA10DENSE,LADA12DENSE}, a single approximately linear relation exists between star formation activity and dense gas tracer emission. This \textit{Gao-Solomon relation} spans a $\gtrsim 10$~dex range in luminosity or mass (Fig.\,\ref{fig:gaosolomon}). Resolved galaxy mapping shows that the relationship also holds over $\sim$\,kpc-sized regions within a diverse sample of galaxies, and recently cloud-scale measurements have also begun to emerge for a few regions in nearby galaxies. These resolved-galaxy and cloud-scale studies fill in the gap between integrated galaxies and Milky Way clouds and cores to form the nearly continuous relation seen in Fig. \ref{fig:gaosolomon} and Table \ref{tab:densescaling}.

Because luminosity scales with area covered and distance to a target when all other things are equal, this impressive correlation mostly results from the wide range of distances, galaxy masses, emission filling factors, and physical areas sampled among the datasets shown. The key physics lies in the ratio of star formation activity to dense gas tracers. We report these ratios in Table \ref{tab:densescaling} as ratios of IR-to-HCN luminosity and SFR-to-HCN luminosity. Assuming a reasonable but uncertain $\alpha_{\rm HCN} \sim 15$~\acounits\ and a characteristic density of $n_{\rm H2}^{\rm dense} \sim 10^4$~cm$^{-3}$, we also report a dense gas depletion time, $\tau_{\rm dep}^{\rm dense} \approx 130$~Myr, and a dense gas star formation efficiency per free-fall time, $\epsilon_{\rm ff}^{\rm dense} \sim 0.3\%$. Despite the uncertainty in $\alpha_{\rm HCN}$, the time for star formation to consume the available dense gas reservoir, $\tau_{\rm dep}^{\rm dense}$, is clearly much shorter than the time to consume all of the available molecular gas, $\tau_{\rm dep}^{\rm mol} \approx 1{-}2$~Gyr (for details see \S\,\ref{sec:sfe}). On the other hand $\epsilon_{\rm ff}^{\rm dense}$ is of the same order as the $\epsilon_{\rm ff} \approx 0.5_{-0.3}^{+0.7}\%$ found for the overall molecular gas reservoir. 

Thus in disk galaxies the SFR per unit $M_{\rm dense}$ is significantly higher than the SFR per unit $M_{\rm mol}$ but the efficiency of star formation relative to gravitational collapse is similarly low in the dense and total molecular material. This low $\epsilon_{\rm ff}^{\rm dense}$ could imply that star formation proceeds slowly within individual dense structures. Alternatively if a single dense core evolves over a free-fall time, $\sim \tau_{\rm ff} \sim 5 \times 10^5$~yr, then the bulk of the dense material must be dispersed, e.g., by stellar feedback, rather than incorporated into stars to produce such a low $\epsilon_{\rm ff}^{\rm dense}$. \citet{MCKEE07REVIEW} point out that for a log-normal density distribution and typical GMC conditions, $M_{\rm dense}/M_{\rm mol} \sim \tau_{\rm ff}^{\rm dense} / \tau_{\rm ff}^{\rm mol}$ so that for fixed SFR, $\epsilon_{\rm ff}^{\rm dense} \sim \epsilon_{\rm ff}^{\rm cloud}$, making this result a natural prediction of turbulence-regulated star formation theory.

If the Gao-Solomon hypothesis of a fixed density threshold for star formation holds, then the SFR per unit total $M_{\rm mol}$, ${\rm SFR} / M_{\rm mol} = (\tau_{\rm dep}^{\rm mol})^{-1}$, should depend primarily on the physical gas density distribution, traced by HCN/CO. The right panel in Fig. \ref{fig:gaosolomon} shows this SFR/CO vs. HCN/CO relationship, which removes the ``more is more'' scaling from the $L_{\rm IR}{-}L_{\rm HCN}$ plot. As shown by \citet{GAO04DENSE}, there is a correlation with the sense that denser, high HCN/CO, systems show higher SFR per $M_{\rm mol}$. That is, in broad strokes stars do form more efficiently in regions where the gas appears denser. However, Fig. \ref{fig:gaosolomon} also displays considerable scatter in SFR/$M_{\rm mol}$ at fixed HCN/CO. This implies a significant variation in the star formation rate per unit dense gas tracer, SFR/HCN. In Fig.\,\ref{fig:gaosolomon} and Table \ref{tab:densescaling}, SFR/HCN has a $\pm 0.47$~dex scatter across a wide range of scales and datasets.
 
Thus, dense gas tracers and SFR tracers correlate well across a broad range of systems, and the SFR per unit $M_{\rm mol}$ appears higher when the physical gas density appears higher. But there are significant variations in both HCN/CO, tracing the physical gas density distribution, and SFR/HCN, tracing the apparent star formation rate per unit dense gas tracer. Recent resolved galaxy mapping of dense gas tracers demonstrates that \textbf{the variations in both HCN/CO and SFR/HCN are systematic in nature and appear directly linked to local environment}, in good agreement with expectations for turbulent clouds.

\subsubsection{Gas density and local environment in galaxy disks}
\label{sec:hcnco}

\begin{figure}[t!]
\begin{center}
\includegraphics[width=1.0 \textwidth,angle=0]{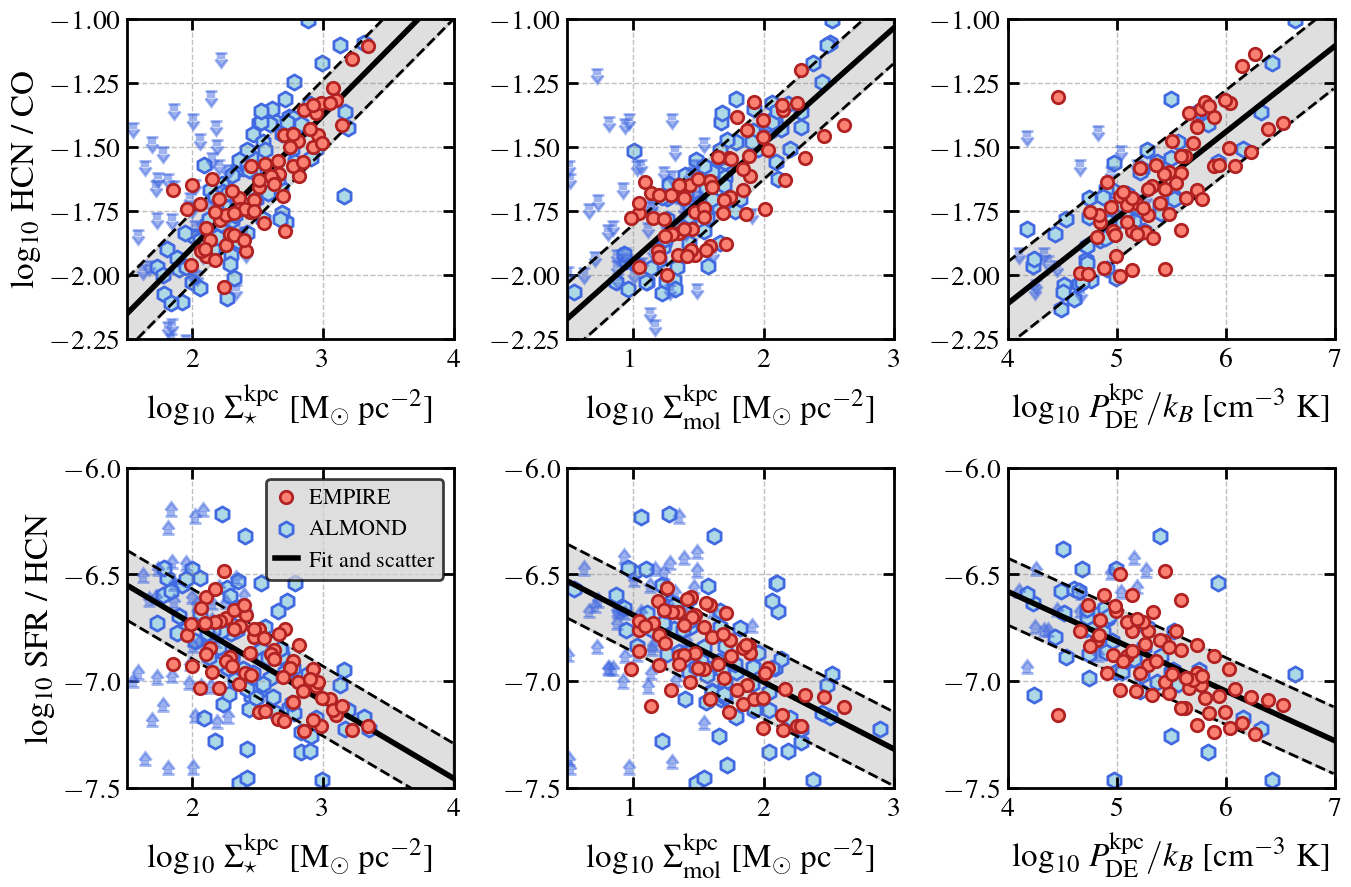}
\caption{Gas density, star formation, and environment from resolved galaxy HCN mapping by EMPIRE \citep{JIMENEZ19DENSE} and ALMOND \citep{NEUMANN23DENSE}. \textit{Top row:} HCN/CO, tracing physical gas density, as a function of tracers of kpc-scale disk structure: stellar surface density $\Sigma_\star^{\rm kpc}$, molecular gas surface density, $\Sigma_{\rm mol}^{\rm kpc}$, and dynamical equilibrium pressure, $P_{\rm DE}^{\rm kpc}$. HCN/CO increases with increasing surface density and $P_{\rm DE}^{\rm kpc}$. \textit{Bottom row:} SFR/HCN, tracing the apparent SFR per dense gas tracer. SFR/HCN declines with increasing surface density and pressure, becoming lower in the high-surface density, high-pressure inner parts of galaxies. Each point in these panels corresponds to a stacked result in one galaxy, conducted at $\approx 2$~kpc resolution. Fits and scatter $\sigma$ (shown as black solid and dashed lines) are reported in Table \ref{tab:densescaling}.
}
\label{fig:densehcnco}
\end{center}
\end{figure}

A main output of recent resolved galaxy surveys is that we have our first high completeness view of how the physical gas density distribution, traced by HCN/CO and similar ratios, varies across the disks of normal galaxies \citep[][]{USERO15DENSE,GALLAGHER18DENSE}. Fig. \ref{fig:densehcnco} and Table \ref{tab:densescaling} summarize the main trends from the two largest resolved galaxy HCN surveys, EMPIRE \citep{JIMENEZ19DENSE} and ALMOND \citep{NEUMANN23DENSE}. The table reports correlations with both $\sim$ kpc scale disk structure, traced by $\Sigma_\star^{\rm kpc}$, $\Sigma_{\rm mol}^{\rm kpc}$, and $P_{\rm DE}^{\rm kpc}$, and correlations with region-averaged cloud-scale gas properties estimated from CO (discussed in \S \ref{sec:moleculargas}).

HCN/CO shows a clear dependence on galaxy environment, becoming higher in regions with high $\Sigma_\star^{\rm kpc}$, high $\Sigma_{\rm mol}^{\rm kpc}$, and high $P_{\rm DE}^{\rm kpc}$. These high-surface density, high-$P_{\rm DE}^{\rm kpc}$ regions also tend to be the inner regions of massive galaxies, including galaxy centers. Merging galaxies and starburst galaxies, which are not the focus of this review, extend these results, with high HCN/CO indicating abundant high density gas in merger-driven starburst and U/LIRG galaxies \citep[e.g.,][]{GAO04DENSE,UEDA21DENSE}. 

Spectroscopic tracers of gas density also exhibit a clear connection to the cloud-scale properties of the molecular gas inferred from CO \citep[][Table\,\ref{tab:densescaling}, Fig.\,\ref{fig:densetracing}]{GALLAGHER18HCNCLOUD,NEUMANN23DENSE}. The mean surface density of clouds, $\left< \Sigma_{\rm mol}^{\rm 150pc} \right>$, and the internal pressure in a region, $\left< P_{\rm int}^{\rm 150pc} \right> \sim \rho \sigma^2$, both correlate with line ratios that trace the physical gas density distribution. Note, though, that \citet{BEMIS23DENSE} find that the relationship between HCN/CO and $P_{\rm int}$ becomes more scattered and complex when combining results for merger-driven U/LIRGs and normal galaxy disks than considering only normal disk galaxies.

These results form a picture consistent with the results in \S \ref{sec:moleculargas}: The distribution of physical gas density varies across galaxy disks in a coherent way, so that when the large-scale average $\Sigma_{\rm mol}^{\rm kpc}$ and the gravitational potential of the disk become high, more gas appears dense \citep[see][for an early version of this hypothesis]{HELFER97DENSE}. We saw in \S \ref{sec:moleculargas} that the cloud-scale mean $\langle \Sigma_{\rm mol}^{\rm cloud} \rangle$ and $\langle \sigma_{\rm mol}^{\rm cloud} \rangle$, both obey similar environmental correlations. And indeed, the density-sensitive line ratios correlate directly with the cloud-scale mean gas properties inferred from CO. One reasonable picture linking all these results is that \textbf{large-scale disk structure sets the cloud-scale gas structure, and then the physical gas density distribution reflects this cloud-scale gas structure.} This might occur naturally, e.g., if the physical density distribution reflects turbulence-driven fluctuations about an environmentally determined mean cloud-scale density.

\subsubsection{Star formation activity, dense gas, and total molecular gas} 
\label{sec:sfrhcn}

SFR/HCN exhibits significant scatter of $\sigma \approx \pm 0.47$ or a FWHM range of $\sim 1$ dex (Fig.\,\ref{fig:gaosolomon}) which is partly explained by systematic trends with environment (Table\,\ref{tab:densescaling} and Fig.\,\ref{fig:densehcnco}). In disk galaxies, SFR/HCN appears lower in high $P_{\rm DE}^{\rm kpc}$, high $\Sigma_{\rm mol}^{\rm kpc}$, and high $\Sigma_\star^{\rm kpc}$ environments, especially galaxy centers \citep[][]{USERO15DENSE,GALLAGHER18DENSE,JIMENEZ19DENSE}. This agrees well with results from the Milky Way, where the Central Molecular Zone (CMZ) has the highest mean density and dense gas fraction of any region in the galaxy \citep{BALLY87DENSE}, but both individual CMZ clouds and the region as a whole appear to exhibit significantly lower SFR relative to their $M_{\rm dense}$ compared to Solar Neighborhood molecular clouds \citep[e.g.,][]{LONGMORE13DENSE,henshaw2023}.

The sense of these results is that SFR/HCN drops under the same circumstances that cause both the mean cloud-scale $\Sigma_{\rm mol}^{\rm cloud}$ and HCN/CO to rise. That is, the SFR per unit dense gas emission drops as the mean gas density rises and dense gas becomes more abundant. This situation could arise naturally if the bulk gas becomes effective at emitting HCN as the mean density of the gas reached high values \citep[e.g., as in][]{KRUMHOLZ07DENSE}. Then, \textbf{the dense gas tracer emission will reflect the density distribution in the bulk molecular gas and not pick out only the overdense, self-gravitating, star-forming phase of the gas.} The importance of this distinction between \textit{dense gas} and \textit{self-gravitating gas} is a main conclusion of \citet{BEMIS23DENSE} and such a distinction emerges naturally from models of star formation in turbulent clouds \citep[e.g.,][]{KRUMHOLZ05EFF,FEDERRATH12EFF}.

A picture like this assumes that the mean dynamical state of the gas is approximately virialized regardless of the mean density, so that only the locally overdense material becomes strongly self-gravitating. The results discussed in \S\,\ref{sec:moleculargas} support this view (to a reasonable extent). Milky Way CMZ work on this topic has emphasized the dynamical state of dense clouds, and similarly suggested that dense clouds still appear approximately virialized despite their high density, though the exact correlation between star formation and $\alpha_{\rm vir}$ remains controversial \citep[e.g.,][]{KRUIJSSEN15CENTER,KAUFFMANN17DENSE}. The handful of recent cloud-scale extragalactic HCN studies in the literature reach related conclusions \citep{querejeta2019,BESLIC21DENSE,eibensteiner2022}, that the dense clouds visible in HCN show variable line width, with high line widths associated with low SFR/HCN, suggestive of a link between the dynamical state of dense gas and its ability to form stars.

Finally, though we do not focus on merging galaxies, we note that studying U/LIRGs, \citet{GARCIABURILLO12DENSE} and \citet{UEDA21DENSE} show variations in SFR/HCN with the opposite sense of those described here. The SFR/$L_{dense}$ seems to \textit{increase} with increasing $\Sigma_{\rm mol}^{\rm kpc}$ and $\Sigma_{\rm SFR}^{\rm kpc}$ in merging systems. Conversion factor effects would amplify rather than explain away these variations. The implication is that in the most extreme systems, SFR/$M_{\rm dense}$ rises as the gas becomes very dense, perhaps consistent with the mean gas density becoming very high and $\tau_{\rm dep}^{\rm mol}$ very short overall in these systems.

\section{STAR FORMATION TIMESCALES AND EFFICIENCIES}
\label{sec:timescales}

Cold, dense gas forms stars. The stars represent a key output of the matter cycle, and measuring the timescale and efficiency with which they form illuminates the physics driving the cycle. Tackling these topics requires estimating the amount of recent star formation activity. Furthermore, associating well-understood star formation tracers with gas measurements from high-resolution observations allows inference of the life cycle of star-forming regions via statistical arguments. Meanwhile, a variety of physical timescales can be estimated based on the gas measurements. Contrasting these physical timescales and comparing them to the inferred evolutionary sequence for star forming regions can help to identify what physical processes govern the internal state and evolution of these regions. Contrasting the inferred evolutionary and physical timescales with the time-averaged rate of star formation per unit gas, accessed via large-scale average measurements of $\tau_{\rm dep}$, also reveals the efficiency with which the gas forms stars, with the efficiency per free-fall time, $\epsilon_{\rm ff}$, being of particular theoretical interest.

\begin{figure}
\begin{center}
\includegraphics[width=1.1\textwidth]{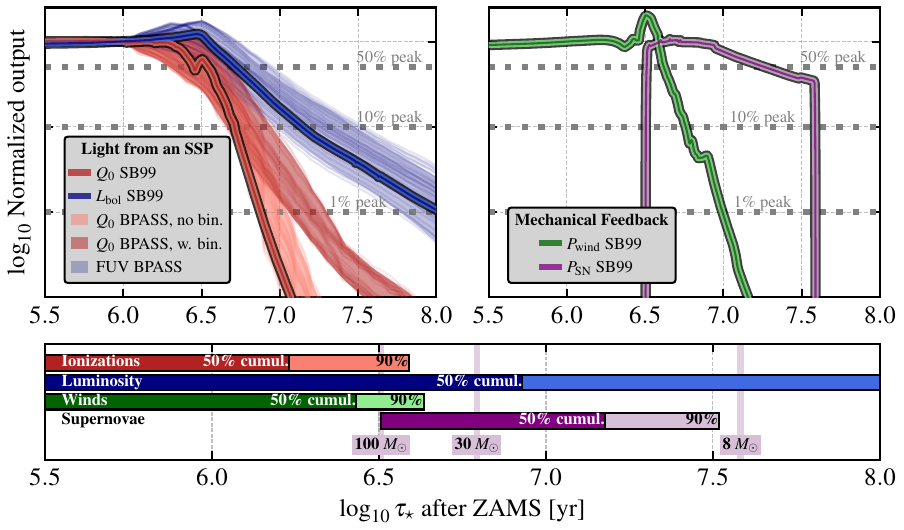}
\end{center}
\caption{Time dependence of emission and mechanical feedback from a simple stellar population (SSP). \textit{Top:} Ionizing photon production $Q_0$ (red), the driver of H$\alpha$ emission, and bolometric luminosity $L_{bol}$ (blue) (left) and power output by stellar winds $P_{\rm wind}$ (green) and supernovae $P_{\rm SN} $(purple) (right) versus time. All curves are normalized in amplitude to their values on the ZAMS (or shortly after the first SN for $P_{\rm SN}$) to emphasize the time behavior. Solid lines are calculated via Starburst99 (SB99), while lighter lines use the BPASS models with and without binaries. \textit{Bottom:} The range of timescales over which 50\% and 90\% of the total emission occurs for each quantity with the time for the first SN explosions of differently massive stars  marked, see also Table \ref{tab:ssptimescale}.
\label{fig:ssptimescale}
}
\end{figure}

\begin{table}[ht]
\tabcolsep7.5pt
\caption{Timescales for emission from simple stellar populations}
\label{tab:ssptimescale}
\begin{center}
\begin{tabular}{|@{}l|c|c|c|c|c@{}|}
\hline
Tracer & Starts & 50\% of peak & 10\% of peak & 50\% cumul. & 90\% cumul. \\ 
 & [Myr] & [Myr] & [Myr] & [Myr] & [Myr] \\ 
\hline
$Q_0$ or $L_{\rm H\alpha}^{\rm corr}$ & 0 & $2.8$ & $5.1$ & $1.7$ & $3.9$ (or $6.9^{\rm a}$) \\
$L_{\rm bol}$ or $L_{\rm MIR}$ & 0 & $5.2$ & $14$ & $8.5$ & $> 100$ \\
$P_{\rm wind}$ & 0 & $4.2$ & $5.8$ & $2.7$ & $4.3$ \\
$P_{\rm SN}$ & $3.2$ (or $6.2^{\rm b}$) & $19$ & $39$ & $15$ & $33$ \\
\hline
\end{tabular}
\end{center}
\begin{tabnote}
$^{\rm a}$ The longer timescale reflects the influence of binaries \citep[BPASS;][]{ELDRIDGE17BPASS}. \\
$^{\rm b}$ The first value assumes all massive stars explode as core-collapse SNe, the second only $< 30$~M$_\odot$ stars explode \citep[see][]{SUKHBOLD16SNE} while the majority of higher mass stars collapse to black holes. \\
\textsc{Notes} --- Timescales for output from a simple stellar population (SSP) mostly using a default Starburst99 model \citep[SB99;][]{STARBURST99,LEITHERER14STARBURST} with a fully populated \citet{KROUPA01IMF} IMF, an upper mass cutoff of $100$~M$_\odot$ and no stellar rotation or binaries (but see Fig. \ref{fig:ssptimescale}). Rows report: H ionizing photon production, $Q_0$, or equivalently extinction corrected H$\alpha$ luminosity, $L_{\rm H\alpha}^{\rm corr}$, assuming gas is present to be ionized; bolometric luminosity, which will closely resemble both $L_{\rm FUV}$ at early times and mid-IR luminosity, $L_{\rm MIR}$ (if dust is present to reprocess the emission); power in stellar winds, $P_{\rm wind}$; and power associated with core-collapse SNe, $P_{\rm SN}$. Columns report times from the ZAMS, $\tau_\star$, for: the output to begin, the output to drop to 50\% and 10\% of its peak value, and for 50\% and 90\% of the total cumulative emission to be emitted (N.B., the cumulative sum is calculated only out to $\tau_\star = 1$~Gyr, which only matters for $L_{\rm bol}$).
\end{tabnote}
\end{table}

\subsection{Inferring star formation activity}
\label{sec:SF}

At large, $\gtrsim$ kpc, scales observations average together many individual regions and as a result a continuous star formation rate (SFR) can be a reasonable approximation. Extensive work has established robust observational approaches to estimate SFR and $\Sigma_{\rm SFR}$ on these scales \citep[see][]{KENNICUTT12REVIEW,SANCHEZ20REVIEW}. At cloud-scales, the idea of an ensemble-average SFR breaks down as observations isolate individual regions in particular evolutionary states. It becomes increasingly appropriate to describe recently formed stars in terms of a single coeval population, often termed a \textit{simple stellar population} (SSP). In this case, star formation activity can be characterized via the mass of stars that has been formed, $M_\star$, and the age of the stellar population, $\tau_\star$, measured relative to the zero age main sequence (ZAMS).

\subsubsection{Approaches to trace star formation}

Beyond the Local Group, tracing star formation still largely depends on observing signatures of massive stars, which are short-lived, hot, and luminous. Three major approaches are widely used. First, outside galaxy centers, production of $h \nu > 13.6$~eV photons is dominated by the most massive, hottest stars. Therefore \textbf{tracking H ionizations} via recombination lines or free-free emission represents perhaps the most unambiguous tracer of recent massive star formation in normal galaxies.

Star-forming regions tend to be dusty. As a result, much of the UV light from massive stars is absorbed by dust and re-emitted as IR emission. \textbf{Observing IR or combined UV and IR continuum emission} therefore also represents a key star formation tracer. This approach is sensitive to lower mass (but still massive) stars than tracing H ionizations.

Alternatively, one can isolate young stellar populations, measure their spectrum or spectral energy distribution (SED) and model these using population synthesis techniques \citep[e.g.,][]{CONROY13REVIEW}. The most common current version of this approach is to \textbf{identify and model the light from young stellar clusters and associations,} though spectroscopy of individual star-forming regions is becoming increasingly common.

\subsubsection{Timescales for star formation tracers} 

Because massive stars are short-lived, their emission evolves rapidly. As a result, star formation tracers each have some associated window for visibility. Table \ref{tab:ssptimescale} summarizes some useful values for commonly used tracers. Ionizing photon production, $Q_0$, and by extension H$\alpha$ luminosity, $L_{\rm H\alpha}$, remain high and nearly constant for $\tau_\star \sim 3$~Myr, then begin to drop rapidly as the most massive stars begin to leave the main sequence. For a population without binaries, $90\%$ of the total emission is emitted before even $\tau_\star \sim5$~Myr, and this extends to only $\sim 7$~Myr when binaries are included. This short timescale and rapid drop illustrate why ionizing photons are often viewed as a gold standard tracer of ongoing or very recent star formation activity.

Both the bolometric luminosity, $L_{\rm bol}$, and non-ionizing far-ultraviolet (FUV) emission also remain initially high and nearly constant for the first few million years. As a result, a SSP emits half of its total FUV emission by $\tau_\star \sim 10$~Myr. After this, emission drops but much more slowly than $Q_0$ drops. With a slope close to $L_\nu \propto \tau^{-1}$, the SSP gives off roughly equal emission per decade of time out to $\gtrsim 100$~Myr. In practice, this means that for continuous star formation, about half the light will come from bright, young ($\lesssim 10$~Myr) regions and half the light from a (more extended, diffuse) set of individually fainter, older ($\lesssim 100$~Myr) regions. This also implies a degeneracy between mass and age when considering only the UV luminosity, so that the precise calibration of UV-based tracers is sensitive to the star formation history \citep[e.g.,][]{BOQUIEN15SFR}.

In practice, ionizing photons and the bolometric luminosity are often observed indirectly via recombination line or dust emission. In these cases gas or dust must be present to reprocess the emission. As we discuss in \S\,\ref{sec:cloudevolution}, observations imply rapid gas clearing times. This may shorten the timescale or increase the spatial scale (e.g., as the reprocessing occurs over a larger region) over which these tracers operate. In practice, this makes empirical validation of the visibility timescale for ISM-based star formation tracers important, e.g., from comparing stellar population modeling to emission for large samples. For ionizing photon tracers the balance of evidence does seem to favor timescales that agree well with the SSP models. HST studies of young clusters find H$\alpha$ emission to disappear by $\tau_\star \sim 4-5$~Myr \citep{HOLLYHEAD15CLUSTERS,HANNON22CLUSTERS} and simulations imply $\lesssim 10$~Myr, and typically $4{-}5$~Myr, timescales for H$\alpha$ \citep[e.g.,][]{FLORESVELAZZQUEZ21SFR,TACHELLA22SFR}. On the other hand, given the longer timescales associated with the UV and bolometric luminosity, the clearing of gas and dust likely affects the visibility time of IR emission.

\subsubsection{Estimating the mass of young stars}

The ionizing photon production, IR luminosity, or SED of an individual region can be used to estimate the mass of recently formed stars. Table \ref{tab:sspnorm} reports relationships relevant to making such estimates for young SSPs.

\begin{table}[ht]
\tabcolsep7.5pt
\caption{Conversions related to star formation activity for single regions}
\label{tab:sspnorm}
\begin{center}
\begin{tabular}{|c|c|c|l|}
\hline
$Y$ & $X$ & C & Description \\ 
\hline
\multicolumn{4}{c}{Ionizing photons and recombination lines assuming Case B$^{\rm a}$} \\
\hline
$Q_0 [{\rm s}^{-1}]$ & $L_{\rm H\alpha}^{\rm corr} [{\rm erg~s}^{-1}]$ & +11.87 & varies by $\pm 0.05$~dex, see \citet{BYLER17SFR} \\
$L ({\rm H}\alpha)$ & $L ({\rm H}\beta)$ & $+0.46$ & 2.86 for $T=10,000$~K; varies with $T$ \\
$L ({\rm H}\alpha)$ & $L ({\rm P}\alpha)$ & +0.93 & 8.51 for $T=10,000$~K; varies with $T$ \\
$L ({\rm H}\alpha)$ & $L ({\rm P}\beta)$ & +1.26 & 18.1 for $T=10,000$~K; varies with $T$ \\
$L ({\rm H}\alpha)$ & $L ({\rm Br}\alpha)$ & +1.56 & 36.1 for $T=10,000$~K; varies with $T$ \\
\hline
\multicolumn{4}{c}{Conversions for an SSP near the zero-age main sequence$^{\rm b}$} \\
\hline
$M_\star~[{\rm M}_\odot]$ & $Q_0 [{\rm s}^{-1}]$ & -46.6 & for $\tau_\star \lesssim 2$~Myr; $+0.5$, $1.4$, $2.7$ dex at $4, 6, 10$ Myr \\
$M_\star~[{\rm M}_\odot]$ & $L_{\rm H\alpha}^{\rm corr} [{\rm erg~s}^{-1}]$ & -34.7 & for $\tau_\star \lesssim 2$~Myr; $+0.5$, $1.4$, $2.7$ dex at $4, 6, 10$ Myr \\
$M_\star~[{\rm M}_\odot]$ & $L_{\rm bol} [{\rm erg~s}^{-1}]$ & -36.5 & for $\tau_\star \lesssim 3$~Myr; $+0.1, 0.4, 0.8$ dex at $4, 6, 10$ Myr \\
$M_\star~[{\rm M}_\odot]$ & $L_{\rm bol} [{\rm L_\odot}]$ & -2.92 & for $\tau_\star \lesssim 3$~Myr; $+0.1, 0.4, 0.8$ dex at $4, 6, 10$ Myr \\
$L_{\rm H\alpha}^{\rm corr}$ & $L_{\rm bol} [{\rm erg~s}^{-1}]$ & -1.8 & or 
$L_{\rm bol} \sim 63 L_{\rm H\alpha}^{\rm corr}$ during the first few Myr\\
$L_{\rm H\alpha}^{\rm corr}$ & $L_{\nu, \rm ff} [{\rm erg~s^{-1}~Hz^{-1}}]$ & $+13.9$ & $+0.45~\log_{10} \left( \frac{T_e}{10^4~{\rm K}}\right) + 0.1 \log_{10} \left(\frac{\nu}{{\rm GHz}}\right)$ $^\textrm{c}$ \\ 
\hline
\multicolumn{4}{c}{Empirical mid-IR star formation activity tracers for JWST filters} \\
\multicolumn{4}{c}{Median empirical conversions from \citet{BELFIORE23SFRJWST}; see there for SFR/M$_\star$ or EW(H$\alpha$) dependence.} \\
\hline
$L_{\rm H\alpha}^{\rm ext} [{\rm erg~s}^{-1}]$ & $\nu L_\nu ({\rm F2100W}) [{\rm erg~s}^{-1}]$ & $-1.51$ & 0.031; median $a_{\rm F2100W}$  \\
$M_\star~[{\rm M}_\odot]$ & $\nu L_\nu ({\rm F2100W}) [{\rm erg~s}^{-1}]$ & $-36.2$ & median $a_{\rm F2100W}$  \\
$L_{\rm H\alpha}^{\rm ext} [{\rm erg~s}^{-1}]$ & $\nu L_\nu ({\rm F1130W}) [{\rm erg~s}^{-1}]$ & $-2.08$ & 0.0084; median $a_{\rm F1130W}$  \\
$M_\star~[{\rm M}_\odot]$ & $\nu L_\nu ({\rm 1130W}) [{\rm erg~s}^{-1}]$ & $-36.8$ & median $a_{\rm F1130W}$  \\
$L_{\rm H\alpha}^{\rm ext} [{\rm erg~s}^{-1}]$ & $\nu L_\nu ({\rm F1000W}) [{\rm erg~s}^{-1}]$ & $-1.56$ & 0.026; median $a_{\rm F1000W}$  \\
$M_\star~[{\rm M}_\odot]$ & $\nu L_\nu ({\rm F1000W}) [{\rm erg~s}^{-1}]$ & $-37.3$ & median $a_{\rm F1000W}$  \\
$L_{\rm H\alpha}^{\rm ext} [{\rm erg~s}^{-1}]$ & $\nu L_\nu ({\rm F770W}) [{\rm erg~s}^{-1}]$ & $-2.14$ & 0.0072; median $a_{\rm F770W}$  \\
$M_\star~[{\rm M}_\odot]$ & $\nu L_\nu ({\rm F770W}) [{\rm erg~s}^{-1}]$ & $-36.8$ & median $a_{\rm F770W}$  \\
$L_{\rm H\alpha}^{\rm ext} [{\rm erg~s}^{-1}]$ & $\nu L_\nu ({\rm F770W}) [{\rm erg~s}^{-1}]$ & $-1.06$ & 0.087; median $a_{\rm F335M}^{\rm PAH}$  \\
$M_\star~[{\rm M}_\odot]$ & $\nu L_\nu ({\rm F335M}^{\rm PAH}) [{\rm erg~s}^{-1}]$ & $-35.7$ & median $a_{\rm F335M}^{\rm PAH}$  \\
\hline
\end{tabular}
\end{center}
\begin{tabnote}
\textsc{Notes--} Useful equations relating star formation tracers to one another and to the mass of recently formed stars, $\mstar$. All equations have the form $\log_{10}(Y)=\log_{10}(X)+C$. \\
$^{\rm a}$ Recombination line ratios from \citet{DRAINE11BOOK} based on \citet{HUMMER87RECOMB}. \\
$^{\rm b}$ These values come from Starburst99 calculations \citep[][]{STARBURST99} that include no stellar rotation, adopt solar metallicity, and use an upper IMF cutoff of 100~M$_\odot$. The most likely systematic variation is that using the equations here might lead to $M_\star$ to be over-estimated by $\sim 0.1{-}0.2$~dex if stellar rotation plays a role or stars much more massive than 120~M$_\odot$ are present. \\
$^{\rm c}$ Free-free luminosity, $L_{\nu , ff}$, to extinction-corrected H$\alpha$ luminosity conversion from \citet{MURPHY11SFR} and \citet{RUBIN68SFR}. Conversion also depends on observed frequency $\nu$ and electron temperature $T_e$.
\end{tabnote}
\end{table}

\smallskip
\textbf{From recombination line emission --} 
For a young SSP with a fully populated IMF, H recombination line or free-free emission from \hii\ regions can be used to infer $Q_0$, and from $Q_0$ the mass of young stars. Table \ref{tab:sspnorm} gives conversions from H$\alpha$ and several other widely used recombination lines to $M_\star$ assuming Case B recombination and electron temperature $T_e = 10,000$~K, but we caution that $T_e$ varies with lower $T_e$ appropriate near Solar metallicity \citep[e.g.,][]{BERG20SFR}. 
These conversions assume no H$\alpha$ extinction or loss of ionizing photons to dust or escape from \textsc{Hii} regions. With modern IFUs it has become common to observe multiple recombination lines. In this case, with an assumed extinction law, Case B recombination, and an estimate of the region temperature one can infer the extinction using approaches like the Balmer decrement \citep[see][]{OSTERBROCK89BOOK}. Radio free-free emission also offers powerful, essentially extinction-free diagnostics of $Q_0$ \citep[e.g.,][]{MURPHY11SFR}.

Hybrid tracers, in which H recombination line tracers are combined with IR tracers, have also become popular \citep[e.g.,][]{CALZETTI07SFR,KENNICUTT12REVIEW}. They are empirically calibrated and shown to work well, but the H recombination line and IR tracers fundamentally have different time and mass sensitivities. The coefficient for the IR portion of the tracer also depends on local and global environment \citep[e.g.,][]{BELFIORE23SFRJWST}. Therefore we suggest to view this as a good option when multiple H recombination lines or free-free measurements are not available. Table \ref{tab:sspnorm} provides recent determinations of coefficients appropriate for JWST bands that have been calibrated to agree with extinction corrections derived from the Balmer decrement for individual \hii\ regions.

\smallskip
\textbf{From mid-IR emission --} 
In most star-forming regions the overwhelming majority of $L_{\rm bol}$ emerges in the IR after being reprocessed by dust. In principle, to capture this emission, one should cover the whole IR range $\lambda \sim 8{-}1000\mu$m. In practice, only JWST can resolve IR emission from individual regions at distances of $d \gtrsim 10$~Mpc. This restricts observations to the mid-IR, $\lambda \lesssim 25\mu$m. Fortunately $\lambda \approx 20{-}40~\mu$m also represents a ``sweet spot'' for tracing star formation activity \citep[e.g.,][]{WHITCOMB23SFR}. The radiation field in star-forming regions is intense enough that bigger grains emit thermally, achieving a high emissivity and emitting a large fraction of $L_{\rm bol}$ at these wavelengths. Meanwhile, at these wavelengths the surrounding diffuse ISM dust emits mainly due to comparatively weaker stochastic heating. Because the mid-IR does not capture the full $L_{\rm bol}$, empirical calibrations represent the best current approach to translate specific mid-IR bands to $M_\star$ and/or $L_{\rm bol}$. Table \ref{tab:sspnorm} reports such median conversions for several JWST bands from \citet{BELFIORE23SFRJWST}.

Below $\lambda \sim 20\mu$m, the situation becomes more complicated, as band and continuum emission from PAHs and very small grains dominate \citep[e.g.,][]{galliano2018}. The abundance of PAHs relative to other dust grains exhibits a strong metallicity dependence \citep[e.g.,][]{LI20DUST} and appears depressed in \textsc{Hii} regions, likely due to preferential destruction of these small, fragile grains \citep[e.g.,][]{CHASTENET23DUST,EGOROV23DUST}. As a result, short wavelength mid-IR emission traces star formation activity but requires more significant environment-dependent corrections to the calibration factors \citep[e.g., see][]{CALZETTI07SFR,BELFIORE23SFRJWST}.

\smallskip
\textbf{Modeling clusters and young stellar populations --}
Once individual star clusters have been identified, their spectra and spectral energy distributions can be modeled to infer $M_\star$ and $\tau_\star$. Recent HST surveys cover from the near-UV to the near-IR \citep{LEGUS15SURVEY,PHANGSHST22SURVEY}, leveraging the bluer (NUV, U, and B) bands to address the age-attenuation degeneracy and the redder bands (V and I) to anchor $M_\star$. Despite this wide wavelength coverage, degeneracies among age, attenuation, and metallicity can confuse old, dust-free objects and young, dusty objects with $10{-}20\%$ of clusters having incorrect ages by a factor of $\gtrsim 10$ \citep[e.g.,][]{whitmore2023b}. The presence or absence of recombination line or dust emission can help break this degeneracy, and as such information becomes more widely available this situation is expected to improve. The latest SED modeling codes even include the ability to model the associated IR emission along with the stellar SED using coupled SSP and physical dust models \citep[e.g.,][]{CIGALE19}. 

Not all stars form in clusters. Estimates of the bound mass fraction at birth vary widely, but in present day disk galaxies, $10{-}25\%$ represents a reasonable current estimate \citep[e.g.,][]{CHANDAR17CLUSTERS,KRUMHOLZ19REVIEW}. The set of recently formed stellar populations can be broadened by identifying unbound young associations that approximately correspond to SSPs. Morphological analysis of blue stellar populations in HST images have been applied to identify such associations and expand our view of recently formed stellar populations beyond only the bound clusters \citep[e.g.,][]{gouliermis2018,larson2023}.

In addition to SED modeling, the spectra of individual young stellar populations are often accessible from IFU mapping. Modeling these spectra also has large potential to reveal $M_\star$ and $\tau_\star$ of individual SSPs and to unravel the recent star formation history of individual regions \citep[e.g.,][]{pessa2023}. In detail, this promise still lies largely in the future because of the need to thoroughly test new SSP libraries that account for very young ages, nebular contributions, and complex underlying continuum emission at cloud-scales. 

\subsubsection{Complications inferring star formation activity at cloud scales}

Isolating the emission associated with an individual region can represent a significant challenge, as both leakage from the region and contamination by extended emission impact the estimation. Moreover, individual regions are not guaranteed to satisfy the sampling assumptions needed for simple conversions from emission to $M_\star$ nor are their ages usually known \textit{a priori}.

\smallskip
\textbf{Diffuse ionized gas and HII region identification --}
Gas-based ionization tracers, including recombination lines and free-free emission, often require assigning emission to individual \textsc{Hii} regions likely to be powered by SSPs. This assignment tends to be morphologically motivated and has many of the same uncertainties as cloud identification based on CO emission (\S \ref{sec:comethods}). Even assuming that the assignment works perfectly, not all $> 13.6$~eV photons ionize gas within \textsc{Hii} regions. Star-forming galaxies also harbor extended reservoirs of diffuse ionized gas (DIG) with lower density but much more mass than \textsc{Hii} regions. In local star-forming galaxies, $40 \pm 20\%$ of the total H$\alpha$ emission arises from the DIG \citep{THILKER02DIG,OEY07DIG,BELFIORE22DIG}. The balance of evidence suggests that this emission mostly reflects ionizing photons leaking from \textsc{Hii} regions with a modest $< 10\%$ contribution from hot evolved stars. Consistent with this, the morphology of the emission resembles extended ``halos'' around \textsc{Hii} regions which are likely powered by escaping photons with typical path length $\sim 1{-}2$ kpc \citep[see][and references therein]{BELFIORE22DIG}. Practically, this means that (1) $\approx 40\%$ of ionizing photons escape from \textsc{Hii} regions and (2) modern instruments detect H$\alpha$ from essentially every line of sight. Correctly quantifying recent star formation activity requires accurately distinguishing \textsc{Hii} regions from diffuse emission and attributing the contribution of the DIG to the appropriate regions. A third source of uncertainty, absorption of ionizing photons by dust, remains poorly studied outside the Galaxy, but within massive star-forming regions in the Milky Way the effect can be significant, up to a $\approx 30\%$ loss of ionizing photons \citep[][]{BINDER18SFR}

\begin{textbox}[ht]
\section{Origin of the infrared cirrus and PAH emission as an ISM tracer}
\label{sec:irtracer}

Outside star-forming regions, the distribution of mid-IR emission closely resembles that of the neutral ISM (e.g., Fig.\,\ref{fig:bubbles}). This mid-IR ``cirrus'' often reflects emission from very small dust grains stochastically heated by the diffuse interstellar radiation field (ISRF). The intensity of such emission varies as

\begin{eqnarray}
\label{eq:ircirrus}
I_\nu &\propto& \Sigma_{\rm gas} \times U \times \frac{D}{G} \left( \times q_{\rm PAH} \right)
\end{eqnarray}

\noindent with $\Sigma_{\rm gas}$ the gas surface density, $U$ the intensity of the ISRF (assumed to have a fixed shape and relatively low intensity), $D/G$ the dust-to-gas mass ratio, and $q_{\rm PAH}$ the abundance of PAHs relative to the overall dust mass (the $q_{\rm PAH}$ term only applies when considering PAH emission).

Away from star-forming regions, the ISRF, $D/G$, and $q_{\rm PAH}$ likely vary weakly while $\Sigma_{\rm gas}$ is highly structured. This leads to a striking resemblance between JWST dust maps and ALMA CO imaging at matched resolution. Quantitatively, CO~(2-1) emission and the diffuse mid-IR emission correlate well. We note the following almost linear correlation between the PAH-dominated F770W filter and CO (2-1) emission tracing molecular gas surface density \citep[updated slightly from][]{LEROY23JWST}:

\begin{eqnarray}
\label{eq:pahtogas}
I_{\rm CO (2-1)}~\left[ {\rm K~km~s}^{-1} \right] &\approx& 0.91~\left( I_{\rm F770W}~\left[ {\rm MJy~sr}^{-1} \right] \right)^{1.06} 
\end{eqnarray}

\noindent Which appears valid between about $I_{\rm 770W} \sim 0.5$~MJy~sr$^{-1}$ and $50$~MJy~sr$^{-1}$. Similar relations hold for the other JWST mid-IR filters, adjusted only moderately to reflect mean band and continuum strength. Even the F2100W emission shows a good correlation with gas outside bright star-forming regions \citep[see][]{LEROY23JWST}.

Though traditionally viewed as a contaminant in SFR measurements, as emphasized by \citet{SANDSTROM23JWST}, the mid-IR cirrus actually represents a major opportunity for ISM studies. The resolution and sensitivity of JWST to ISM emission in the mid-IR make it immediately competitive with ALMA as an ISM-mapping instrument. Over the next decade we expect this to yield a sharp new view of the ISM that contributes to many of the key topics discussed throughout this article.
\end{textbox}

\smallskip
\textbf{Infrared cirrus --} In addition to compact emission from star-forming regions, galaxies exhibit widespread infrared cirrus emission. In the disks of normal star-forming galaxies, about $60{-}70\%$ of the mid-IR emission at both continuum and PAH bands appears diffuse, i.e. originates outside \textsc{Hii} regions \citep[][]{BELFIORE23SFRJWST,PATHAK24MIDIR}. This extended emission emerges because dust is generally well-mixed with the pervasive ISM gas and even relatively weak radiation fields can excite dust enough to generate detectable emission. The resolution and sensitivity of JWST and the direct dependence of this IR cirrus on ISM column density mean that the emission is both a contaminant for tracing star formation activity and a potential powerful tracer of the ISM (see text box near Eq. \ref{eq:ircirrus}).

\smallskip
\textbf{Stochastic sampling of the IMF --} The most massive stars contribute heavily to $Q_0$, rendering ionizing photon tracers particularly sensitive to the effects of stochasticially sampling the IMF. Even if the IMF remains universal on average, stochastic sampling can lead to the case where an individual stellar population does not harbor stars of all masses. This, in turn, will cause the ratios among $M_\star$, $Q_0$, $L_{\rm FUV}$, and $L_{\rm bol}$ to deviate from the predictions of idealized calculations like those in Fig. \ref{fig:ssptimescale}. Simulations suggest that stochasticity primarily affects stellar populations with $M_\star \lesssim 10,000$~M$_\odot$ \citep[e.g.,][]{KRUMHOLZ19REVIEW} and only becomes negligible once $M_\star \gtrsim 50,000$~M$_\odot$ \citep{FOUESNEAU10SFR,POPESCU12SFR}. For lower masses, individual $M_\star$ and $\tau_{\star}$ estimates have thus large uncertainties. IFU observations of \textsc{Hii} regions and HST studies frequently access individual regions or clusters with mass $\sim 10^2{-}10^3$~M$_\odot$, so that stochasticity represents an important consideration in this regime \citep[e.g.,][]{JUNG23IMF}. Fortunately, statistical tools have emerged that can treat populations of regions and account for stochastic sampling and a range of discrete ages for individual regions \citep[e.g.,][]{DASILVA14SFR,KRUMHOLZ15SFR}. However, any individual estimate will still have a significant systematic uncertainty.

Beyond only stochasticity, the maximum stellar mass formed may depend on the properties of the parent cloud or another environmental factor. Moreover, the simplifying assumption that all stars in a population form simultaneously may break down, especially in large, complex regions. In such a scenario, the visibility timescales for ionizing photon tracers and UV or mid-IR emission can be extended compared to Fig. \ref{fig:ssptimescale} and Table \ref{tab:ssptimescale}, which can also lead to more complexity when modeling the SEDs of regions. 

\subsection{Time evolution of molecular clouds and star-forming regions}
\label{sec:cloudevolution}

At resolution $\theta \lesssim 200$\,pc, a spatial de-correlation between tracers of molecular gas and star formation becomes evident \citep[][see Fig. \ref{fig:cloudlifetimes_methods}]{SCHRUBA10SFGAS}, i.e., \ha\ and CO emission show distinct distributions \citep[e.g., Fig. \ref{fig:sketch_cloud};][]{KRECKEL18SFR}. This has been widely interpreted to reflect the visible evolution of star-forming regions, e.g., from molecular cloud to \textsc{Hii} region to exposed star cluster (\S \ref{sec:intro_SF} and Fig. \ref{fig:sketch_cloud}). Here we review the observational inference of timescales associated with molecular cloud evolution based on such data. We focus on two main approaches, which both require cloud-scale data: (a) statistical arguments based on comparison of tracers sensitive to different phases of the evolution of a star-forming region, and (b) kinematic arguments that take advantage of large-scale dynamics or gas flows. 

\subsubsection{Required Resolution, Assumptions, and Methods}
\label{sec:cloudevolution_methods}

\begin{figure}[t]
\includegraphics[width=1.0\textwidth]{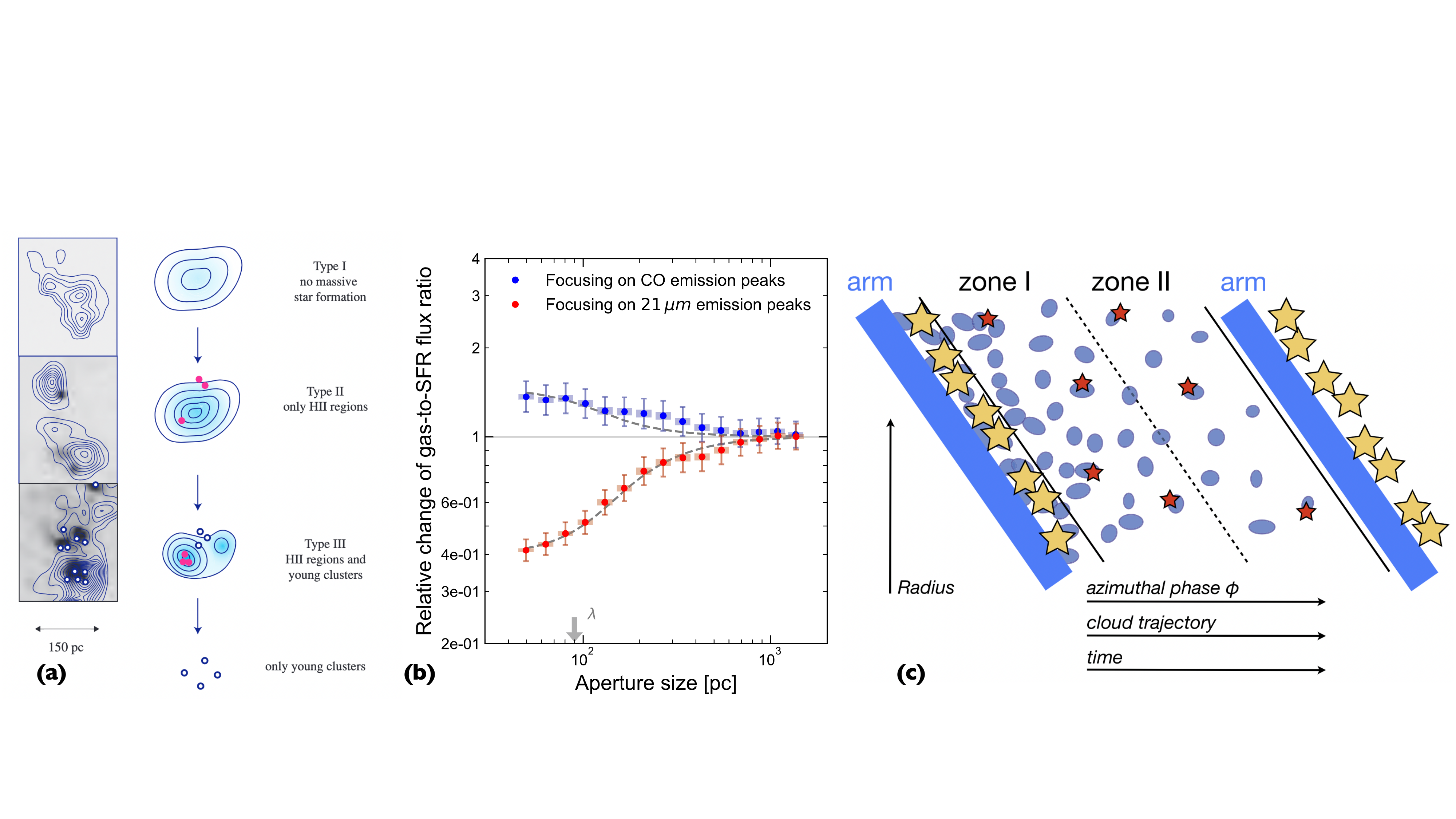}
\caption{Schematic illustration of different methods to estimate cloud lifetimes. \textit{Left \& middle:} Examples of methods adopting stochastic sampling by associating molecular clouds with different tracers of star formation activity \citep[from][]{KAWAMURA09TIMES} and modeling the break-down of the Kennicutt-Schmidt relation as a function of scale and type of peak \citep[also sometimes referred to as a ``tuning fork'' diagram, here from][and see \citealt{SCHRUBA10SFGAS}]{KIM23TIMES}. \textit{Right:} Schematic example of a method leveraging the known gas flow through an inter-arm region to estimate cloud lifetimes by evaluating varying cloud populations along the orbit \citep[from][]{meidt2015}.}
\label{fig:cloudlifetimes_methods}
\end{figure}

When estimating timescales, most of the following assumptions are typically made, either explicitly or implicitly. First, once formed, all clouds will collapse and form stars in a similar fashion. Second, the formation of (massive) stars is quasi-instantaneous, i.e., there are not multiple episodes present within a region, e.g., due to triggering. Third, the efficiency of the conversion of gas into stars is constant and similar for all clouds. And fourth, there is no accretion of gas mass onto the cloud after the cloud formed. Further, the conditions leading to cloud collapse are assumed to be identical across the sample of clouds studied so that there is no dependence on initial cloud mass or location within the region analyzed. 

Broadly speaking there are two avenues in the literature to estimate cloud lifetimes and related timescales (Fig.\,\ref{fig:cloudlifetimes_methods}): statistical modeling of the correlation observed between tracers of different evolutionary stages or utilization of large-scale dynamics or gas flows.

\smallskip
\textbf{Stochastic sampling —} Assuming that star formation is a stochastic process and that cloud-scale observations capture individual regions in random, independent evolutionary states, several approaches have been used to estimate timescales associated with various stages of molecular cloud evolution. These include (a) analysis of sight line statistics of gas and star formation tracers \citep[][]{SCHINNERER19TIMES,PAN22TIMES}, (b) modeling of the scale-dependence of the flux bias about gas and star formation peaks \citep[e.g.,][Fig.~\ref{fig:cloudlifetimes_methods} middle]{KRUIJSSEN19TIMES,KIM22TIMES}, (c) measurement and modeling of the two-point correlation between gas and star formation peaks \citep[e.g.,][]{GRASHA18TIMES,TURNER22TIMES,peltonen2023}, and (d) direct association of gas peaks with different star formation tracers \citep[e.g.,][Fig. \ref{fig:cloudlifetimes_methods} left]{KAWAMURA09TIMES,FUKUI10REVIEW,CORBELLI17TIMES}. Fundamentally, these all work by measuring the spatial correlation (or decorrelation) of molecular clouds with tracers of star formation or different star formation tracers with one another, e.g., mid-IR emission, H$\alpha$ from \textsc{Hii} regions, or young star clusters. Then the results of the various analyses can be modeled to infer the relative duration of each observed phase.

In order to assign absolute timescales, rather than only relative durations of different stages, these methods require that the visibility timescale for one star formation tracer is known, i.e., one of the tracers is typically used as a ``clock.'' The studies mentioned above either use the visibility timescale for H$\alpha$ emission from \textsc{Hii} regions (\S \ref{sec:SF}, Fig. \ref{fig:ssptimescale}; methods: a, b) or the estimated ages of star clusters (methods: c, d). Further, one typically needs to apply sensible mass limits to the clouds, star clusters, and/or \hii\ regions that enter the analysis in order to ensure that the progenitor clouds being analyzed will plausibly evolve into the observed \hii\ regions or star clusters.

\smallskip
\textbf{Dynamical flow-lines —} Other methods to estimate the lifetime of molecular clouds utilize information on the expected orbits of gas clouds in a galactic disk potential. Examples include calculating the travel time between different gas phases (e.g., \citet{engargiola2003} measured the centroid velocity offset between atomic and molecular phases in clouds) or assuming that clouds are formed at certain locations in the galactic disk. For example, utilizing the fact that spiral arms are well-defined locations in galactic disks that are efficient in producing clouds \citep[][]{COLOMBO14GMCS,querejeta2021,MEIDT21ARMS}, \citet{meidt2015} developed a formalism to estimate lifetimes based on cloud statistics in the inter-arm region, where streaming motions are negligible. The advantage of this method is that it does not require the presence of star formation and avoids making assumptions on the uncertain amount of time that clouds spent within the arm. Note that \citet{koda2021} expanded the formalism to start from a location of enhanced cloud production like a spiral arm where streaming motions need to be taken into account.

\subsubsection{Timescales for Cloud Evolution Stages}
\label{sec:timescales_cloudevolution}

\begin{table}[ht]
\tabcolsep7.5pt
\caption{Typical timescales for different stages in cloud evolution}
\label{tab:cloudevolution}
\begin{center}
\begin{tabular}{|l|c|c|c|}
\hline
Stage$^{\rm a}$ &  Tracer & Timescale & References \\
\hline
cloud lifetime & any CO emission & 5-30\,Myr & (1)--(12)\\
molecular cloud, dense gas formation$^\textrm{b}$ & only CO emission & 4-20\,Myr & (3),(13),(14) \\
onset of star formation & only mid-IR \& CO emission & 1-3\,Myr & 
(3),(13)--(15), \S\,\ref{sec:timescales_cloudevolution} \\
stellar feedback$^{\rm c}$ & H$\alpha$ \& mid-IR \& CO line emission & 1-6\,Myr & (1),(3),(7),(16)--(19) \\
\hline
\end{tabular}
\end{center}
\begin{tabnote}
$^{\rm a}$ Following the nomenclature presented in Figure \ref{fig:sketch_cloud}. \\\ 
$^{\rm b}$  We also refer to this as the cloud collapse or ramp-up phase. \\
$^{\rm c}$ Observations only probe the presence of gas/dust coincident with \hii\ regions and/or young star clusters.
See text for details. \\
\textsc{References}:
(1) \citet{CHEVANCE20TIMES},  
(2) \citet{choi2023},  
(3) \citet{CORBELLI17TIMES}, 
(4) \citet{demachi2023}, 
(5) \citet{engargiola2003}, 
(6) \citet{KAWAMURA09TIMES},  
(7) \citep{KIM22TIMES},    
(8) \citet{KRUIJSSEN19TIMES},   
(9) \citep{meidt2015},   
(10) \citet{MIURA12TIMES}, 
(11) \citet[][assuming a \ha\ timescale of 5\,Myr]{PAN22TIMES},  
(12) \citet{ward2022},    
(13) \citet{KIM21TIMES},
(14) \citet{KIM23TIMES},
(15) \citet{HOLLYHEAD15CLUSTERS},
(16) \citet{GRASHA18TIMES},   
(17) \citet{HANNON22CLUSTERS},   
(18) \citet{peltonen2023},  
(19) \citet{TURNER22TIMES}.

\end{tabnote}
\end{table}

Table \ref{tab:cloudevolution} summarizes timescales inferred using these methods. The methods discussed here all report cloud lifetimes, defined as the time from molecular cloud formation to cloud destruction, of about 5-30\,Myr \citep[e.g.,][but see \citealt{koda2023} for an alternative view]{KIM22TIMES,PAN22TIMES}. The large spread partly stems from the choice of tracers used \citep[which can yield results different by a factor of $\sim 2$ for the same galaxy, e.g.,][]{MIURA12TIMES,CORBELLI17TIMES,demachi2023,CHEVANCE20TIMES}, but also reflects real physical differences in cloud lifetimes as a function of environment \citep{PAN22TIMES,KIM22TIMES}. For example, analyzing the statistical properties of sight lines, \citet{PAN22TIMES} report a strong correlation between the fraction of molecular gas-only sight lines and galaxy stellar mass and an additional dependence on Hubble type and galaxy morphology. This implies longer lifetimes in high-mass galaxies and is in excellent agreement with the results of ``tuning fork'' modeling by \citet{KIM22TIMES}. Molecular gas-only sight lines probe surface densities $\Sigma_{\rm mol}^{\rm 150pc}$ of 10 to a few $\rm 10^4\,\msun\,pc^{-2}$ and also appear to be more common in galaxy centers than outer disks. 

From the spatial (and thus temporal) overlap of molecular gas with different SFR tracers, these analyses also yield typical timescales for each stage of the evolution. The time from molecular cloud formation to the onset of star formation --- also referred to as ramp-up or collapse time --- has been estimated from observing molecular clouds without evidence for even highly embedded star formation \citep[i.e., no coincident mid-IR emission; e.g.,][]{CORBELLI17TIMES,KIM21TIMES,KIM23TIMES}. This seems to last $\approx 4{-}20$~Myr, more than half (and perhaps as high as $\sim 80\%$) of the total cloud lifetime. The onset of star formation, i.e., embedded star formation with no star cluster or \textsc{Hii} region visible in the optical, is traceable via mid-IR continuum emission. This phase appears to be short, $\approx 1{-}3$~Myr based on
spatial decorrelation analysis \citep{CORBELLI17TIMES,KIM21TIMES,KIM23TIMES}. Reinforcing this point, in early JWST imaging, out of $>1,000$ compact 21$\rm \mu$m sources, the vast majority ($>$90\%) are associated with an \hii\ region or young stellar associations \citep{HASSANI23JWST}. Assuming \ha\ is visible out to $\tau_\star \sim4$\,Myr (Table \ref{tab:ssptimescale}, Fig. \ref{fig:ssptimescale}), this would imply as low as $\lesssim 0.5$\,Myr for this embedded phase.

The stellar feedback timescale, during which molecular clouds overlap with \textsc{Hii} regions and/or young stellar clusters appears to be $\approx 1-6$ Myr \citep[e.g.,][]{KIM22TIMES,GRASHA18TIMES,peltonen2023}. In a recent direct demonstration of this fast clearing, \citet{hannon2023} analyzed the HST-based \ha\ morphology of star clusters with modeled ages. The youngest clusters, with $\tau_\star \approx 1{-}2$ Myr, show concentrated \ha\ morphology, consistent with little ionized gas clearing. Older clusters, with $\tau_\star \approx 2{-}3$ Myr, show a partially exposed appearance. Clusters older than $\tau_\star \approx 5$\,Myr no longer show any \ha\ emission, in good agreement with the clearing or feedback times in Table \ref{tab:cloudevolution}.


\begin{figure}[t]
\includegraphics[width=\textwidth]{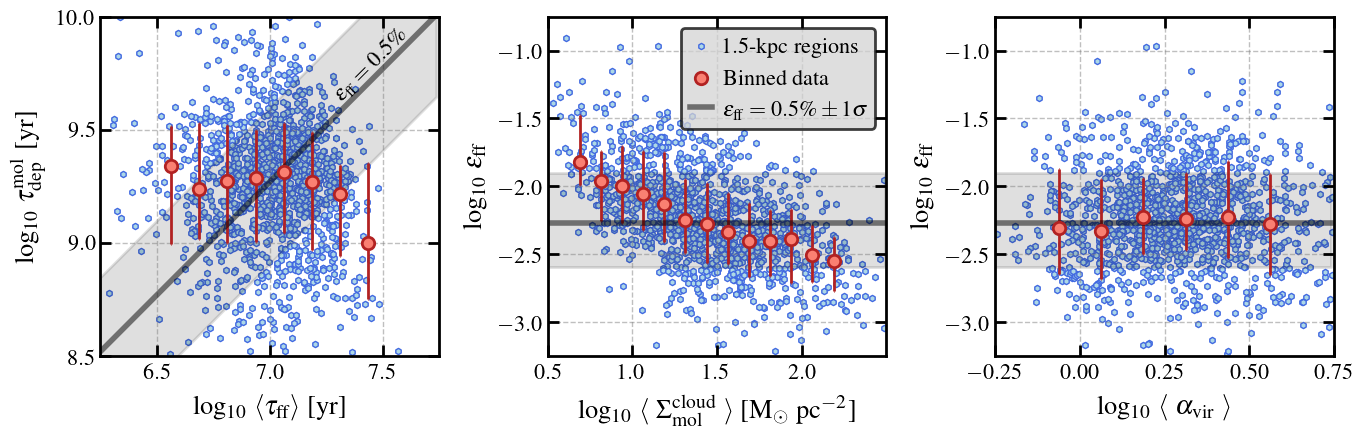}
\caption{Gravitational free-fall time and star formation in PHANGS--ALMA galaxies using data from \citet{SUN22CLOUDS}. \textit{Left:} Molecular gas depletion time $\tau_{\rm dep}^{\rm mol}$ as a function of gravitational free-fall time $\tau_{\rm ff}$ estimated from cloud-scale CO imaging. \textit{Middle:} star formation efficiency per gravitational free-fall time, $\epsilon_{\rm ff}$, as a function of cloud-scale $\Sigma_{\rm mol}^{\rm 150pc}$; \textit{Right:} $\epsilon_{\rm ff}$ as a function of gravitational boundedness, expressed by the virial parameter $\alpha_{\rm vir}$. In all panels, blue points show results for individual 1.5~kpc hexagonal apertures red point show the median and $1\sigma$ scatter in $y$ as a function of $x$, and the black line and shaded region show the median and $1\sigma$ scatter in $\epsilon_{\rm ff}$ (i.e., these show the median ratio of $y$ to $x$; they are not a fit to the data).}
\label{fig:phystime}
\end{figure}

\begin{table}[t!]
\tabcolsep7.5pt
\caption{Molecular gas timescales and efficiencies in galaxy disks}
\label{tab:phystimes}
\begin{center}
\begin{tabular}{|l|c|c|c|c|}
\hline
Timescale & Formula & Mass-weighted range & $\epsilon$ & Rank corr. \\
\hline
\multicolumn{5}{c}{Turbulent cloud timescales} \\
\hline
Free-fall time & 
$\tau_{\rm ff} = \sqrt{\frac{3 \pi}{32 G \rho}} = 4.4 \times 10^6 \left( \frac{n_{\rm H2}}{100~{\rm cm^{-3}}} \right)^{-0.5}$
&  $11^{+5.5}_{-4.3}$ Myr & $0.5_{-0.3}^{+0.7}~\%$ & $-0.11$ \\
Crossing time & $\tau_{\rm cr} \simeq \sigma / R \sim 2 \sigma / H_{\rm mol} \sim \sigma / \theta$
& $15^{+7.2}_{-4.5}$ Myr & $0.8^{+0.8}_{-0.4}~\%$ & $-0.04$ \\
\hline
\multicolumn{5}{c}{Kinematic timescales} \\
\hline
Orbital time & $\tau_{orb} = \frac{2 \pi}{\Omega_{\rm circ}} = \frac{2 \pi r_{\rm gal}}{v_{\rm circ}}$ & $160^{+70}_{-64}$ Myr & $8^{+8}_{-4}~\%$ & $0.14$ \\
Shearing time & $\tau_{shear} = A_{\rm Oort}^{-1} = \frac{2}{\Omega_{\rm circ}~(1 - \beta)}$ & $59^{+30}_{-23}$ Myr & $3^{+3}_{-1.5}\%$ & $0.18$ \\
Collision time & $\tau_{\rm coll} \sim 4 N_{\rm cl} R_{\rm cl}^2 / \tau_{\rm shear}$ & $91_{-38}^{+77}$ Myr & $5^{+7}_{-2.5}~\%$ & $0.05$ \\
\hline
\multicolumn{5}{c}{Molecular gas depletion time} \\
\hline
Molecular gas & $\tau_{\rm dep}^{\rm mol} = \frac{M_{\rm mol}}{{\rm SFR}}$ & $1.9_{-0.9}^{+1.3}$ Gyr & --- & --- \\
Dense gas$^a$ & $\tau_{\rm dep}^{\rm dense} = \frac{M_{\rm dense}}{{\rm SFR}}$ & $\approx 0.15_{-0.1}^{+0.3}$ Gyr & --- & --- \\
\hline
\multicolumn{5}{c}{Cloud evolutionary timescales} \\
\hline
Cloud lifetime & From Table \ref{tab:cloudevolution} & $\sim 5{-}30$~Myr & $\sim 0.3{-}1.5$~\% & --- \\
\hline
\end{tabular}
\end{center}
\begin{tabnote}
$^a$ Dense gas depletion time adopting $\alpha_{\rm HCN} \sim 15$~\acounits\ and $\sim 0.5$ dex scatter, see \S\,\ref{sec:denseresults}. \\
\textsc{Notes} --- Timescales calculated using mass-weighted cloud-scale 150~pc gas properties within 1.5~kpc apertures following \citet{SUN22CLOUDS} and using the database presented in that paper (as in Fig. \ref{fig:phystime}). Efficiency $\epsilon$ calculated following Eq. \ref{eq:efficiency} and \citet{SUN22CLOUDS,SUN23SFGAS} by contrasting the relevant timescale with $\tau_{\rm dep}^{\rm mol}$ in each 1.5-kpc aperture. The quoted timescales and efficiencies report the median and 16{-}84\% range for all apertures in the \citet{SUN22CLOUDS} database that have $>50\%$ completeness in the high resolution CO data. The last column reports the rank correlation coefficient relating the timescale to $\tau_{\rm dep}^{\rm mol}$ region-by-region with. These are all statistically significant, but also all indicate only weak correlations. 
\end{tabnote}
\end{table}

\subsection{Efficiencies and physical timescales}
\label{sec:sfe}

Clouds evolve, and both the properties of the turbulent gas and the larger galactic environment set physical timescales that should relate to this evolution. The cloud-scale molecular gas observations described in \S \ref{sec:moleculargas} allow systematic estimates of many of these timescales. These are of interest on their own and can also be contrasted with the ramp-up and gas clearing times to infer the underlying physical drivers for cloud evolution. More, because the molecular gas depletion time is known at large scales, each of these timescales implies a corresponding star formation efficiency, $\epsilon$, via

\begin{equation}
\label{eq:efficiency}
\epsilon \equiv {\rm SFR} \left( \frac{M_{\rm mol}}{\langle \tau \rangle } \right)^{-1} = \frac{\langle \tau \rangle}{\tau_{\rm dep}^{\rm mol}}~.
\end{equation}

\noindent with $\langle \tau \rangle$ the timescale of interest weighted appropriately and averaged over a region large enough to access the mean $\tau_{\rm dep}^{\rm mol}$ unbiased by stochasticity associated with the evolutionary state of individual regions \citep[see][]{LEROY16GMCS,SUN22CLOUDS}\footnote{In Milky Way studies, it is common to refer to the star formation efficiency as the fraction of gas converted to stars per cloud lifetime, equivalent to the last line in Table \ref{tab:phystimes}.}.

Table \ref{tab:phystimes} reports expressions, ranges of estimated values, and characteristic efficiencies associated with several timescales of interest ($\epsilon$) from \citet{SUN22CLOUDS} based on PHANGS--ALMA. These represent the current best estimates for these timescales at cloud-scales in normal star-forming galaxies. We also note the degree to which each of these timescales correlate with $\tau_{\rm dep}^{\rm mol}$ region-by-region providing an indirect indication of the degree they might reflect the controlling process for star formation. 

\smallskip

\textbf{Gravitational free-fall and turbulent crossing time.}
The gravitational free-fall time, $\tau_{\rm ff}$, expresses the time for a gas cloud to collapse in the absence of other support. Meanwhile, the turbulent crossing time, $\tau_{\rm cr}$, expresses the time for a parcel of material moving at the one dimensional turbulent velocity dispersion, $\sigma_{\rm turb}$, to cross a cloud. Both capture characteristic timescales for a turbulent or bound cloud to evolve and depend on quantities ($\rho$, $\sigma$, $R$) accessible from cloud-scale CO imaging.

For a large set of regions in main sequence star-forming galaxies, \citet{SUN22CLOUDS} find median $\tau_{\rm ff} \sim 11$~Myr and  $\tau_{\rm cr} \sim 15$~Myr. Contrasting these with the typical $\tau_{\rm dep}^{\rm mol}$ implies $\epsilon \sim 0.005$, that is $\sim 0.5\%$ of the gas is converted to stars per $\tau_{\rm ff}$ or $\tau_{\rm cr}$. This $\epsilon_{\rm ff}$ agrees well with a large suite of observational estimates across many scales using a variety of techniques \citep[e.g.,][]{KRUMHOLZ07EFF,EVANS14DENSE,UTOMO18EFF,KRUMHOLZ19REVIEW}. To our knowledge the determination of $\epsilon_{\rm ff}$ by \citet{SUN22CLOUDS,SUN23SFGAS} across 80 nearby galaxies represents the most systematic determination to date, and their $\epsilon_{\rm ff} = 0.5_{-0.3}^{+0.7}\%$ should be viewed as the current best cloud-scale value and a modern, clear expression of the inefficiency of star formation relative to direct gravitational collapse in such systems. 

Both the free-fall time and the crossing time are comparable to the cloud collapse time (or dense gas formation time; $\sim 4{-}20$~Myr) discussed in \S\,\ref{sec:timescales_cloudevolution}. This offers strong support for the idea of ``star formation in a crossing time'' posited by \citet{ELMEGREEN00CROSSING} and for the frequent adoption of $\tau_{\rm ff}$ as a key fiducial timescale \citep[e.g.,][]{KRUMHOLZ05EFF,PADOAN12EFF}. In simple terms: the various lines of evidence from comparing signatures of star formation and molecular gas discussed above suggest a relatively CO-bright, star formation-poor phase of order a free-fall or a cloud crossing time. After this time, stellar feedback (\S \ref{sec:feedback}) determines the next steps in cloud evolution. This picture agrees with the idea of accelerating star formation, which has also been inferred from the demographics of Milky Way star-forming regions \citep{MURRAY11LIFETIMES,LEE16LIFETIMES} and simulations of individual low-mass clouds \citep{STARFORGE22,STARFORGE23}. 

\smallskip 
\textbf{Orbital, shearing, and cloud collision timescale.}
The efficiency of star formation and evolution of clouds will also be shaped by galaxy-scale processes. Large-scale dynamical features evolve over an orbital time, and shear may disrupt self-gravitating clouds \citep[e.g.,][]{LIU21GMCS,meidt2015} or it may help induce collisions between clouds. Cloud collisions have been highlighted as an effective way to bring gas to high densities and induce high-mass star formation, with a number of likely colliding clouds identified especially within Milky Way CO data \citep[e.g.,][]{TAN00COLLIDE,INOUE13COLLIDE,FUKUI21COLLIDE}. 

Leveraging kinematic modeling by \citet{LANG20KINEMATICS}, \citet{SUN22CLOUDS} also calculated characteristic kinematic timescales. These include the orbital time ($\tau_{\rm orb}$) and the shearing time ($\tau_{\rm shear}$), which expresses the rate at which differential rotation moves material together or apart, and the time between cloud collisions ($\tau_{\rm coll}$) induced by differential rotation \citep[following][]{TAN00COLLIDE}. These kinematic timescales are longer than the cloud internal timescales, with $\tau_{\rm orb} \sim 160$~Myr and $\tau_{\rm shear}$, $\tau_{\rm coll} \sim 70$ and $90$~Myr (Table \,\ref{tab:phystimes}). Contrasting with $\tau_{\rm dep}^{\rm mol}$ implies efficiencies of $\approx 3{-}8\%$ relative to these timescales. These kinematic timescales appear a few times longer than the cloud collapse time described above, which provides evidence against the idea that the kinematic timescales govern the build-up to star formation. That said, both the cloud collapse time and the precise kinematic timescales relevant to star formation have a fair bit of uncertainty, so this is not a conclusive comparison. In particular, we note that $\tau_{\rm coll}$ likely represents an overestimate, as streaming motions and intersections between gas flow lines will increase the rate of collisions.

Perhaps more confusing, \citet{SUN22CLOUDS} show that in fact \textit{none} of these timescales, including $\tau_{\rm ff}$ and $\tau_{\rm cr}$, show a particularly strong correlation with the local $\sim$kpc averaged gas depletion time (see Table \ref{tab:phystimes}). If one of these timescales captured the critical environmental effect for cloud collapse or dense gas formation, we would expect a strong correlation. Instead, the rank correlation coefficients $\rho$ linking $\tau_{\rm dep}^{\rm mol}$ to the other timescales range between $-0.1$ and $0.2$ with large scatter. Is this some observational bias or perhaps a reflection of low dynamic range in the local galaxy population? It might be both or either, but certainly these observations had the opportunity to clearly establish a link between $\tau_{\rm dep}$, expressing the star formation rate per unit gas, and these physical timescales. Yet no strong relation of this sort has emerged so far.

\smallskip
\textbf{Efficiency per free-fall time and cloud-scale gas properties.}
The star formation efficiency per free-fall time, $\epsilon_{\rm ff}$, has emerged as a central prediction of turbulence-regulated star formation theories \citep[e.g.,][]{KRUMHOLZ05EFF,PADOAN12EFF,FEDERRATH12EFF}. As described above, the median value of $\epsilon_{\rm ff}$ from cloud-scale studies agrees well with previous work. Thus it likely represents a general reference value and shows that the overall normalization of these models appears reasonable. 

Fig. \ref{fig:phystime} provides a more detailed view on how $\epsilon_{\rm ff}$ varies as a function of the regional average cloud-scale gas properties. Many turbulent theories of star formation make specific predictions about how $\epsilon_{\rm ff}$ should vary as a function of the initial conditions in a molecular cloud, the most of which is that as the virial parameter of the gas increases, $\epsilon_{\rm ff}$ should decrease \citep[e.g.,][]{HENNEBELLE11EFF,PADOAN12EFF}. Indeed, lower star formation activity in more weakly bound gas would seem to be a natural outcome. Despite this expectation, in a synthetic analysis of the Milky Way and several nearby galaxies, \citet{SCHRUBA19GMCS} found no clear correlation between the regional average $\epsilon_{\rm ff}$ and $\alpha_{\rm vir}$. Nor is such a correlation evident in the extensive measurements from \citet{SUN22CLOUDS} shown in Fig. \ref{fig:phystime}. Instead, $\epsilon_{\rm ff}$ is clearly anti-correlated with the region-average cloud-scale $\Sigma_{\rm mol}^{\rm cloud}$ and velocity dispersion $\sigma_{\rm mol}$ \citep[Fig.\,\ref{fig:phystime} and][]{LEROY17SFGAS,SCHRUBA19GMCS}. 

Note that Fig. \ref{fig:phystime} shows region averages of current cloud properties, while theory often predicts star formation activity based on the initial conditions in a cloud. \citet{KIM21FEEDBACK} highlight how cloud properties can appear dramatically different after the effects of feedback, so that an instantaneous comparison of individual cloud properties to star formation activity fails to capture the initial conditions. On the other hand, as discussed in the previous section, the longer collapse time relative to the stellar feedback timescale implies that most of the clouds in any given region are not expected to have be strongly processed by feedback, so the effect of such concerns on the region average cloud properties may be minimal, or at least not immediately clear. Our best assessment is that $\tau_{\rm dep}^{\rm mol}$ appears to vary more weakly than na\"ively expected as a function of $\alpha_{\rm vir}$ and to vary too strongly as a function of $\Sigma_{\rm mol}$ and $\sigma_{\rm mol}$ (not shown, but similar to the middle panel in Fig. \ref{fig:phystime}).

The observational estimates of $\alpha_{\rm vir}$ and mean density both have significant uncertainties, especially related to the small-scale geometry that sets the gravitational potential. To address the uncertainties related to both the ensemble average and the small-scale geometry, the dual paths forward seem to be (a) systematic higher resolution surveys of the molecular gas that significantly improve current estimates of $\alpha_{\rm vir}$ and the volume density, and (b) careful matched mock observations applying the same averaging and measurement techniques used for observations to numerical simulations \citep[e.g., like][]{KIM21FEEDBACK}. Given that turbulent star formation models are now widely adopted, including implementation into numerical simulations, such tests seem like a critical next step and this strikes us as an area where more effort is urgently needed.

\section{STELLAR FEEDBACK}
\label{sec:feedback}

Once stars form, they exert feedback on the surrounding ISM. Photoionization, along with energy and momentum input from stellar winds, supernovae, radiation pressure, protostellar jets, and thermal heating of the gas all reshape the ISM. Fig. \ref{fig:ssptimescale} and Table \ref{tab:ssptimescale} illustrate the time dependence of some key feedback modes \citep[following][]{AGERTZ13FEEDBACK}. Stellar winds and radiation act immediately, while supernovae take longer. If all massive stars explode, the first SN occurs at $\tau_\star \sim 3{-}4$~Myr, but if the most massive stars collapse directly to black holes \citep[e.g.,][]{SUKHBOLD16SNE} the first SN may not occur until $\gtrsim 6$~Myr. Because feedback is multi-mode, plays out across multiple time and length scales, and involves a complex coupling between different feedback mechanisms and the ISM, multiple distinct types of observations are needed to constrain its impact. Key observations that have emerged over the last decade are the net effect of feedback on resolved galaxy scales, the implications from fast gas clearing for feedback, the evolving strength and dominant mechanisms driving \textsc{Hii} region expansion, and the location and visible impact of SN explosions on the neutral ISM.

\begin{figure}[ht]
\includegraphics[width=0.8\textwidth]{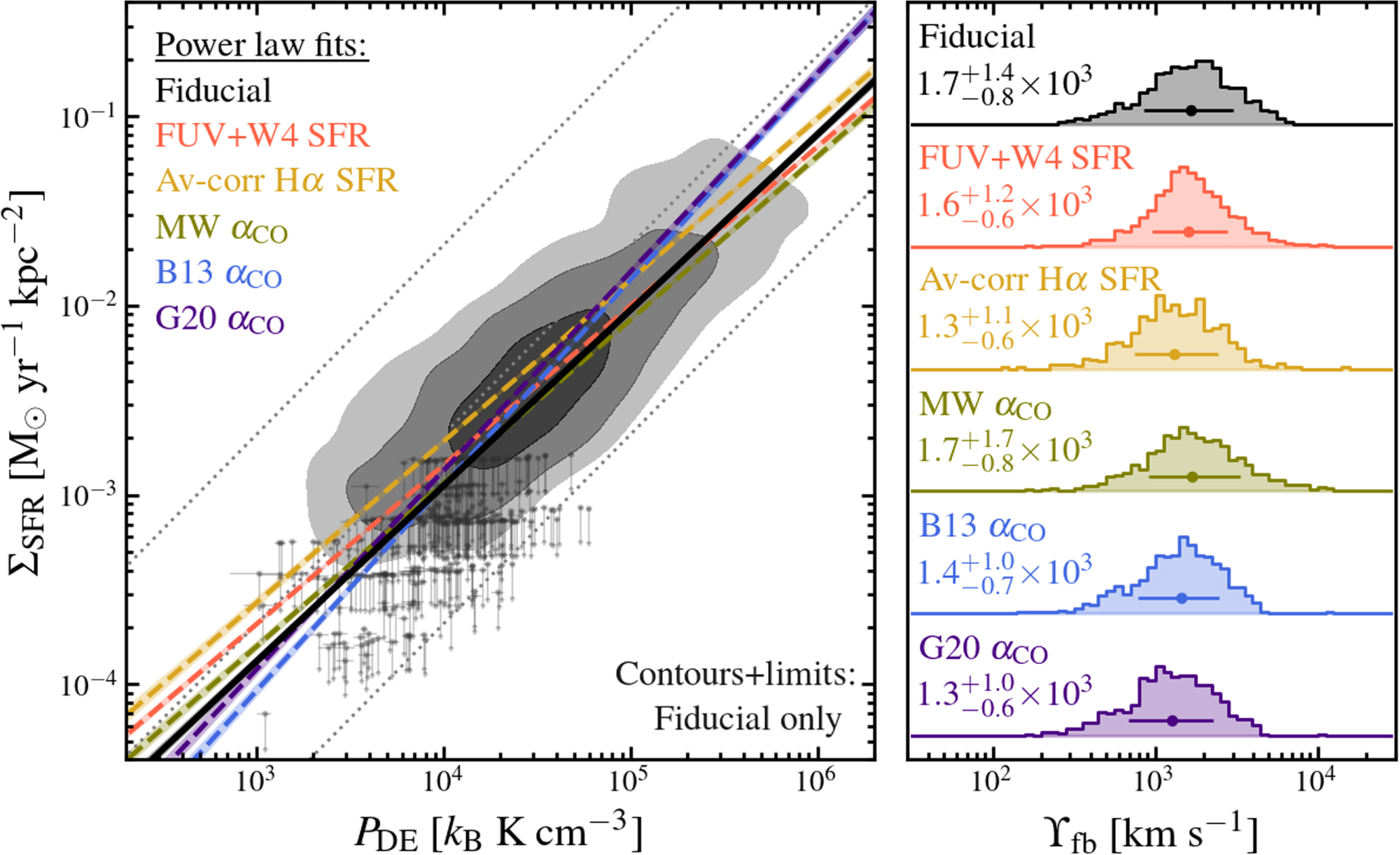} 
\caption{Comparison between SFR surface density, ultimately tracing the source of stellar feedback, and the dynamic equilibrium pressure needed to support the disk in vertical equilibrium \citep[from][]{SUN23SFGAS}. The coefficient linking the two, $\Upsilon_{\rm fb} \approx 1,300{-}1,700$~km~s$^{-1}$, gives an important empirical measure of the effective yield of stellar feedback on $\sim$ kpc scales. The right panels show the implied distribution of $\Upsilon_{\rm fb}$ and its variation when changing assumptions for the conversion of observations into physical quantities.}
\label{fig:sfrvspde}
\end{figure}

\subsection{Feedback yield constraints from pressure balance}
\label{sec:pdefeedback}

The requirement to maintain vertical dynamical equilibrium (see Eq. \ref{eq:pde} and surrounding) has important implications for the overall strength of stellar feedback. In vertical self-regulation models, stellar feedback supplies the pressure that balances the weight of gas in the galaxy potential. Thus, by estimating and then comparing $P_{\rm DE}$ to the star formation rate surface density, $\Sigma_{\rm SFR}$, for a region in equilibrium, one can solve for the \textbf{feedback yield}, $\Upsilon_{\rm fb} \equiv P_{\rm DE}/\Sigma_{\rm SFR}$ \citep[see][and references therein]{OSTRIKER22PRESS}, where the total yield is a sum of terms from the different pressure components. For support of the disk by turbulent pressure, the yield captures the momentum injected in the ISM per stellar mass formed (primarily by supernovae), and for support by thermal pressure, the yield represents the ratio of heating and cooling rate coefficients multiplied by a stellar UV energy production efficiency. As a specific momentum, $\Upsilon_{\rm fb}$ can be expressed in units of km~s$^{-1}$, with values of $\sim 1,000{-}3,000$~km~s$^{-1}$ predicted from simulations \citep[e.g.,][]{KIM13PRESS,OSTRIKER22PRESS}. As a measure of ``net'' support at fairly large scales, $\Upsilon_{\rm fb}$ should in principle depend on a wide variety of factors including stellar clustering, ISM density, and ISM structure, and measurements are needed to determine its actual value. Following this logic, the observed $\Sigma_{\rm SFR}{-}P_{\rm DE}$ relationship has emerged over the last few years as an important scaling relationship in resolved galaxy studies  \citep[e.g.,][]{SUN20PRESS,BARRERA21PRESS,FISHER22PRESS}.

Fig. \ref{fig:sfrvspde} shows an extensive set of recent measurements from \citet{SUN23SFGAS}. They find mean $\Upsilon_{\rm fb} \approx 1,300{-}1,700$~km~s$^{-1}$ with about $\pm 0.3$~dex scatter across several thousand $1.5$-kpc sized regions in $\sim 80$ galaxies. Recent observational studies vary in the exact observed slope relating $P_{\rm DE}$ and $\Sigma_{\rm SFR}$, with \citet{SUN23SFGAS} finding a slope of $\sim 0.85{-}1.05$ depending on the adopted assumptions. Simulations return slightly steeper slopes, and exploring the environmental variation of $\Upsilon_{\rm fb}$ in detail will be an important next step.

\subsection{Constraints on gas-clearing mechanism from short stellar feedback times}
\label{sec:clearingtimefb}

As detailed in \S \ref{sec:timescales_cloudevolution}, joint statistical analysis of gas and star formation tracers show that molecular gas decorrelates from star formation tracers at high spatial resolution, independent of the exact methodology used. This implies that once massive star formation begins, the cold gas is strongly affected and probably cleared from the region within $\sim 6$\,Myr (see Table \ref{tab:cloudevolution}). The fact that tracers of star formation and molecular gas decorrelate has major implications for the mechanisms responsible for gas clearing. As discussed in \S \ref{sec:SF}, ionizing photon production drops significantly even by $\tau_\star \sim 4{-}5$~Myr. Meanwhile, core-collapse supernovae do not begin exploding until $3{-}6$~Myr with most SN feedback occuring later. Thus the fast feedback timescale implies that stellar winds, radiation, and heating, often collectively referred to as ``pre-supernova'' feedback, substantially disrupt clouds (see \S\,\ref{sec:hiifeedback}).

\begin{figure}[ht]
\includegraphics[width=\textwidth]{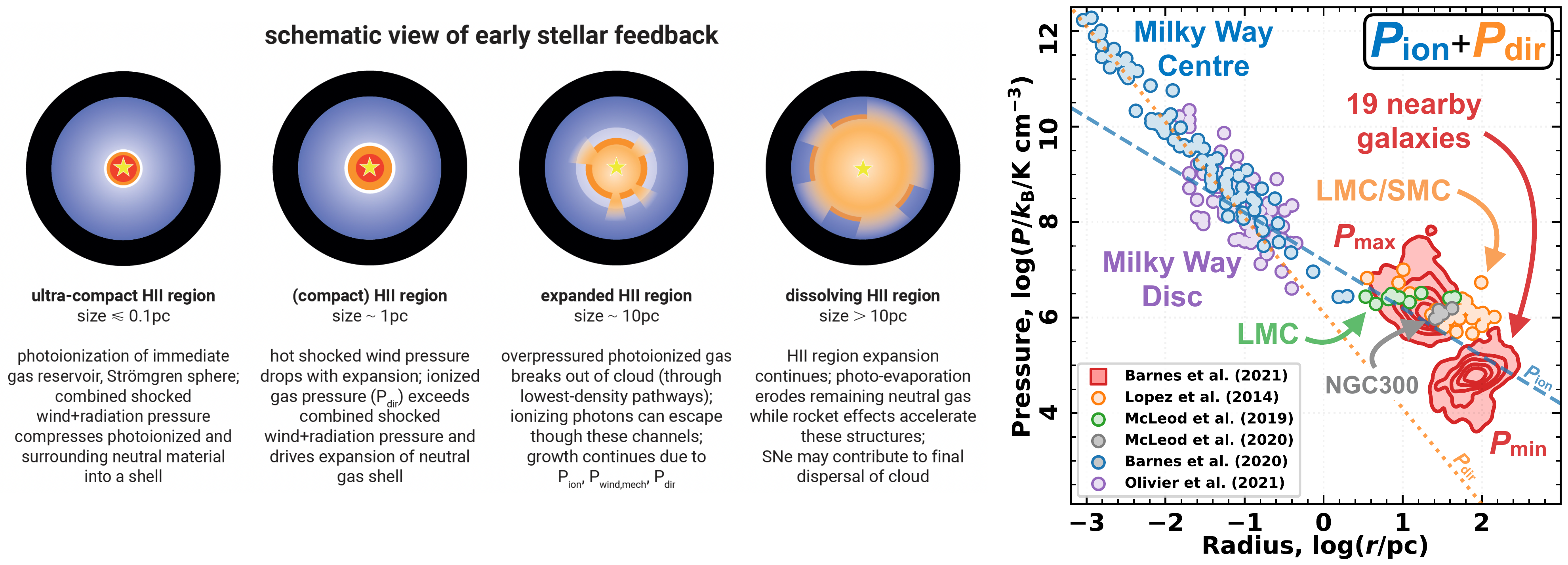}
\caption{Evolution of \textsc{Hii} regions. \textit{Left:} Schematic view of the impact of (pre-SN) stellar feedback caused by massive forming stars in a spherical cloud showing the evolution from a deeply embedded ultra-compact to a large dissolving \hii\ region (see text for details); the quoted numbers are representative for a $\rm 10^4\,\msun$ stellar cluster. The exact sizes are indicative only and depend on both the number of massive stars and the density of the surrounding medium. The yellow star represents massive stars emitting ionizing radiation, white to dark blue colors represent the decreasing density of the neutral gas, red to orange colors the decreasing temperature of the ionized gas, and increasing transparency decreasing ionized gas density.
\textit{Right:} Estimates of two forms of pre-SN stellar feedback associated with massive forming stars and their \textsc{Hii} region over several orders of size \citep[adapted from][see also text for details]{barnes2021}.}
\label{fig:feedback}
\end{figure}

\subsection{Mechanisms for gas clearing}
\label{sec:hiifeedback}

Given the potential key role of pre-SN feedback in gas clearing, much effort has focused on assessing the drivers for the expansion of \textsc{Hii} regions at different evolutionary stages. These regions represent the likely sites of much of this pre-SN feedback. Thanks to these efforts \citep[summarized in recent reviews by][]{chevance2022,girichidis2020,KRUMHOLZ14REVIEW} and because of the challenge of resolving \textsc{Hii} regions across the spectrum at extragalactic distances, the theoretical underpinning of \textsc{Hii} region evolution remains well ahead of the observational understanding obtained from cloud-scale observations. 

\smallskip
\textbf{Idealized physical picture --}
Fig. \ref{fig:feedback} (\textit{left}) illustrates the expected stages for the evolution of an individual \hii\ region. 
At the earliest stages, while the powering cluster is still heavily embedded in its birth cloud, wind-driven shocks from newly formed massive stars will heat gas immediately surrounding the stars to very high temperature ($\rm T \sim 10^6-10^8\,K$). This very hot gas will produce X-ray emission. Meanwhile, extreme UV photons photoionize gas to produce a Str\"omgren sphere. At these early stages, this photoionized gas is expected to be strongly compressed by momentum injection from winds \citep[][]{lancaster2021a} and radiation \citep[][]{krumholz2009,MURRAY10FEEDBACK,DRAINE11BOOK}. As long as the gas and dust shell remains optically thick to UV photons, the radiation can efficiently couple to the gas and transfer momentum, leading to expansion of the region.

At later stages, when the shell radius exceeds a characteristic radius proportional to the ionizing luminosity of the source (see Eq.\,5 of \citealt{krumholz2009} or Eq.\,16 of \citealt{kim2016}), the evolution transitions to resemble the \citet{spitzer1978} solution for an expanding neutral shell driven by the pressure of photoionized gas. The over-pressured photoionized gas expands fastest through low-density portions of the inhomogeneous surrounding cloud, which will often lead it to break out of the cloud \citep[][]{kim2018}. This can create a blister \hii\ region \citep{whitworth1979} or champagne-flow \citep[][]{tenorio-tagle1979} of trans-sonic outflowing ionized gas. In this case, the ionizing photons that leak from the region will contribute to the diffuse ionized gas component on larger scales \citep[][]{kim2019}. If feedback is strong enough, the neutral ISM shell can freely expand into the lower density environment around the birth cloud until its density thins out and it becomes indistinguishable from the diffuse ISM. At some point during this expansion, the first SNe should explode, ending the purely ``pre-SN'' feedback phase. 

In reality this picture will be complicated by clustering of massive stars as well as potential triggering of star formation along the rim of the swept-up shell. The relative importance of photo-dissociation versus dispersal of neutral gas for molecular cloud disruption is also not yet well constrained from observations, although theory and simulations suggest the latter will be more important for clouds with high surface density \citep[e.g.,][]{krumholz2009,fall2010,kim2018}.

\smallskip
\textbf{Observational assessments of pre-SN feedback --}
Observational studies have used multi-wavelength observations to estimate pressures for samples of \textsc{Hii} regions \citep[following][]{LOPEZ11FEEDBACK,LOPEZ14FEEDBACK} to test aspects of a scenario like the one described above. One commonly adopted framework, illustrated in Fig. \ref{fig:feedback} (\textit{right}), is to consider and combine different pressure terms as a function of region size. This treats the size as a proxy for age or evolutionary state, though note that this will be uncertain when mixing regions with different luminosities and different ambient ISM conditions.

Emission lines from ionized gas allow estimates of electron density, $n_{\rm e}$, and temperature, $T_{\textsc{Hii}}$ \citep[e.g.,][]{kewley2019}. These together allow one to estimate the \textbf{thermal pressure} in the warm ionized gas, $P_{\rm ion} = (n_{\rm e} + n_{\rm H} + n_{\rm He}) k_B T_{\textsc{Hii}}$ or $P_{\rm ion} \approx 2 n_{\rm e} k_B T_{\textsc{Hii}}$, which assumes singly ionized He. Ideally, $T_{\rm \textsc{Hii}}$ would be estimated from spectroscopy but assuming $T_{\rm \textsc{Hii}} \sim 10^4$~K is also common. $n_{\rm e}$ can also be derived from spectroscopy, which yields the physical densities responsible for producing the line luminosity (but note that this can be biased high if emission from the ionization front rather than the volume-filling gas dominates), or from an assumed geometry and integrated H$\alpha$ or free-free emission, e.g., taking $n_{\rm e} \propto (Q^O/R^3)^{1/2}$.

From the region luminosity and size one can estimate the ``direct'' \textbf{radiation pressure} due to the momentum imparted by stellar photons, $P_{\rm dir} \approx f L_{\rm bol}/ (4 \pi R^2 c)$, with the pre-factor $f \approx 3$ as a geometric factor intended to capture the average throughout the region \citep[e.g.,][]{LOPEZ14FEEDBACK}. Calculation of $P_{\rm dir}$ typically assumes that effectively all of the (mostly UV) emission from young stars imparts momentum to gas or dust. In highly optically thick regions, IR photons re-emitted by the dust after it absorbs the initial starlight can also impart momentum \citep[e.g.,][]{THOMPSON05PRESS,MURRAY10FEEDBACK}. In theory, this can amplify the radiation pressure by a factor up to $\sim \tau_{\rm IR}$ as each scattering or absorption contributes an additional $\sim L_{\rm bol}/c$. In practice, achieving high $\tau_{\rm IR}$ requires very high $\Sigma_{\rm mol}$, so this term appears most relevant in starburst galaxies and young massive clusters.

The \textbf{pressure associated with stellar winds} remains highly uncertain. The influence of winds can be seen most directly by X-ray observations of $\sim 10^7$~K hot, shocked gas, but such observations are scarce. Lacking these, many studies use indirect methods to estimate $P_{\rm wind}$, most commonly leveraging theoretical predictions for mass loss rate and mechanical luminosity of the powering population (e.g., Fig. \ref{fig:ssptimescale}). Formally, the pressure due to winds goes as $P_{\rm wind,mech} \propto L_{\rm wind} / (4 \pi R^2 v_{\rm wind})$ with $L_{\rm wind}=1/2 \dot{M} v_{\rm wind}^2$ the mechanical luminosity, $\dot{M}$ the mass loss rate, and $v_{\rm wind}$ the wind speed. In practice, because $\dot{M}$ and $v_{\rm wind}$ will be fixed in the stellar models, the wind term ends up proportional to $P_{\rm wind,mech} \propto L_{\rm bol} / (4 \pi R^2 c)$ with a prefactor that depends on both the physical coupling of the winds to the shell and the time variation of the ratio of mechanical luminosity to $L_{\rm bol}$. The similarity of this term to the $P_{\rm dir}$ term then means that the importance of winds in a given study will depend on the assumptions made, highlighting the importance of both theoretical work and observations that refine our assessment of the impact of winds. Fortunately such simulations are proceeding \citep[e.g.,][]{lancaster2021a,lancaster2021b,lancaster2021c}, and we highlight that folding X-ray data into cloud-scale analyses represents a key next step.

At large $\sim 50{-}100$~pc scales, the thermal pressure, $P_{\rm ion}$, and direct radiation pressure, $P_{\rm dir}$, have been estimated for $\gtrsim 5{,}000$ \textsc{Hii} regions based on VLT/MUSE observations \citep[][]{MCLEOD21HII,barnes2021,DELLABRUNA22HIIREGION}. These studies, illustrated in Fig. \ref{fig:feedback}, generally find $P_{\rm ion}$ and $P_{\rm dir}$ of the same order. However, determining which term represents the dominant term depends on uncertain size measurements and geometric assumptions. The combined $P_{\rm ion}$ and $P_{\rm dir}$ appear typically higher than the estimated $P_{\rm DE}$ in the surrounding disk, implying that even these large \textsc{Hii} regions are still expanding. 

Capturing the earliest phases of cloud disruption is critical to test the picture above. This requires assessing the pressures in small, young, still-embedded regions, including compact \textsc{Hii} regions and young massive clusters (YMCs). \citet{olivier2021} present the most complete assessment of pressures for small, $< 1$~pc, Milky Way \textsc{Hii} regions and find high $\tau_{\rm IR}$ to be common, suggesting IR-reprocessed radiation pressure to play an important role in the earliest stages of region evolution and again noting large uncertainties in assessing the effects of winds. They suggest a transition to thermal gas pressure-driven expansion at scales of a few pc. Similar measurements have now expanded to dozens of more massive forming clusters in the nearest starburst galaxies, which similarly suggest a strong role for radiation pressure at early times, high $\tau_{\rm IR}$, and evidence of powerful feedback driven outflows at early $\tau_\star \lesssim 1$~Myr times \citep[e.g.,][]{leroy2018,LEVY21CLUSTERS}.

\begin{textbox}[ht]
\section{Metal mixing as a feedback outcome and a feedback tracer}
\label{sec:metalmixing}

\noindent
Stars produce and distribute heavy elements and recycled matter into the surrounding ISM. This represents another form of stellar feedback, one that plays a critical role in the evolution of galaxies \citep[e.g.,][]{MAIOLINO19REVIEW}. Thanks to multi-object spectroscopy and IFU mapping, region-by-region estimates of the oxygen abundance have become common over the last few years. These have the potential to constrain the amount of enrichment due to individual star formation events and to illuminate how heavy elements produced by star formation mix into the surrounding ISM. Recent observations find typical azimuthal scatter about the large-scale radial gradient, in $12+\log_{10} {\rm O/H}$ of $\approx 0.02-0.05$~dex \citep[e.g.,][]{williams2022,li2023}. They find larger scatter in massive galaxies and grand-design spiral galaxies and higher enrichment often, but not always, in arm regions compared to inter-arm regions \citep{kreckel2019,sanchez-menguiano2020,li2023} and associated with higher gas velocity dispersion \citep{GROVES23SFR}. Several studies have also constructed region-to-region correlation functions and so constrained the length scales over which the oxygen abundance is correlated among regions to be $\sim 0.5-1$~kpc \citep[e.g.,][]{KRECKEL20METALS,li2023}. Different mixing models try to explain these azimuthal variations from simple stochastic force diffusion models \citep[especially][]{krumholz2018}, to a chemical evolution carousel coordinated by spiral arms \citep{ho2017} or large-scale inflow and outflow driven by spiral arms \citep[e.g.,][]{grand2016} with so far mixed observational support \citep{kreckel2019,li2023,metha2021}. Both the measurements and theory remain novel and represent areas of future growth, and connecting these small-scale enrichment patterns to other diagnostics of feedback and gas flows in galaxies represents a major opportunity for both observations and simulation.
\end{textbox}

\begin{figure}[ht]
\includegraphics[width=\textwidth]{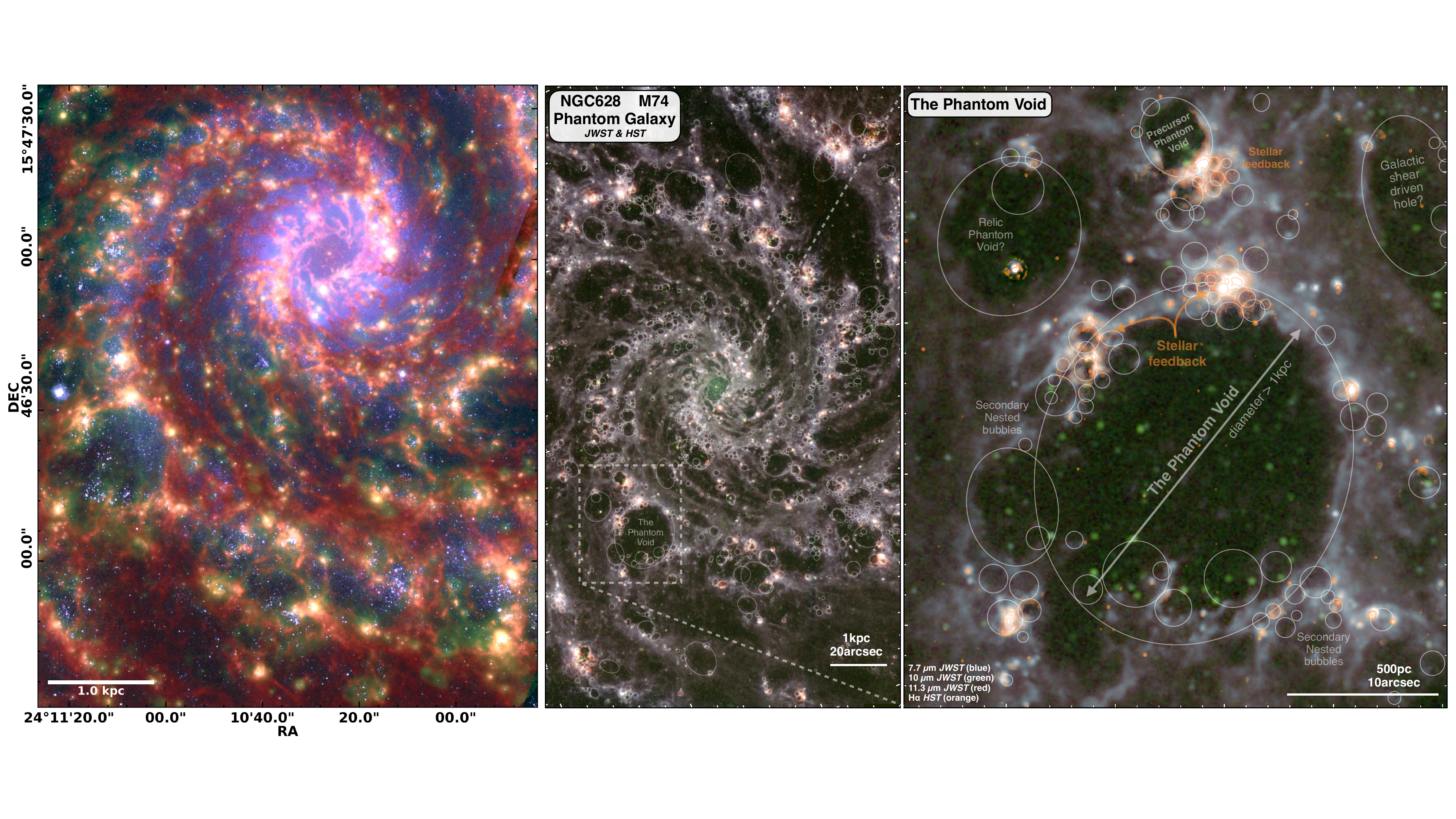}
\caption{JWST observations of the neutral ISM (via its PAH emission) of nearby galaxies reveal abundant bubble-like structures which are likely caused by stellar feedback and might have even triggered new star-forming events at their rims. Panels show the PAH distribution in contrast to other star formation tracers in the inner star-forming disk of NGC\,628. \textit{Left}: RGB image showing the distribution of PAH 7.7$\mu$m emission from JWST/MIRI (red), \ha\ emission from VLT/MUSE (green), and HST B-band continuum tracing young stars \citep[taken from][]{watkins2023a}. \textit{Middle \& right} \citep[taken from][]{barnes2023}: The central 11\,kpc$\times$6\,kpc map and a zoom-in on the largest cavity dubbed 'Phantom Void' show the PAH distribution from JWST (MIRI filters: F1130W in red, F1000W in green, F770W in blue) with continuum-subtracted HST \ha\ imaging overlaid in orange and identified star-forming bubbles from \citet{watkins2023a} as white ellipses.
}
\label{fig:bubbles}
\end{figure}

\subsection{Location and impact of supernovae}
\label{sec:snefeedback}

SNe deposit large amounts of energy and momentum into the ISM. They stir turbulence, support the galactic disk, launch winds, and add new elements to the ISM. Numerical simulations show that the location where SNe explode determines their impact on ISM morphology and sets which ISM phase feels the brunt of the feedback \citep[e.g.,][]{WALCH15FEEDBACK}. For example, the cooling radius $R_{\rm c}$ for a SN in a simulated turbulent medium is approximately $R_{\rm c} \sim 6.3~(n / 100~{\rm cm}^{-3})^{-0.4}$~pc \citep{MARTIZZI15FEEDBACK}.

\subsubsection{Location of supernova explosions}

The results discussed in \S\,\ref{sec:clearingtimefb} suggest that pre-SN feedback often clears the cold gas before the first core-collapse SNe occur. Thus, many SNe should occur in lower density regions and so impact a large physical volume. SN locations should also be clustered, so that later core-collapse SNe at $\tau_\star \sim 20{-}30$~Myr should explode into regions where previous SNe have already occurred. Even allowing for a prompt component, white-dwarf SNe will generally trace an older stellar population and so statistically sample the overall distribution of ISM densities in a region.

These expectations can be tested by observations, and this represents an area of rapid growth for understanding stellar feedback. Most simply, the locations where SN explode can be observed using high resolution CO imaging. Because recent ($\lesssim 100$~yr) SN explosions will not have substantially reshaped their environments on $\sim 10{-}100$~pc scales, cloud-scale measurements probe the gaseous environment likely to be affected by the SN in the future. \citet[][]{mayker2023} show that in PHANGS--ALMA, $\approx 40\%$ of the $40$ core-collapse SNe to explode recently within the survey area lack a nearby massive molecular cloud at $150$~pc resolution, \textbf{in good agreement with significant pre-SN gas clearing and suggesting that many core-collapse SNe exert their feedback directly on the diffuse ISM.} The SNe that are near molecular gas often appear displaced from CO peaks, appearing near cloud edges. This suggests that higher resolution observations may show them to lie in cleared regions on smaller $\ll 100$~pc scales \citep[e.g.,][]{SARBADHICARY23SNR}. These results agree with the observation that $\sim 20{-}30\%$ of Galactic SN remnants (SNR) exhibit some interaction with molecular clouds \citep[e.g.,][]{JIANG10SNR}. They also agree with the generally modest extinctions observed towards type II SNe \citep[e.g.,][]{PEJCHA15SNR}, which imply low line of sight column densities to these objects. Throughout these studies, there is a tendency for stripped envelope SNe, associated with more massive stars and more rapid explosions, to be more associated with nearby molecular gas.

The small samples of well-localized SNe in galaxies with high quality ISM maps currently limit such studies. To overcome this, \citet{SARBADHICARY23SNR} recently proposed that similar information can be obtained by measuring the gaseous environments of near-future core-collapse SN progenitors, i.e., red supergiants and Wolf-Rayet stars. Such stars can be individually picked in space telescope observations in galaxies out to $\sim 20$~Mpc, yielding samples 100s to 1000s times larger than for recent SNe. Initial results from this approach in M33 appear highly consistent with significant pre-clearing of gas. Similarly, as IFU mapping unveils larger and larger samples of supernova remnants \citep{LONG22SNR}, the gaseous environments of these remnants will represent another key constraint, though in these cases the SN shocks \textit{have} significantly affected their environments.

\subsubsection{ISM shells, bubbles, and holes}

The cold ISM displays a strikingly structured morphology. \textsc{Hi} observations reveal large-scale holes and shells in the atomic gas \citep[e.g.,][]{bagetakos2011A,POKHREL20HOLES}. \textsc{Hii} regions and supernova remnants represent expanding, feedback-driven shells and bubbles on small scales (see above). However, the resolution of \textsc{Hi} data remains limited, while the ionized nebulae present a view of gas already affected by feedback and also often  suffer from limited resolution. Very recently, JWST imaging of PAH emission has reinvigorated this field, revealing an intricate network of shells and bubble-like features that pervade galaxy disks \citep[e.g.,][]{watkins2023a,barnes2023}. PAH emission traces the neutral ISM (see Eq.~\ref{eq:ircirrus} and surrounding) with high $\sim 0.3'' \lesssim 30$~pc resolution (at $d \lesssim 20$~Mpc) and sensitivity to even low column density material. Many of these bubbles are visibly associated with star formation activity, and so represent good candidates to trace the impact of feedback on the ISM at $\sim 10{-}1,000$~pc scales.

Comparison between observed bubble populations and those found in simulations offers a promising direction to constrain the overall impact of feedback on the ISM \citep[e.g.,][]{nath2020}. \citet{watkins2023a} demonstrate the potential of such studies in some of the first JWST imaging (see Fig. \ref{fig:bubbles}). In a single star-forming galaxy, they identify $\sim$1,700 bubbles with sizes $\approx 5{-}500$~pc (median $35$~pc) specifically selected based on association with signatures of star formation activity. They compare their inferred size distribution to theoretical expectations and infer that merging of bubbles must significantly affect the observations. Such complex, multi-generation drivers are also observed in case studies of individual shells \citep{barnes2023} and \textsc{Hi} holes in dwarf galaxies \citep{WEISZ09HOLES}.

When kinematics and stellar ages can be obtained, they add essential information to such studies. The observed age, size, and expansion velocity of shells can be compared to predictions from simulations using realistic ISM conditions \citep[e.g.,][]{MARTIZZI15FEEDBACK,WALCH15SN,IFFRIG15SNE,KIM15SNE,KIM17PRESS} to help distinguish the mechanism driving the expansion. Unfortunately, kinematics are not available from JWST imaging and \textsc{Hi} imaging cannot isolate small shells at $d \sim 10{-}20$~Mpc. Recently, \citet{watkins2023b} have identified star formation-driven expanding shells visible in ALMA CO data, while \citet{egorov2023} have identified hundreds of candidates for rapidly expanding shells in the ionized gas. Both studies place empirical constraints on driving mechanism and coupling efficiencies, implying a strong role for SNe in driving the expansions. The \citet{watkins2023b} work suggests significant accumulation of swept up gas in the shells. Moving forward, combining the JWST PAH imaging with kinematic information from ALMA, VLT/MUSE, and \textsc{Hi} represents one of the most promising directions to constrain the combined impact of SNe and pre-SN feedback on the ISM.

\section{GALAXY CENTERS}
\label{sec:centers}

The physical conditions in the inner regions of galaxies can be much more extreme than those found at larger radii. Galaxy centers have high $\Sigma_\star$, short orbital timescales, and high rotational shear. They often host intense massive star formation and/or an active galactic nucleus (AGN), and both phenomena may exert intense feedback and drive large-scale outflows. As discussed in \S\,\ref{sec:cloudprops} and \S\,\ref{sec:denseresults}, the density, temperature, and opacity of molecular clouds vary significantly between centers and galaxy disks, reflecting the influence of these extreme environments. The contrast between galaxy centers and disks is particularly strong for barred spiral galaxies, where cloud-scale $\Sigma_{\rm mol}^{\rm cloud}$ may be $\gtrsim 10\times$ higher in the centers compared to the disks and the cloud-scale internal pressure, $P_{\rm int}$, is on average $\gtrsim300\times$ higher than that in the disks \citep[][see \S\,\ref{sec:cloudprops}]{SUN20GMCS}. 

Major mergers can produce similar conditions but are less common than barred galaxies in the local universe \citep[e.g.,][]{kormendy2004}. As a result, galaxy centers offer the most accessible high-density, high-activity environment in nearby galaxies. Multi-wavelength observations that survey galaxy disks often cover the central region. As a result, the last decade has seen a large increase in multi-wavelength, high-resolution observations of these key regions. We discuss the statistical properties of cloud-scale CO emission and dense gas tracers in these regions above. Here we expand on the nature of these key environments and the likely physical drivers of the distinct gas and star formation properties found in galaxy centers.

\begin{figure}[ht]
\includegraphics[width=\textwidth]{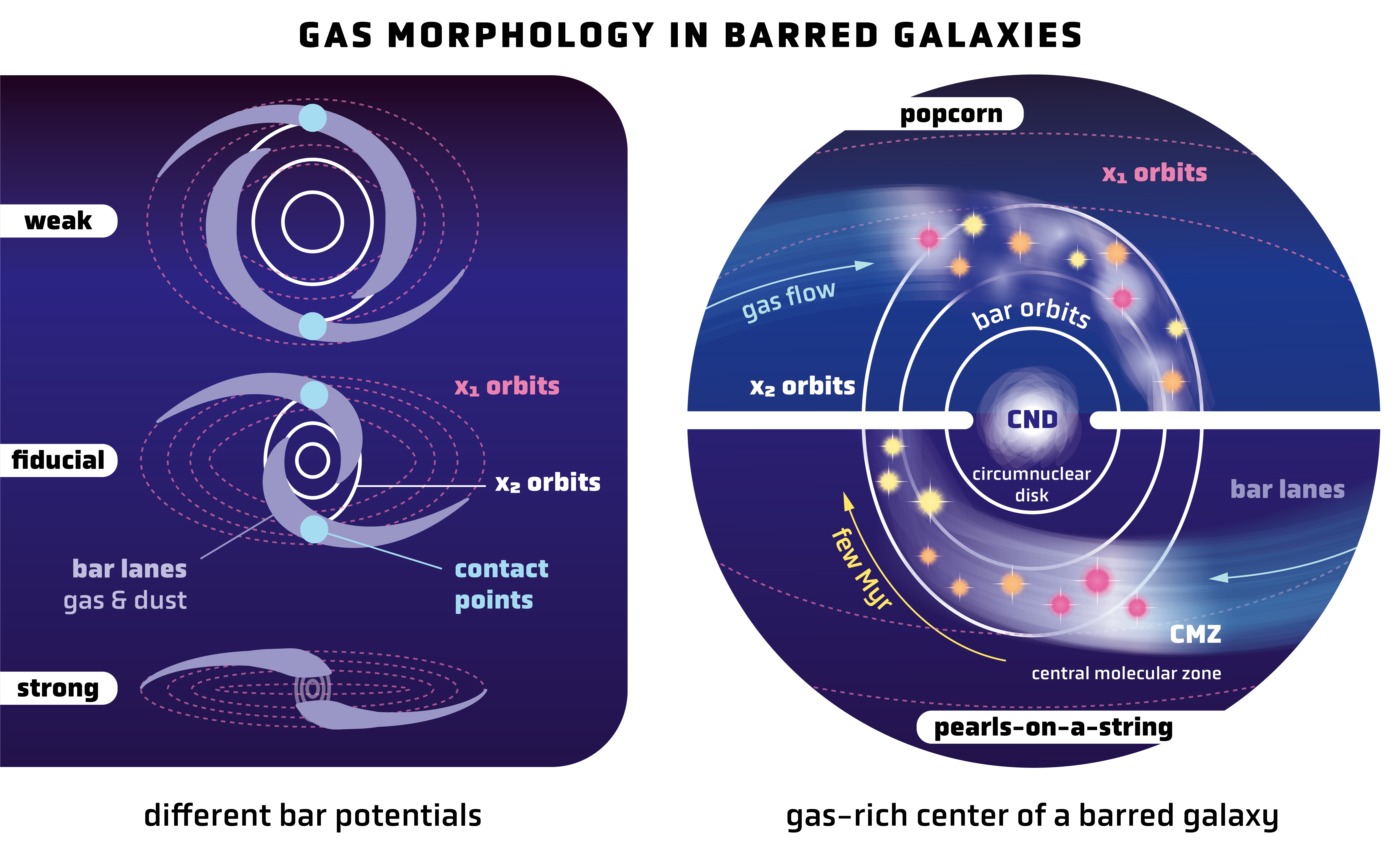}
\caption{Gas morphology in barred galaxies. \textit{Left:} The geometry of the bar potential is expected to strongly affect the ISM distribution. Gas and dust lanes (shown in pale blue) form along the leading side of the bar (clockwise rotation) and may form a spiral- or ring-like structure in the galaxy center \citep[based on][]{athanassoula1992}. While the $x_1$ orbits along the bar major axis (dashed pink ellipses) always exist, the presence of the perpendicular $x_2$ orbit family oriented along the bar minor axis (white ellipses) depends on the exact bar potential. \textit{Right:} Schematic illustrating two scenarios proposed for the star formation process in the central rings of barred galaxies \citep[following][]{boeker2008}. In the \textit{popcorn} scenario (top half), gas density fluctuations result from small-scale, star formation-driven processes. This leads to a stochastic gas density distribution and, hence, randomly distributed stellar ages. In the \textit{pearls-on-a-string} scenario (bottom half), gas density builds up at preferred locations due to large-scale gas flows. This leads to a systematic age gradient along the ring. In analogy to the Milky Way center, the spiral- or ring-like structure can be referred to as CMZ (central molecular zone) or central ring. If present, the inner small gas disk can be referred to as CND (circumnuclear disk). Gas density is represented from high to low in bright to dark colors, while the relative age of (and the attenuation affecting) the star clusters increases (decreases) from red to yellow.}
\label{fig:center_schematic}
\end{figure}

\begin{marginnote}[]
\entry{CMZ}{central molecular zone, defined in the Milky Way as ring-like molecular gas overdensity with radius $\lesssim 300$\,pc. }
\entry{CND or CNR}{circumnuclear disk or ring. A gas disk or ring close to the nucleus, occurring within the CMZ and having radii $\rm \lesssim 100\,pc$ ($\lesssim 10$~pc in the Milky Way). }
\end{marginnote}

\begin{textbox}[ht]
\section{Central Molecular Zones (CMZs) and Circumnuclear Disks (CNDs)}
\label{sec:cmz}

Recent publications often refer to molecular gas-rich centers of (barred) spiral galaxies as \textit{central molecular zones} \citep[CMZs;][]{serabyn1996}. These are defined in loose analogy to the Milky Way's CMZ, a ring-like accumulation of molecular gas with a diameter of $\sim$200{-}400\,pc \citep[e.g.,][]{henshaw2023} generated by inward matter transport due to the Milky Way’s large stellar bar. Inside the CMZ, the Milky Way also harbors a smooth circumnuclear disk or ring (CND or CNR) of molecular gas with diameter $\sim$10-15\,pc. 

The Galactic CMZ appears similar to the ring- or spiral-like star-forming or dust rings seen in the inner regions of nearby galaxies. Though sometimes referred to as ``nuclear'' rings, we suggest to maintain a clear analogy to the Milky Way and refer to these as \textit{central} rings. Studying a large set of nearby galaxies, \citet{comeron2014} find that about $35\%$ of galaxies host such central rings. Analogs to the CMZ are also visible in the molecular gas, where they were originally identified as \textit{twin peaks} of CO emission in the centers of barred galaxies \citep[][]{kenney1992}. Such features are present in about $30\%$ of nearby massive star-forming galaxies, with most CMZs ($\sim 60\%$) found in barred hosts \citep[][]{stuber2023}. Note that not every barred galaxy has a clear ring- or spiral-like molecular gas feature in its center. In Fig.\,\ref{fig:center_schematic} we provide a schematic of the main characteristics in the center of a barred galaxy with a proposed naming convention.
\end{textbox}

\subsection{Formation and Evolution of Gas-Rich Centers of Barred Galaxies}
\label{sec:centers_gas}

Ring- or spiral-like concentrations of molecular gas and young stars are common in the inner regions of barred disk galaxies \citep[e.g.,][]{stuber2023}. These rings, or \textit{central molecular zones} (CMZs, see text box), are widely agreed to be created by gas inflows resulting from the highly non-axisymmetric, tumbling gravitational potential of stellar bars. The exact physical mechanisms by which rings or CMZs form remains a topic of active research \citep[e.g.,][]{athanassoula1992,SORMANI23BAR}, but observations show that the magnitude of gas flow to the center can easily reach $\gtrsim 1$~M$_\odot$~yr$^{-1}$ \citep[e.g.,][]{haan2009}. 

Two main scenarios have been proposed for how gas collapses and star formation proceeds in rings \citep[][see Fig.\,\ref{fig:center_schematic}]{boeker2008}. In the \textit{popcorn} scenario, gas accumulates along the ring and then collapses to form massive stars due to the influence of local processes. This results in star formation occurring at random locations along the ring.

In the \textit{pearls-on-a-string} scenario, gas enters the ring from the gas and dust lanes along the bar. Gas accumulates at these positions, often called contact points, undergoes star formation, and then the newly formed massive star clusters travel along the ring. This results in a visible age sequence, with the youngest clusters near the contact points and older ones closer to the opposite contact point Modifications of this scenario have been suggested in which star formation is triggered at preferred locations along the gas orbit, e.g., pericenter, rather than at the contact points \citep[see][]{henshaw2023}.

Regardless of location, there is good reason to expect star formation in central rings to be bursty in nature. The rate of star formation in a CMZ should depend on the rate at which matter is transported inward by stellar bars or spiral arms to the galaxy center. CMZs may also experience a cycle of star formation and feedback, in which the energy deposited due to feedback from the previous generation of star formation must be dissipated before a new generation of stars can form
\citep[see discussion in][]{henshaw2023}. 

\smallskip
\textbf{Theoretical expectations --} Hydrodynamical simulations of barred galaxies suggest a close link between the properties of the gas inflow and the resulting patterns of star formation. For example, \citet{seo2013} report that stellar cluster age gradients only emerge for low gas inflow rates, appearing when the gas densities are just high enough for star formation to occur at the contact points. \citet{moon2022} show that asymmetric inflow rates appear to be required to create lop-sided distributions of star formation. Simulations also suggest bursty star formation in CMZs due to stellar feedback \citep[e.g.,][]{armillotta2019}, as supernovae from previous bursts can increase the turbulent kinetic energy in the gas and temporarily suppress star formation. Meanwhile, the formation of an inner CND appears sensitive to the physics included in the simulation. For example, \citet{tress2020} notice that CNDs form only in simulations that include both gas self-gravity and stellar feedback, while \citet{moon2023} find that they only develop in the presence of magnetic fields. More broadly, simulations of CMZs still vary dramatically as the set-up of the gravitational potential, gas flows, and physical prescriptions implemented for, e.g., stellar feedback, gas self-gravity, and magnetic fields can differ markedly between studies.

Simulations also demonstrate a large impact of fast dynamical times on observations of CMZs. Because the time evolution of the gas and star formation proceeds rapidly, \citet{sormani2020} note that patterns in the stellar output may only emerge when averaged over time and not be accessible in observations of any single system. \citet{schinnerer2023} pushed this conclusion even farther forward, carrying out comparisons between tailored hydrodynamical simulations and observations of the prominent CMZ in NGC\,1365. They showed that any visible age trends among formed stars tend to be transient and to vary dramatically as a function of time. As a result, observed age trends in CMZs (or at least NGC\,1365's CMZ) should be viewed with suspicion. Their work also demonstrates that efforts combining observations with dedicated simulations for a larger number of CMZs will be critical to settle the issue of how star formation proceeds within CMZs. 

\smallskip
\textbf{Observational examples --} There have been many mm- and radio-wave studies of gas-rich galaxy centers. These have had varied scientific goals, including understanding the fueling and impact of AGN \citep[e.g., NUGA, GATOS,][]{GARCIABURILLO03GMCS,garcia-burillo2021}, measuring supermassive black hole masses \citep[e.g., WISDOM,][]{davis2018}, or probing astrochemistry via spectral line surveys \citep[e.g., ALCHEMI,][]{ALCHEMI21SURVEY}. Because of the broad range of science goals associated with the observations and the heterogeneous observing strategies, there is still relatively little synthetic work on samples of CMZs as such. The work mentioned in \S\,\ref{sec:cloudmorph} and \S\,\ref{sec:denseresults} represents, to our knowledge, the best synthetic statistical work on the properties of gas in galaxy centers.

Fortunately, the literature is rich with individual case studies of galaxy centers. To illustrate the diversity seen, we highlight recent studies of CMZs in three barred galaxies. NGC\,253, NGC\,1365, and NGC\,5806 have similarly high central SFRs of $\rm 2-5\,\msun\,yr^{-1}$, about $\sim 20{-}50\times$ higher than that of the Milky Way's CMZ. Their central molecular gas reservoirs differ in  $\sim$1\,dex steps from one another and the Milky Way \citep[MW CMZ: $\sim 10^6$~M$_\odot$, NGC\,253: $\sim 10^8$~M$_\odot$,  NGC\,5806: $\rm 10^9\,\msun$, NGC\,1365: $\rm 10^{10}\,\msun$;][]{henshaw2023,BENDO15SFR,choi2023,schinnerer2023}. The common SFR with widely varying $M_{\rm mol}$ illustrates the wide range in star formation activity found in CMZs discussed below. Within all three CMZs, some age trend is apparent from the distribution of forming massive star clusters and the properties of the clouds themselves \citep{levy2022,schinnerer2023,choi2023}, but note the caveats related to rapid evolution of CMZs discussed above. Variations of molecular gas properties are also visible in all three CMZs and appear related to different physical drivers: massive stellar feedback in NGC\,253 \citep[][]{krieger2020}; inter-cloud motions in converging large-scale flows in NGC\,1365 \citep[][]{schinnerer2023}; and cloud-cloud collisions and shear in NGC\,5806 \citep[][]{choi2023}. 

Moving closer towards the nucleus, analogs to the 10-15\,pc diameter MW CND have also been observed in barred galaxies. We highlight the impressive CNDs in NGC\,1808 \citep[diameter $\sim$100\,pc,][]{salak2017} and NGC\,1365 \citep[diameter $\lesssim$1\,kpc,][]{schinnerer2023}. The significant difference in size appears to be driven by the difference in length of the stellar bars (6\,kpc vs. 28\,kpc)\footnote{The length of the bar in the Milky Way is debated with older studies reporting $\sim$7\,kpc while newer analyses suggest $\sim$10\,kpc \citep[e.g., see \S1 of][]{hilmi2020}.}. Both CNDs lack star formation and might be similar in structure to the smooth molecular gas disks seen in early-type galaxies where shear or tidal forces are thought to suppress cloud collapse \citep[see \S\,\ref{sec:centers_activity}; also][]{DAVIS22GMCS}.

CMZs also tend to produce the brightest molecular line emission from galaxies, allowing a wide range of molecular transitions to be observed at high physical resolution \citep[e.g.,][]{meier2004,MEIER05LINES}. As a result, mm-wave spectroscopy of CMZs has been used to diagnose the variable physical conditions found in these regions, assessing density, excitation, molecular abundance \citep[e.g., key goals of the ALCHEMI ALMA large program,][]{ALCHEMI21SURVEY}, and even the presence of shocks \citep[e.g.,][]{harada2019}. These studies are key to understand astrochemistry and sharpen our ability to infer physical conditions from molecular spectroscopy; increasingly they are also key to understand the physics of star formation and cloud evolution within CMZs \citep[e.g.,][]{eibensteiner2022,liu2023}.

\subsection{Star Formation in Gas-Rich Centers of Barred Galaxies}
\label{sec:centers_activity}

Thanks to their high concentrations of gas, gas-rich centers host some of the most intense star formation activity in the local universe. They also represent a prime location for interactions between AGN and the surrounding ISM and the main launching sites of galaxy-scale outflows. Gas-rich centers exhibit a wide spread in gas depletion times \citep[][]{querejeta2021}. Despite the larger uncertainty on their CO conversion factor \citep[see \S\,\ref{sec:alphaco}; e.g.,][]{TENG22XCO}, they sometimes appear strikingly inefficient, forming significantly fewer stars than expected given their abundant, high-density gas (see \S\,\ref{sec:hcnco}).

\smallskip
\textbf{Heavily embedded star formation activity --}
Due to their high gas surface densities and small sizes, CMZs represent the most challenging locations within local galaxies to account for the effects of extinction on star formation tracers. Thus CMZs, especially in highly inclined galaxies, also represent the most likely locations for optical or UV tracers to miss embedded formation of massive stars. As a result, IR and radio star formation tracers have proven critical to study CMZs. JWST promises to revolutionize this topic by offering an unbiased view of dust, stars, and recombination lines from young stellar populations. Demonstrating this promise, \citet{whitmore2023} already doubled the number of known young ($\leq$10\,Myr), massive ($\rm \mstar \geq 10^6\,\msun$) stellar clusters in the star-bursting center of NGC\,1365 to $\sim$30 using JWST near-IR imaging.

\smallskip
\textbf{Key sites to study young massive cluster formation --} 
CMZs often harbor young massive clusters (YMCs) and even superstar clusters (SSCs). Evidence remains mixed regarding whether the cluster mass function or maximum cluster mass change as a function of environment \citep{KRUMHOLZ19REVIEW}, but in practice their high $\Sigma_{\rm SFR}$ and large gas reservoirs make CMZs the best local environments to find and characterize proto-YMCs. In the nearest CMZs, ALMA sub-mm, JWST IR, and VLA radio observations can achieve resolutions of a few parsec and better, sufficient to resolve forming clusters at wavelengths largely robust to extinction. ALMA has proven particularly powerful in this regard, e.g., by identifying $> 40$ proto-SSCs via recombination line, free-free, and dust emission in the CMZs of NGC 253 and NGC 4945, the two nearest barred massive spirals \citep[e.g.,][]{leroy2018,EMIG20SFR}. In these systems, these proto-SSCs are so bright that detailed ionized and molecular line spectroscopy capture outflow signatures, e.g., via P-Cygni line profiles, and recover the density and temperature structure of the forming clusters, e.g., via vibrationally excited molecular lines \citep[e.g.,][]{LEVY21CLUSTERS,bellocchi2023}.

\smallskip
\textbf{Star formation suppression --}
Observations also reveal significant reservoirs of molecular gas in some galaxy centers that harbor little or no star formation activity \citep[e.g.,][]{LIU21GMCS,DAVIS22GMCS,schinnerer2023}. The central regions of early-type galaxies, in particular, often exhibit lower star formation activity than expected from their gas reservoirs \citep{davis2014}. The morphology of gas reservoirs associated with suppressed star formation often appears remarkably smooth \citep[][]{DAVIS22GMCS}, consistent with dynamical stabilization of the gas against collapse. Shear has been invoked as an important mechanism to achieve such stabilization \citep[e.g.,][]{martig2009,davis2014,meidt2015}. Case studies using high spatial resolution ALMA data suggest that galactic rotational shear likely does stabilize the smooth CO disk in the lenticular galaxy NGC\,4429 \citep{LIU21GMCS}, while either shear or tidal forces could explain the lack of star formation in the CND of NGC\,1365  \citep[][]{schinnerer2023}. Hydrodynamical simulations also suggest dynamical stabilization to be important; e.g., \citet[][]{gensior2020} find smooth gas morphologies in simulations when a significant bulge component is present as long as they implement a star formation threshold that depends on $\alpha_{\rm vir}$ (\S\,\ref{sec:cloudprops}, \S\,\ref{sec:sfe}). 

CMZs also often host outflows driven by either AGN or star formation \citep[see reviews by][Thompson \& Heckman in this volume]{veilleux2020}. Most directly, cold outflows deplete the molecular gas reservoir, often several times faster than star formation itself \citep[e.g.,][]{krieger2019}. This removes fuel for future star formation and may contribute to the burstiness of CMZs mentioned above. AGN jets and outflows may also exert negative feedback on the gas in the central disk, rendering it more turbulent, hotter, and overall less prone to star formation (e.g., M\,51: \citealt{querejeta2016} and NGC\,1068: \citealt{saito2022}).

\section{SUMMARY AND OUTLOOK}
\label{sec:summary}

Over the last decade, large areas in representative samples of star-forming main sequence galaxies have been observed with overlapping multi-wavelength surveys that have resolution matched to the scale of individual molecular clouds, \hii\ regions, and young star clusters. These \textit{cloud-scale} observations spatially separate the different evolution phases of the gas--star formation--feedback matter cycle (Fig.\,\ref{fig:sketch_cloud}). This allows for measurements of the demographics of multiple phases of the matter cycle, a clear view of how these demographics depend on galactic environment, and statistical inferences of the timescales and efficiencies associated with star formation and stellar feedback processes.

\begin{summary}[SUMMARY POINTS]

\begin{enumerate}

\item Major investments by ALMA have led to a highly complete view of cloud-scale CO emission from local, massive main sequence star-forming galaxies (\S \ref{sec:moleculargas}). The cloud-scale molecular gas surface density ($\Sigma_{\rm mol}$), line width ($\sigma_{\rm mol}$), and internal pressure ($P_{\rm int}$) vary systematically. These variations correlate with the local galactic environment, so that $\Sigma_{\rm mol}$, $\sigma_{\rm mol}$, and $P_{\rm int}$ are all higher in regions where the large-scale dynamical equilibrium pressure ($P_{\rm DE}$) and stellar and gas surface density are high. Cloud-scale conditions also vary between galactic morphological environments, with higher $\Sigma_{\rm mol}$, $\sigma_{\rm mol}$, and $P_{\rm int}$ in galaxy centers, mergers, and early-type galaxies (\S \ref{sec:cloudmorph}). Most molecular gas approximately obeys the \textit{Heyer-Keto relation} between $\Sigma_{\rm mol}$ and $\sigma_{\rm mol}$ at fixed size scale (Eq. \ref{eq:heyerketo}) expected for objects virialized or bound with respect to their own self-gravity.

\smallskip

\item Mm-wave spectroscopy can constrain the physical density distribution in the molecular gas (\S \ref{sec:densegasmeaning}). Dedicated surveys have mapped several dozen resolved galaxies in the dense gas tracer HCN. Consistent with the cloud-scale CO results, gas density, traced, e.g., by HCN/CO, appears higher in regions with high large-scale mean stellar surface density and high $P_{\rm DE}$ (\S \ref{sec:hcnco}). Individual parts of galaxies fall on the \textit{Gao-Solomon relation} relating star formation activity and dense gas tracer luminosity (\S \ref{sec:denseresults}). Observations also reveal systematic variations in the ratio of SFR/HCN as a function of local environment, with lower SFR/HCN in high $P_{\rm DE}$, high stellar surface density regions. This suggests a picture in which HCN/CO and similar ratios trace gas density but they do not always access over-dense, self-gravitating structures, especially in environments with high mean gas density.

\smallskip

\item High resolution observations of gas and star formation tracers allow the statistical inference of timescales associated with the matter cycle (\S \ref{sec:cloudevolution}). Massive clouds appear to experience a ramp-up period of $\sim 5{-}20$~Myr before the onset of intense high-mass star formation. This is comparable to the gravitational free-fall or turbulent crossing times inferred for the gas. After high-mass star formation begins, multiple methods find evidence for fast, $\lesssim 5$~Myr, clearing of gas from these regions before the first SNe.

\smallskip

\item This rapid gas clearing implies a strong role for winds, radiation, and thermal gas pressure (often collectively called pre-supernova feedback) in disrupting molecular clouds (\S \ref{sec:clearingtimefb}, \ref{sec:hiifeedback}). This conclusion is supported by the large fraction of core-collapse SNe and SN progenitors observed without associated molecular gas (\S\,\ref{sec:snefeedback}). Apparently, pre-SN feedback ``clears the way'' for SNe to explode into a lower density medium and exert a large influence on the low density, extended ISM.

\smallskip

\item On large scales, the net effect of feedback can be inferred by invoking the need for feedback to support the weight of the ISM in the galaxy potential (\S \ref{sec:pdefeedback}). Current observations of main sequence star-forming galaxies imply large-scale feedback yields of $\Upsilon_{\rm fb} \approx 1,300{-}1,700$~km~s$^{-1}$. SNe also appear likely to cause many of the shells and holes visible in tracers of cold gas and dust. Characterizing these represents an important next step for feedback studies (\S \ref{sec:snefeedback}).

\smallskip

\item High resolution observations of tracers of gas, gas kinematics, and star formation also allow for the statistical inference of key timescales and the efficiency of star formation relative to these timescales (\S \ref{sec:sfe}). The distributions of cloud-scale gravitational free-fall times, turbulent crossing times, and shearing times are now determined across representative samples of nearby galaxies (\S \ref{sec:sfe}). Contrasting these with the large-scale molecular gas depletion time, $\tau_{\rm dep}^{\rm mol}$, yields observational estimates of the star formation efficiency. The theoretically important star formation efficiency per free-fall time is $\epsilon_{\rm ff} \approx 0.5_{-0.3}^{+0.7}\%$ at cloud-scales, and the implied value for the dense gas is similar $\epsilon_{\rm ff}^{\rm dense} \approx 0.3\%$. However, the correlations between $\epsilon_{\rm ff}$ and cloud-scale gas properties predicted by turbulent star formation models remain elusive, and this may pose a challenge to these models (\S \ref{sec:sfe}).

\smallskip

\item In galaxy centers, molecular gas and star formation properties can drastically differ from those seen in galactic disks. Centers offer easy access to extreme environments, where the impact of large-scale dynamics and high star-formation and/or AGN activity can be studied and our understanding of the matter cycle can be tested (\S\,\ref{sec:centers}). 

\smallskip

\item All of this work leverages continued improvements in our ability to infer the amount of molecular gas from CO (\S \ref{sec:alphaco}), physical gas density from mm-wave spectroscopy (\S \ref{sec:densegasmeaning}), the mass and likely age of young stars from star formation tracers (\S \ref{sec:SF}), and the impact of stellar feedback from multiwavelength data (\S \ref{sec:hiifeedback},\ref{sec:snefeedback}). The last decade has seen advances in all of these topics, with the combination of numerical modeling, extragalactic and Galactic observations being the key to progress.

\end{enumerate}

\end{summary}

These new surveys represent significant investments by major facilities including ALMA, HST, VLT/MUSE, and JWST. Their combination clearly represents more than the sum of their individual parts, and the joint datasets are becoming one of the main tools to understand the physics of the matter cycle. Ultimately, the hope is that such aligned cloud-scale surveys will play a role in understanding region evolution analogous to that played by the deep field surveys illuminating galaxy evolution. Hence, one immediate recommendation is to continue to obtain and exploit these data, compiling the best possible cloud-scale view of star formation in disk galaxies. Another key step will be to establish more concrete comparisons to numerical simulations, which are critical tools to express physical hypotheses given the multi-scale, multi-phase, complex nature of the topic. Beyond this, we highlight several clear next steps to make observational progress understanding the matter cycle.

\begin{issues}[FUTURE ISSUES \& DIRECTIONS]

\begin{enumerate}

\item Observations that \textbf{resolve individual regions in tracers of gas, dust, and star formation activity} are critical to test the emerging picture described here. Such observations will access the density and kinematic structure of the ISM on the scales where self-gravity dominates and allow for a resolved view of the process (or absence) of cloud disruption by stellar feedback. For these studies to provide general answers, they should target diverse galaxy samples and cover large, contiguous areas at the key $\approx 5{-}20$~pc resolution. ALMA, HST, JWST, NOEMA, and VLT/MUSE are capable of such observations targeting very nearby galaxies now and a next generation VLA (ngVLA) could push such studies out to beyond the Virgo Cluster, accessing an even more diverse sample of galaxies.

\smallskip

\item Measuring the \textbf{cloud-scale structure of the atomic gas and hot gas} also represents an important next step. X-ray emission from hot, shocked gas is a key tracer of feedback, but there has been relatively little work to integrate this view with observations of the colder neutral and ionized gas. Meanwhile, \textsc{Hi} represents most of the ISM mass in galaxy disks, and its physical state is critical to the formation of molecular clouds. Progress can be  achieved with current facilities, including \textit{Chandra}, MeerKAT, ASKAP, and the Karl G. Jansky Very Large Array (VLA) by targeting the closest galaxies. In the longer term, \textit{Athena}, the Square Kilometer Array (SKA), and the ngVLA promise to make major progress.

\smallskip

\item We have emphasized results for relatively massive, main sequence star-forming galaxies and galaxy centers, which illuminate the core physics driving the matter cycle. Extending this multi-phase, cloud-scale view to \textbf{representatives samples of merger-driven starbursts, dwarf galaxies, early-type galaxies, members of galaxy clusters, and edge-on systems} will be critical to understand how the matter cycle in galaxies \textit{varies} across the universe.

\smallskip

\item We advocate increased focus on the galaxy scaling relations between \textbf{(a) $\Sigma_{\rm SFR}$ and $P_{\rm DE}$ and (b) density sensitive line ratios like HCN/CO and environment}, because they constrain the feedback yield ($\Upsilon_{\rm fb}$) and the cloud-scale gas properties. The sample sizes and diversity of systems surveyed remain limited for both relations because of the difficulty assembling the extensive multi-wavelength view needed and signal-to-noise limitations for the dense gas tracers.

\smallskip

\item The widespread availability of modeling for individual young stellar populations represents an important step forward. It will be important to \textbf{better link results from cluster SED modeling and spectral fitting of young stellar populations to the multi-phase view of the ISM.} With results from \textit{Euclid} and \textit{Roman} imminent, our view of resolved clusters in galaxies will expand dramatically and work to make best use of these data (which lack UV coverage) will also be key.

\smallskip

\item Almost all of the work described here depends on \textbf{accurate translation of observables into physical quantities}. Continued progress in this area is essential, especially work to (a) sharpen our inference of gas mass, density, and dynamical state from mm-line and mid-IR dust emission observations, (b) model the masses and ages of individual young stellar populations from multi-wavelength data, and (c) assess the strength and coupling to the ISM of the various stellar feedback mechanisms. The path forward mixes increasing realistic numerical experiments with expensive observational case studies, and producing realistic uncertainties appropriate for use in statistical analysis will be as important as improved estimators.

\end{enumerate}

\end{issues}

\section*{DISCLOSURE STATEMENT}

The authors are not aware of any affiliations, memberships, funding, or financial holdings that might be perceived as affecting the objectivity of this review. 

\section*{ACKNOWLEDGMENTS}

We gratefully acknowledge the support of many long-term collaborators who contributed key work reviewed here. These include the members of the PHANGS, PAWS, and EMPIRE teams. We gratefully acknowledge Ashley Barnes, I-Da Chiang, Maria Jesus Jimenez Donaire, Kathryn Kreckel, Justus Neumann, Lukas Neumann, Sophia Stuber, Jiayi Sun, and Antonio Usero for their assistance with the figures, tables, and calculations presented here as well as their valuable input on the content of the review. We also thank Ashley Barnes, Jaeyeon Kim, Akiko Kawamura, Kathryn Kreckel, Sharon Meidt, Jiayi Sun, and Elizabeth Watkins for permission to include figures from their works. In addition, we thank Francesco Belfiore, Mederic Boquien, Alberto Bolatto, Eric Emsellem, Simon Glover, Brent Groves, Jonathan Henshaw, Kathryn Kreckel, Janice Lee, Sharon Meidt, Miguel Querejeta, Karin Sandstrom, and Mattia Sormani for their input on the text. We also gratefully acknowledge the editorial input of Eve Ostriker and Robert Kennicutt. This arxiv version benefited immensely from a careful review and input by Todd Thompson. E.S. thanks the Caltech astronomy department and IPAC for their hospitality while writing this article and acknowledges helpful conversations with Lee Armus and Philipp Hopkins. E.S. gratefully acknowledges support via the Deutsche Forschungsgemeinschaft and the European Union's Horizon 2020 programme that enabled some of the research presented. A.K.L. gratefully acknowledges support from a Humboldt Research award and the National Science Foundation.

\bibliography{araa_refs}{}

\begin{thebibliography}{}
\expandafter\ifx\csname natexlab\endcsname\relax\def\natexlab#1{#1}\fi

\bibitem[{{Accurso} et~al.(2017){Accurso}, {Saintonge}, {Catinella}, {Cortese},
  {Dav{\'e}} et~al.}]{ACCURSO17XCO}
{Accurso} G, {Saintonge} A, {Catinella} B, {Cortese} L, {Dav{\'e}} R, et~al.
  2017.
\textit{\mnras} 470:4750--4766

\bibitem[{{Agertz} et~al.(2013){Agertz}, {Kravtsov}, {Leitner} \&
  {Gnedin}}]{AGERTZ13FEEDBACK}
{Agertz} O, {Kravtsov} AV, {Leitner} SN, {Gnedin} NY. 2013.
\textit{\apj} 770:25

\bibitem[{{Armillotta} et~al.(2019){Armillotta}, {Krumholz}, {Di Teodoro} \&
  {McClure-Griffiths}}]{armillotta2019}
{Armillotta} L, {Krumholz} MR, {Di Teodoro} EM, {McClure-Griffiths} NM. 2019.
\textit{\mnras} 490:4401--4418

\bibitem[{{Armus} et~al.(2023){Armus}, {Lai}, {U}, {Larson}, {Diaz-Santos}
  et~al.}]{armus2023}
{Armus} L, {Lai} T, {U} V, {Larson} KL, {Diaz-Santos} T, et~al. 2023.
\textit{\apjl} 942:L37

\bibitem[{{Armus} et~al.(2009){Armus}, {Mazzarella}, {Evans}, {Surace},
  {Sanders} et~al.}]{GOALS09SURVEY}
{Armus} L, {Mazzarella} JM, {Evans} AS, {Surace} JA, {Sanders} DB, et~al. 2009.
\textit{\pasp} 121:559

\bibitem[{{Athanassoula}(1992)}]{athanassoula1992}
{Athanassoula} E. 1992.
\textit{\mnras} 259:345--364

\bibitem[{{Bagetakos} et~al.(2011){Bagetakos}, {Brinks}, {Walter}, {de Blok},
  {Usero} et~al.}]{bagetakos2011A}
{Bagetakos} I, {Brinks} E, {Walter} F, {de Blok} WJG, {Usero} A, et~al. 2011.
\textit{\aj} 141:23

\bibitem[{{Ballesteros-Paredes} et~al.(2011){Ballesteros-Paredes}, {Hartmann},
  {V{\'a}zquez-Semadeni}, {Heitsch} \& {Zamora-Avil{\'e}s}}]{BALLESTEROS11GMCS}
{Ballesteros-Paredes} J, {Hartmann} LW, {V{\'a}zquez-Semadeni} E, {Heitsch} F,
  {Zamora-Avil{\'e}s} MA. 2011.
\textit{\mnras} 411:65--70

\bibitem[{{Bally} et~al.(1987){Bally}, {Stark}, {Wilson} \&
  {Henkel}}]{BALLY87DENSE}
{Bally} J, {Stark} AA, {Wilson} RW, {Henkel} C. 1987.
\textit{\apjs} 65:13

\bibitem[{{Barnes} et~al.(2021){Barnes}, {Glover}, {Kreckel}, {Ostriker},
  {Bigiel} et~al.}]{barnes2021}
{Barnes} AT, {Glover} SCO, {Kreckel} K, {Ostriker} EC, {Bigiel} F, et~al. 2021.
\textit{\mnras} 508:5362--5389

\bibitem[{{Barnes} et~al.(2020){Barnes}, {Kauffmann}, {Bigiel}, {Brinkmann},
  {Colombo} et~al.}]{BARNES20DENSE}
{Barnes} AT, {Kauffmann} J, {Bigiel} F, {Brinkmann} N, {Colombo} D, et~al.
  2020.
\textit{\mnras} 497:1972--2001

\bibitem[{{Barnes} et~al.(2023){Barnes}, {Watkins}, {Meidt}, {Kreckel},
  {Sormani} et~al.}]{barnes2023}
{Barnes} AT, {Watkins} EJ, {Meidt} SE, {Kreckel} K, {Sormani} MC, et~al. 2023.
\textit{\apjl} 944:L22

\bibitem[{{Barrera-Ballesteros} et~al.(2021){Barrera-Ballesteros},
  {S{\'a}nchez}, {Heckman}, {Wong}, {Bolatto} et~al.}]{BARRERA21PRESS}
{Barrera-Ballesteros} JK, {S{\'a}nchez} SF, {Heckman} T, {Wong} T, {Bolatto} A,
  et~al. 2021.
\textit{\mnras} 503:3643--3659

\bibitem[{{Belfiore} et~al.(2023){Belfiore}, {Leroy}, {Williams}, {Barnes},
  {Bigiel} et~al.}]{BELFIORE23SFRJWST}
{Belfiore} F, {Leroy} AK, {Williams} TG, {Barnes} AT, {Bigiel} F, et~al. 2023.
\textit{\aap} 678:A129

\bibitem[{{Belfiore} et~al.(2022){Belfiore}, {Santoro}, {Groves}, {Schinnerer},
  {Kreckel} et~al.}]{BELFIORE22DIG}
{Belfiore} F, {Santoro} F, {Groves} B, {Schinnerer} E, {Kreckel} K, et~al.
  2022.
\textit{\aap} 659:A26

\bibitem[{{Bellocchi} et~al.(2023){Bellocchi}, {Mart{\'\i}n-Pintado},
  {Rico-Villas}, {Mart{\'\i}n} \& {Jim{\'e}nez-Sierra}}]{bellocchi2023}
{Bellocchi} E, {Mart{\'\i}n-Pintado} J, {Rico-Villas} F, {Mart{\'\i}n} S,
  {Jim{\'e}nez-Sierra} I. 2023.
\textit{\mnras} 519:L68--L73

\bibitem[{{Bemis} \& {Wilson}(2023)}]{BEMIS23DENSE}
{Bemis} AR, {Wilson} CD. 2023.
\textit{\apj} 945:42

\bibitem[{{Bendo} et~al.(2015){Bendo}, {Beswick}, {D'Cruze}, {Dickinson},
  {Fuller} \& {Muxlow}}]{BENDO15SFR}
{Bendo} GJ, {Beswick} RJ, {D'Cruze} MJ, {Dickinson} C, {Fuller} GA, {Muxlow}
  TWB. 2015.
\textit{\mnras} 450:L80--L84

\bibitem[{{Berg} et~al.(2020){Berg}, {Pogge}, {Skillman}, {Croxall},
  {Moustakas} et~al.}]{BERG20SFR}
{Berg} DA, {Pogge} RW, {Skillman} ED, {Croxall} KV, {Moustakas} J, et~al. 2020.
\textit{\apj} 893:96

\bibitem[{{Bergin} \& {Tafalla}(2007)}]{BERGIN07REVIEW}
{Bergin} EA, {Tafalla} M. 2007.
\textit{\araa} 45:339--396

\bibitem[{{Bertoldi} \& {McKee}(1992)}]{BERTOLDI92GMCS}
{Bertoldi} F, {McKee} CF. 1992.
\textit{\apj} 395:140

\bibitem[{{Be{\v{s}}li{\'c}} et~al.(2021){Be{\v{s}}li{\'c}}, {Barnes},
  {Bigiel}, {Puschnig}, {Pety} et~al.}]{BESLIC21DENSE}
{Be{\v{s}}li{\'c}} I, {Barnes} AT, {Bigiel} F, {Puschnig} J, {Pety} J, et~al.
  2021.
\textit{\mnras} 506:963--988

\bibitem[{{Bigiel} et~al.(2008){Bigiel}, {Leroy}, {Walter}, {Brinks}, {de Blok}
  et~al.}]{BIGIEL08SFGAS}
{Bigiel} F, {Leroy} A, {Walter} F, {Brinks} E, {de Blok} WJG, et~al. 2008.
\textit{\aj} 136:2846--2871

\bibitem[{{Binder} \& {Povich}(2018)}]{BINDER18SFR}
{Binder} BA, {Povich} MS. 2018.
\textit{\apj} 864:136

\bibitem[{{Blitz}(1993)}]{BLITZ93GMCS}
{Blitz} L. 1993.
\textit{{Giant Molecular Clouds}}. In \textit{Protostars and Planets III}, eds.
  EH~{Levy}, JI~{Lunine}

\bibitem[{{B{\"o}ker} et~al.(2008){B{\"o}ker}, {Falc{\'o}n-Barroso},
  {Schinnerer}, {Knapen} \& {Ryder}}]{boeker2008}
{B{\"o}ker} T, {Falc{\'o}n-Barroso} J, {Schinnerer} E, {Knapen} JH, {Ryder} S.
  2008.
\textit{\aj} 135:479--495

\bibitem[{{Bolatto} et~al.(2008){Bolatto}, {Leroy}, {Rosolowsky}, {Walter} \&
  {Blitz}}]{BOLATTO08GMCS}
{Bolatto} AD, {Leroy} AK, {Rosolowsky} E, {Walter} F, {Blitz} L. 2008.
\textit{\apj} 686:948--965

\bibitem[{{Bolatto} et~al.(2013){Bolatto}, {Wolfire} \&
  {Leroy}}]{BOLATTO13REVIEW}
{Bolatto} AD, {Wolfire} M, {Leroy} AK. 2013.
\textit{\araa} 51:207--268

\bibitem[{{Boquien} et~al.(2019){Boquien}, {Burgarella}, {Roehlly}, {Buat},
  {Ciesla} et~al.}]{CIGALE19}
{Boquien} M, {Burgarella} D, {Roehlly} Y, {Buat} V, {Ciesla} L, et~al. 2019.
\textit{\aap} 622:A103

\bibitem[{{Boquien} et~al.(2015){Boquien}, {Calzetti}, {Aalto}, {Boselli},
  {Braine} et~al.}]{BOQUIEN15SFR}
{Boquien} M, {Calzetti} D, {Aalto} S, {Boselli} A, {Braine} J, et~al. 2015.
\textit{\aap} 578:A8

\bibitem[{{Brunetti} et~al.(2021){Brunetti}, {Wilson}, {Sliwa}, {Schinnerer},
  {Aalto} \& {Peck}}]{BRUNETTI21GMCS}
{Brunetti} N, {Wilson} CD, {Sliwa} K, {Schinnerer} E, {Aalto} S, {Peck} AB.
  2021.
\textit{\mnras} 500:4730--4748

\bibitem[{{Byler} et~al.(2017){Byler}, {Dalcanton}, {Conroy} \&
  {Johnson}}]{BYLER17SFR}
{Byler} N, {Dalcanton} JJ, {Conroy} C, {Johnson} BD. 2017.
\textit{\apj} 840:44

\bibitem[{{Calzetti} et~al.(2005){Calzetti}, {Kennicutt}, {Bianchi}, {Thilker},
  {Dale} et~al.}]{CALZETTI05SFR}
{Calzetti} D, {Kennicutt} R.~C. J, {Bianchi} L, {Thilker} DA, {Dale} DA, et~al.
  2005.
\textit{\apj} 633:871--893

\bibitem[{{Calzetti} et~al.(2007){Calzetti}, {Kennicutt}, {Engelbracht},
  {Leitherer}, {Draine} et~al.}]{CALZETTI07SFR}
{Calzetti} D, {Kennicutt} RC, {Engelbracht} CW, {Leitherer} C, {Draine} BT,
  et~al. 2007.
\textit{\apj} 666:870--895

\bibitem[{{Calzetti} et~al.(2015){Calzetti}, {Lee}, {Sabbi}, {Adamo}, {Smith}
  et~al.}]{LEGUS15SURVEY}
{Calzetti} D, {Lee} JC, {Sabbi} E, {Adamo} A, {Smith} LJ, et~al. 2015.
\textit{\aj} 149:51

\bibitem[{{Cao} et~al.(2023){Cao}, {Wong}, {Bolatto}, {Leroy}, {Rosolowsky}
  et~al.}]{cao2023}
{Cao} Y, {Wong} T, {Bolatto} AD, {Leroy} AK, {Rosolowsky} E, et~al. 2023.
\textit{\apjs} 268:3

\bibitem[{{Cappellari} et~al.(2011){Cappellari}, {Emsellem}, {Krajnovi{\'c}},
  {McDermid}, {Scott} et~al.}]{ATLAS3DSURVEY11}
{Cappellari} M, {Emsellem} E, {Krajnovi{\'c}} D, {McDermid} RM, {Scott} N,
  et~al. 2011.
\textit{\mnras} 413:813--836

\bibitem[{{Chandar} et~al.(2017){Chandar}, {Fall}, {Whitmore} \&
  {Mulia}}]{CHANDAR17CLUSTERS}
{Chandar} R, {Fall} SM, {Whitmore} BC, {Mulia} AJ. 2017.
\textit{\apj} 849:128

\bibitem[{{Chastenet} et~al.(2023){Chastenet}, {Sutter}, {Sandstrom},
  {Belfiore}, {Egorov} et~al.}]{CHASTENET23DUST}
{Chastenet} J, {Sutter} J, {Sandstrom} K, {Belfiore} F, {Egorov} OV, et~al.
  2023.
\textit{\apjl} 944:L11

\bibitem[{{Chevance} et~al.(2020){Chevance}, {Kruijssen}, {Hygate}, {Schruba},
  {Longmore} et~al.}]{CHEVANCE20TIMES}
{Chevance} M, {Kruijssen} JMD, {Hygate} APS, {Schruba} A, {Longmore} SN, et~al.
  2020.
\textit{\mnras} 493:2872--2909

\bibitem[{{Chevance} et~al.(2022){Chevance}, {Kruijssen}, {Krumholz}, {Groves},
  {Keller} et~al.}]{chevance2022}
{Chevance} M, {Kruijssen} JMD, {Krumholz} MR, {Groves} B, {Keller} BW, et~al.
  2022.
\textit{\mnras} 509:272--288

\bibitem[{{Chevance} et~al.(2023){Chevance}, {Krumholz}, {McLeod}, {Ostriker},
  {Rosolowsky} \& {Sternberg}}]{chevance2023}
{Chevance} M, {Krumholz} MR, {McLeod} AF, {Ostriker} EC, {Rosolowsky} EW,
  {Sternberg} A. 2023.
\textit{{The Life and Times of Giant Molecular Clouds}}. In
  \textit{Astronomical Society of the Pacific Conference Series}, eds.
  S~{Inutsuka}, Y~{Aikawa}, T~{Muto}, K~{Tomida}, M~{Tamura}, vol. 534 of
  \textit{Astronomical Society of the Pacific Conference Series}

\bibitem[{{Chiang} et~al.(2023){Chiang}, {Sandstrom}, {Chastenet}, {Bolatto},
  {Koch} et~al.}]{CHIANG23XCO}
{Chiang} ID, {Sandstrom} KM, {Chastenet} J, {Bolatto} AD, {Koch} EW, et~al.
  2023.
\textit{arXiv e-prints} :arXiv:2311.00407

\bibitem[{{Choi} et~al.(2023){Choi}, {Liu}, {Bureau}, {Cappellari}, {Davis}
  et~al.}]{choi2023}
{Choi} W, {Liu} L, {Bureau} M, {Cappellari} M, {Davis} TA, et~al. 2023.
\textit{\mnras} 522:4078--4097

\bibitem[{{Colombo} et~al.(2014){Colombo}, {Hughes}, {Schinnerer}, {Meidt},
  {Leroy} et~al.}]{COLOMBO14GMCS}
{Colombo} D, {Hughes} A, {Schinnerer} E, {Meidt} SE, {Leroy} AK, et~al. 2014.
\textit{\apj} 784:3

\bibitem[{{Comer{\'o}n} et~al.(2014){Comer{\'o}n}, {Salo}, {Laurikainen},
  {Knapen}, {Buta} et~al.}]{comeron2014}
{Comer{\'o}n} S, {Salo} H, {Laurikainen} E, {Knapen} JH, {Buta} RJ, et~al.
  2014.
\textit{\aap} 562:A121

\bibitem[{{Conroy}(2013)}]{CONROY13REVIEW}
{Conroy} C. 2013.
\textit{\araa} 51:393--455

\bibitem[{{Corbelli} et~al.(2017){Corbelli}, {Braine}, {Bandiera}, {Brouillet},
  {Combes} et~al.}]{CORBELLI17TIMES}
{Corbelli} E, {Braine} J, {Bandiera} R, {Brouillet} N, {Combes} F, et~al. 2017.
\textit{\aap} 601:A146

\bibitem[{{Cormier} et~al.(2018){Cormier}, {Bigiel}, {Jim{\'e}nez-Donaire},
  {Leroy}, {Gallagher} et~al.}]{cormier2018}
{Cormier} D, {Bigiel} F, {Jim{\'e}nez-Donaire} MJ, {Leroy} AK, {Gallagher} M,
  et~al. 2018.
\textit{\mnras} 475:3909--3933

\bibitem[{{da Silva} et~al.(2014){da Silva}, {Fumagalli} \&
  {Krumholz}}]{DASILVA14SFR}
{da Silva} RL, {Fumagalli} M, {Krumholz} MR. 2014.
\textit{\mnras} 444:3275--3287

\bibitem[{{Dalcanton} et~al.(2012){Dalcanton}, {Williams}, {Lang}, {Lauer},
  {Kalirai} et~al.}]{PHAT12SURVEY}
{Dalcanton} JJ, {Williams} BF, {Lang} D, {Lauer} TR, {Kalirai} JS, et~al. 2012.
\textit{\apjs} 200:18

\bibitem[{{Davis} et~al.(2018){Davis}, {Bureau}, {Onishi}, {van de Voort},
  {Cappellari} et~al.}]{davis2018}
{Davis} TA, {Bureau} M, {Onishi} K, {van de Voort} F, {Cappellari} M, et~al.
  2018.
\textit{\mnras} 473:3818--3834

\bibitem[{{Davis} et~al.(2022){Davis}, {Gensior}, {Bureau}, {Cappellari},
  {Choi} et~al.}]{DAVIS22GMCS}
{Davis} TA, {Gensior} J, {Bureau} M, {Cappellari} M, {Choi} W, et~al. 2022.
\textit{\mnras} 512:1522--1540

\bibitem[{{Davis} et~al.(2014){Davis}, {Young}, {Crocker}, {Bureau}, {Blitz}
  et~al.}]{davis2014}
{Davis} TA, {Young} LM, {Crocker} AF, {Bureau} M, {Blitz} L, et~al. 2014.
\textit{\mnras} 444:3427--3445

\bibitem[{{Della Bruna} et~al.(2022){Della Bruna}, {Adamo}, {McLeod}, {Smith},
  {Savard} et~al.}]{DELLABRUNA22HIIREGION}
{Della Bruna} L, {Adamo} A, {McLeod} AF, {Smith} LJ, {Savard} G, et~al. 2022.
\textit{\aap} 666:A29

\bibitem[{{Demachi} et~al.(2023){Demachi}, {Fukui}, {Yamada}, {Tachihara},
  {Hayakawa} et~al.}]{demachi2023}
{Demachi} F, {Fukui} Y, {Yamada} RI, {Tachihara} K, {Hayakawa} T, et~al. 2023.
\textit{arXiv e-prints} :arXiv:2305.19192

\bibitem[{{den Brok} et~al.(2021){den Brok}, {Chatzigiannakis}, {Bigiel},
  {Puschnig}, {Barnes} et~al.}]{DENBROK21LINES}
{den Brok} JS, {Chatzigiannakis} D, {Bigiel} F, {Puschnig} J, {Barnes} AT,
  et~al. 2021.
\textit{\mnras} 504:3221--3245

\bibitem[{{Denis-Alpizar} et~al.(2020){Denis-Alpizar}, {Stoecklin}, {Dutrey} \&
  {Guilloteau}}]{DENISALPIZAR20COLLIDE}
{Denis-Alpizar} O, {Stoecklin} T, {Dutrey} A, {Guilloteau} S. 2020.
\textit{\mnras} 497:4276--4281

\bibitem[{{Denis-Alpizar} et~al.(2018){Denis-Alpizar}, {Stoecklin},
  {Guilloteau} \& {Dutrey}}]{DENISAPLIZAR18COLLIDE}
{Denis-Alpizar} O, {Stoecklin} T, {Guilloteau} S, {Dutrey} A. 2018.
\textit{\mnras} 478:1811--1817

\bibitem[{{Dieter} \& {Goss}(1966)}]{DIETER66REVIEW}
{Dieter} NH, {Goss} WM. 1966.
\textit{Reviews of Modern Physics} 38:256--297

\bibitem[{{Dobbs} et~al.(2014){Dobbs}, {Krumholz}, {Ballesteros-Paredes},
  {Bolatto}, {Fukui} et~al.}]{DOBBS14REVIEW}
{Dobbs} CL, {Krumholz} MR, {Ballesteros-Paredes} J, {Bolatto} AD, {Fukui} Y,
  et~al. 2014.
\textit{{Formation of Molecular Clouds and Global Conditions for Star
  Formation}}. In \textit{Protostars and Planets VI}, eds. H~{Beuther},
  RS~{Klessen}, CP~{Dullemond}, T~{Henning}

\bibitem[{{Donovan Meyer} et~al.(2013){Donovan Meyer}, {Koda}, {Momose},
  {Mooney}, {Egusa} et~al.}]{DONOVANMEYER13GMCS}
{Donovan Meyer} J, {Koda} J, {Momose} R, {Mooney} T, {Egusa} F, et~al. 2013.
\textit{\apj} 772:107

\bibitem[{{Downes} \& {Solomon}(1998)}]{DOWNES98XCO}
{Downes} D, {Solomon} PM. 1998.
\textit{\apj} 507:615--654

\bibitem[{{Draine}(2011)}]{DRAINE11BOOK}
{Draine} BT. 2011.
\textit{{Physics of the Interstellar and Intergalactic Medium}}

\bibitem[{{Dumouchel} et~al.(2010){Dumouchel}, {Faure} \&
  {Lique}}]{DUMOUCHEL10COLLIDE}
{Dumouchel} F, {Faure} A, {Lique} F. 2010.
\textit{\mnras} 406:2488--2492

\bibitem[{{Dunne} et~al.(2022){Dunne}, {Maddox}, {Papadopoulos}, {Ivison} \&
  {Gomez}}]{DUNNE22XCO}
{Dunne} L, {Maddox} SJ, {Papadopoulos} PP, {Ivison} RJ, {Gomez} HL. 2022.
\textit{\mnras} 517:962--999

\bibitem[{{Egorov} et~al.(2023{\natexlab{a}}){Egorov}, {Kreckel}, {Glover},
  {Groves}, {Belfiore} et~al.}]{egorov2023}
{Egorov} OV, {Kreckel} K, {Glover} SCO, {Groves} B, {Belfiore} F, et~al.
  2023{\natexlab{a}}.
\textit{\aap} 678:A153

\bibitem[{{Egorov} et~al.(2023{\natexlab{b}}){Egorov}, {Kreckel}, {Sandstrom},
  {Leroy}, {Glover} et~al.}]{EGOROV23DUST}
{Egorov} OV, {Kreckel} K, {Sandstrom} KM, {Leroy} AK, {Glover} SCO, et~al.
  2023{\natexlab{b}}.
\textit{\apjl} 944:L16

\bibitem[{{Eibensteiner} et~al.(2022){Eibensteiner}, {Barnes}, {Bigiel},
  {Schinnerer}, {Liu} et~al.}]{eibensteiner2022}
{Eibensteiner} C, {Barnes} AT, {Bigiel} F, {Schinnerer} E, {Liu} D, et~al.
  2022.
\textit{\aap} 659:A173

\bibitem[{{Eldridge} et~al.(2017){Eldridge}, {Stanway}, {Xiao}, {McClelland},
  {Taylor} et~al.}]{ELDRIDGE17BPASS}
{Eldridge} JJ, {Stanway} ER, {Xiao} L, {McClelland} LAS, {Taylor} G, et~al.
  2017.
\textit{\pasa} 34:e058

\bibitem[{{Elmegreen}(1989)}]{ELMEGREEN89PRESS}
{Elmegreen} BG. 1989.
\textit{\apj} 338:178

\bibitem[{{Elmegreen}(2000)}]{ELMEGREEN00CROSSING}
{Elmegreen} BG. 2000.
\textit{\apj} 530:277--281

\bibitem[{{Elmegreen} \& {Scalo}(2004)}]{ELMEGREEN04REVIEW}
{Elmegreen} BG, {Scalo} J. 2004.
\textit{\araa} 42:211--273

\bibitem[{{Emig} et~al.(2020){Emig}, {Bolatto}, {Leroy}, {Mills}, {Jim{\'e}nez
  Donaire} et~al.}]{EMIG20SFR}
{Emig} KL, {Bolatto} AD, {Leroy} AK, {Mills} EAC, {Jim{\'e}nez Donaire} MJ,
  et~al. 2020.
\textit{\apj} 903:50

\bibitem[{{Emsellem} et~al.(2022){Emsellem}, {Schinnerer}, {Santoro},
  {Belfiore}, {Pessa} et~al.}]{PHANGSMUSE22SURVEY}
{Emsellem} E, {Schinnerer} E, {Santoro} F, {Belfiore} F, {Pessa} I, et~al.
  2022.
\textit{\aap} 659:A191

\bibitem[{{Engargiola} et~al.(2003){Engargiola}, {Plambeck}, {Rosolowsky} \&
  {Blitz}}]{engargiola2003}
{Engargiola} G, {Plambeck} RL, {Rosolowsky} E, {Blitz} L. 2003.
\textit{\apjs} 149:343--363

\bibitem[{{Erroz-Ferrer} et~al.(2019){Erroz-Ferrer}, {Carollo}, {den Brok},
  {Onodera}, {Brinchmann} et~al.}]{MAD19SURVEY}
{Erroz-Ferrer} S, {Carollo} CM, {den Brok} M, {Onodera} M, {Brinchmann} J,
  et~al. 2019.
\textit{\mnras} 484:5009--5027

\bibitem[{{Evans} et~al.(2014){Evans}, {Heiderman} \&
  {Vutisalchavakul}}]{EVANS14DENSE}
{Evans} Neal~J. I, {Heiderman} A, {Vutisalchavakul} N. 2014.
\textit{\apj} 782:114

\bibitem[{{Evans} et~al.(2020){Evans}, {Kim}, {Wu}, {Chao}, {Heyer}
  et~al.}]{EVANS20DENSE}
{Evans} Neal~J. I, {Kim} KT, {Wu} J, {Chao} Z, {Heyer} M, et~al. 2020.
\textit{\apj} 894:103

\bibitem[{{Fall} et~al.(2010){Fall}, {Krumholz} \& {Matzner}}]{fall2010}
{Fall} SM, {Krumholz} MR, {Matzner} CD. 2010.
\textit{\apjl} 710:L142--L146

\bibitem[{{Faure} et~al.(2007){Faure}, {Varambhia}, {Stoecklin} \&
  {Tennyson}}]{FAURE07LINES}
{Faure} A, {Varambhia} HN, {Stoecklin} T, {Tennyson} J. 2007.
\textit{\mnras} 382:840--848

\bibitem[{{Federrath} \& {Klessen}(2012)}]{FEDERRATH12EFF}
{Federrath} C, {Klessen} RS. 2012.
\textit{\apj} 761:156

\bibitem[{{Fern{\'a}ndez-L{\'o}pez} et~al.(2014){Fern{\'a}ndez-L{\'o}pez},
  {Arce}, {Looney}, {Mundy}, {Storm} et~al.}]{FERNANDEZLOPEZ14DENSE}
{Fern{\'a}ndez-L{\'o}pez} M, {Arce} HG, {Looney} L, {Mundy} LG, {Storm} S,
  et~al. 2014.
\textit{\apjl} 790:L19

\bibitem[{{Field} et~al.(2011){Field}, {Blackman} \& {Keto}}]{FIELD11GMCS}
{Field} GB, {Blackman} EG, {Keto} ER. 2011.
\textit{\mnras} 416:710--714

\bibitem[{{Fisher} et~al.(2022){Fisher}, {Bolatto}, {Glazebrook}, {Obreschkow},
  {Abraham} et~al.}]{FISHER22PRESS}
{Fisher} DB, {Bolatto} AD, {Glazebrook} K, {Obreschkow} D, {Abraham} RG, et~al.
  2022.
\textit{\apj} 928:169

\bibitem[{{Flores Vel{\'a}zquez} et~al.(2021){Flores Vel{\'a}zquez}, {Gurvich},
  {Faucher-Gigu{\`e}re}, {Bullock}, {Starkenburg}
  et~al.}]{FLORESVELAZZQUEZ21SFR}
{Flores Vel{\'a}zquez} JA, {Gurvich} AB, {Faucher-Gigu{\`e}re} CA, {Bullock}
  JS, {Starkenburg} TK, et~al. 2021.
\textit{\mnras} 501:4812--4824

\bibitem[{{Flower}(1999)}]{FLOWER99COLLIDE}
{Flower} DR. 1999.
\textit{\mnras} 305:651--653

\bibitem[{{Fouesneau} \& {Lan{\c{c}}on}(2010)}]{FOUESNEAU10SFR}
{Fouesneau} M, {Lan{\c{c}}on} A. 2010.
\textit{\aap} 521:A22

\bibitem[{{Fukui} et~al.(2021){Fukui}, {Habe}, {Inoue}, {Enokiya} \&
  {Tachihara}}]{FUKUI21COLLIDE}
{Fukui} Y, {Habe} A, {Inoue} T, {Enokiya} R, {Tachihara} K. 2021.
\textit{\pasj} 73:S1--S34

\bibitem[{{Fukui} \& {Kawamura}(2010)}]{FUKUI10REVIEW}
{Fukui} Y, {Kawamura} A. 2010.
\textit{\araa} 48:547--580

\bibitem[{{Gallagher} et~al.(2018{\natexlab{a}}){Gallagher}, {Leroy}, {Bigiel},
  {Cormier}, {Jim{\'e}nez-Donaire} et~al.}]{GALLAGHER18DENSE}
{Gallagher} MJ, {Leroy} AK, {Bigiel} F, {Cormier} D, {Jim{\'e}nez-Donaire} MJ,
  et~al. 2018{\natexlab{a}}.
\textit{\apj} 858:90

\bibitem[{{Gallagher} et~al.(2018{\natexlab{b}}){Gallagher}, {Leroy}, {Bigiel},
  {Cormier}, {Jim{\'e}nez-Donaire} et~al.}]{GALLAGHER18HCNCLOUD}
{Gallagher} MJ, {Leroy} AK, {Bigiel} F, {Cormier} D, {Jim{\'e}nez-Donaire} MJ,
  et~al. 2018{\natexlab{b}}.
\textit{\apjl} 868:L38

\bibitem[{{Galliano} et~al.(2018){Galliano}, {Galametz} \&
  {Jones}}]{galliano2018}
{Galliano} F, {Galametz} M, {Jones} AP. 2018.
\textit{\araa} 56:673--713

\bibitem[{{Gao} \& {Solomon}(2004{\natexlab{a}})}]{GAO04SURVEY}
{Gao} Y, {Solomon} PM. 2004{\natexlab{a}}.
\textit{\apjs} 152:63--80

\bibitem[{{Gao} \& {Solomon}(2004{\natexlab{b}})}]{GAO04DENSE}
{Gao} Y, {Solomon} PM. 2004{\natexlab{b}}.
\textit{\apj} 606:271--290

\bibitem[{{Garc{\'\i}a-Burillo} et~al.(2021){Garc{\'\i}a-Burillo},
  {Alonso-Herrero}, {Ramos Almeida}, {Gonz{\'a}lez-Mart{\'\i}n}, {Combes}
  et~al.}]{garcia-burillo2021}
{Garc{\'\i}a-Burillo} S, {Alonso-Herrero} A, {Ramos Almeida} C,
  {Gonz{\'a}lez-Mart{\'\i}n} O, {Combes} F, et~al. 2021.
\textit{\aap} 652:A98

\bibitem[{{Garc{\'\i}a-Burillo} et~al.(2003){Garc{\'\i}a-Burillo}, {Combes},
  {Hunt}, {Boone}, {Baker} et~al.}]{GARCIABURILLO03GMCS}
{Garc{\'\i}a-Burillo} S, {Combes} F, {Hunt} LK, {Boone} F, {Baker} AJ, et~al.
  2003.
\textit{\aap} 407:485--502

\bibitem[{{Garc{\'\i}a-Burillo} et~al.(2012){Garc{\'\i}a-Burillo}, {Usero},
  {Alonso-Herrero}, {Graci{\'a}-Carpio}, {Pereira-Santaella}
  et~al.}]{GARCIABURILLO12DENSE}
{Garc{\'\i}a-Burillo} S, {Usero} A, {Alonso-Herrero} A, {Graci{\'a}-Carpio} J,
  {Pereira-Santaella} M, et~al. 2012.
\textit{\aap} 539:A8

\bibitem[{{Gensior} et~al.(2020){Gensior}, {Kruijssen} \&
  {Keller}}]{gensior2020}
{Gensior} J, {Kruijssen} JMD, {Keller} BW. 2020.
\textit{\mnras} 495:199--223

\bibitem[{{Genzel} et~al.(2012){Genzel}, {Tacconi}, {Combes}, {Bolatto}, {Neri}
  et~al.}]{GENZEL12XCO}
{Genzel} R, {Tacconi} LJ, {Combes} F, {Bolatto} A, {Neri} R, et~al. 2012.
\textit{\apj} 746:69

\bibitem[{{Girichidis} et~al.(2020){Girichidis}, {Offner}, {Kritsuk},
  {Klessen}, {Hennebelle} et~al.}]{girichidis2020}
{Girichidis} P, {Offner} SSR, {Kritsuk} AG, {Klessen} RS, {Hennebelle} P,
  et~al. 2020.
\textit{\ssr} 216:68

\bibitem[{{Glover} \& {Clark}(2012)}]{GLOVER12XCO}
{Glover} SCO, {Clark} PC. 2012.
\textit{\mnras} 426:377--388

\bibitem[{{Glover} \& {Clark}(2016)}]{GLOVER16XCO}
{Glover} SCO, {Clark} PC. 2016.
\textit{\mnras} 456:3596--3609

\bibitem[{{Glover} \& {Mac Low}(2011)}]{GLOVER11XCO}
{Glover} SCO, {Mac Low} MM. 2011.
\textit{\mnras} 412:337--350

\bibitem[{{Goldsmith} \& {Kauffmann}(2017)}]{GOLDSMITH17DENSE}
{Goldsmith} PF, {Kauffmann} J. 2017.
\textit{\apj} 841:25

\bibitem[{{Gong} et~al.(2020){Gong}, {Ostriker}, {Kim} \& {Kim}}]{GONG20XCO}
{Gong} M, {Ostriker} EC, {Kim} CG, {Kim} JG. 2020.
\textit{\apj} 903:142

\bibitem[{{Gouliermis}(2018)}]{gouliermis2018}
{Gouliermis} DA. 2018.
\textit{\pasp} 130:072001

\bibitem[{{Graci{\'a}-Carpio} et~al.(2008){Graci{\'a}-Carpio},
  {Garc{\'\i}a-Burillo}, {Planesas}, {Fuente} \& {Usero}}]{GRACIACARPIO08DENSE}
{Graci{\'a}-Carpio} J, {Garc{\'\i}a-Burillo} S, {Planesas} P, {Fuente} A,
  {Usero} A. 2008.
\textit{\aap} 479:703--717

\bibitem[{{Grand} et~al.(2016){Grand}, {Springel}, {Kawata}, {Minchev},
  {S{\'a}nchez-Bl{\'a}zquez} et~al.}]{grand2016}
{Grand} RJJ, {Springel} V, {Kawata} D, {Minchev} I, {S{\'a}nchez-Bl{\'a}zquez}
  P, et~al. 2016.
\textit{\mnras} 460:L94--L98

\bibitem[{{Grasha} et~al.(2018){Grasha}, {Calzetti}, {Bittle}, {Johnson},
  {Donovan Meyer} et~al.}]{GRASHA18TIMES}
{Grasha} K, {Calzetti} D, {Bittle} L, {Johnson} KE, {Donovan Meyer} J, et~al.
  2018.
\textit{\mnras} 481:1016--1027

\bibitem[{{Groves} et~al.(2023){Groves}, {Kreckel}, {Santoro}, {Belfiore},
  {Zavodnik} et~al.}]{GROVES23SFR}
{Groves} B, {Kreckel} K, {Santoro} F, {Belfiore} F, {Zavodnik} E, et~al. 2023.
\textit{\mnras} 520:4902--4952

\bibitem[{{Grudi{\'c}} et~al.(2022){Grudi{\'c}}, {Guszejnov}, {Offner},
  {Rosen}, {Raju} et~al.}]{STARFORGE22}
{Grudi{\'c}} MY, {Guszejnov} D, {Offner} SSR, {Rosen} AL, {Raju} AN, et~al.
  2022.
\textit{\mnras} 512:216--232

\bibitem[{{Grudic} et~al.(2023){Grudic}, {Offner}, {Guszejnov},
  {Faucher-Gigu{\`e}re} \& {Hopkins}}]{STARFORGE23}
{Grudic} MY, {Offner} SSR, {Guszejnov} D, {Faucher-Gigu{\`e}re} CA, {Hopkins}
  PF. 2023.
\textit{The Open Journal of Astrophysics} 6:48

\bibitem[{{Gurvich} et~al.(2020){Gurvich}, {Faucher-Gigu{\`e}re}, {Richings},
  {Hopkins}, {Grudi{\'c}} et~al.}]{GURVICH20PRESS}
{Gurvich} AB, {Faucher-Gigu{\`e}re} CA, {Richings} AJ, {Hopkins} PF,
  {Grudi{\'c}} MY, et~al. 2020.
\textit{\mnras} 498:3664--3683

\bibitem[{{Haan} et~al.(2009){Haan}, {Schinnerer}, {Emsellem},
  {Garc{\'\i}a-Burillo}, {Combes} et~al.}]{haan2009}
{Haan} S, {Schinnerer} E, {Emsellem} E, {Garc{\'\i}a-Burillo} S, {Combes} F,
  et~al. 2009.
\textit{\apj} 692:1623--1661

\bibitem[{{Hacar} et~al.(2023){Hacar}, {Clark}, {Heitsch}, {Kainulainen},
  {Panopoulou} et~al.}]{HACAR23FILAMENTS}
{Hacar} A, {Clark} SE, {Heitsch} F, {Kainulainen} J, {Panopoulou} GV, et~al.
  2023.
\textit{{Initial Conditions for Star Formation: a Physical Description of the
  Filamentary ISM}}. In \textit{Astronomical Society of the Pacific Conference
  Series}, eds. S~{Inutsuka}, Y~{Aikawa}, T~{Muto}, K~{Tomida}, M~{Tamura},
  vol. 534 of \textit{Astronomical Society of the Pacific Conference Series}

\bibitem[{{Hannon} et~al.(2022){Hannon}, {Lee}, {Whitmore}, {Mobasher},
  {Thilker} et~al.}]{HANNON22CLUSTERS}
{Hannon} S, {Lee} JC, {Whitmore} BC, {Mobasher} B, {Thilker} D, et~al. 2022.
\textit{\mnras} 512:1294--1316

\bibitem[{{Hannon} et~al.(2023){Hannon}, {Whitmore}, {Lee}, {Thilker}, {Deger}
  et~al.}]{hannon2023}
{Hannon} S, {Whitmore} BC, {Lee} JC, {Thilker} DA, {Deger} S, et~al. 2023.
\textit{\mnras}

\bibitem[{{Harada} et~al.(2019){Harada}, {Sakamoto}, {Mart{\'\i}n}, {Watanabe},
  {Aladro} et~al.}]{harada2019}
{Harada} N, {Sakamoto} K, {Mart{\'\i}n} S, {Watanabe} Y, {Aladro} R, et~al.
  2019.
\textit{\apj} 884:100

\bibitem[{{Hassani} et~al.(2023){Hassani}, {Rosolowsky}, {Leroy}, {Boquien},
  {Lee} et~al.}]{HASSANI23JWST}
{Hassani} H, {Rosolowsky} E, {Leroy} AK, {Boquien} M, {Lee} JC, et~al. 2023.
\textit{\apjl} 944:L21

\bibitem[{{Heiderman} et~al.(2010){Heiderman}, {Evans}, {Allen}, {Huard} \&
  {Heyer}}]{HEIDERMAN10DENSE}
{Heiderman} A, {Evans} Neal~J. I, {Allen} LE, {Huard} T, {Heyer} M. 2010.
\textit{\apj} 723:1019--1037

\bibitem[{{Helfer} \& {Blitz}(1997)}]{HELFER97DENSE}
{Helfer} TT, {Blitz} L. 1997.
\textit{\apj} 478:162--171

\bibitem[{{Hennebelle} \& {Chabrier}(2011)}]{HENNEBELLE11EFF}
{Hennebelle} P, {Chabrier} G. 2011.
\textit{\apjl} 743:L29

\bibitem[{{Henshaw} et~al.(2023){Henshaw}, {Barnes}, {Battersby}, {Ginsburg},
  {Sormani} \& {Walker}}]{henshaw2023}
{Henshaw} JD, {Barnes} AT, {Battersby} C, {Ginsburg} A, {Sormani} MC, {Walker}
  DL. 2023.
\textit{{Star Formation in the Central Molecular Zone of the Milky Way}}. In
  \textit{Astronomical Society of the Pacific Conference Series}, eds.
  S~{Inutsuka}, Y~{Aikawa}, T~{Muto}, K~{Tomida}, M~{Tamura}, vol. 534 of
  \textit{Astronomical Society of the Pacific Conference Series}

\bibitem[{{Henshaw} et~al.(2016){Henshaw}, {Longmore}, {Kruijssen}, {Davies},
  {Bally} et~al.}]{HENSHAW16GMCS}
{Henshaw} JD, {Longmore} SN, {Kruijssen} JMD, {Davies} B, {Bally} J, et~al.
  2016.
\textit{\mnras} 457:2675--2702

\bibitem[{{Heyer} \& {Dame}(2015)}]{HEYER15REVIEW}
{Heyer} M, {Dame} TM. 2015.
\textit{\araa} 53:583--629

\bibitem[{{Heyer} et~al.(2009){Heyer}, {Krawczyk}, {Duval} \&
  {Jackson}}]{HEYER09GMCS}
{Heyer} M, {Krawczyk} C, {Duval} J, {Jackson} JM. 2009.
\textit{\apj} 699:1092--1103

\bibitem[{{Heyer} \& {Brunt}(2004)}]{HEYER04GMCS}
{Heyer} MH, {Brunt} CM. 2004.
\textit{\apjl} 615:L45--L48

\bibitem[{{Hilmi} et~al.(2020){Hilmi}, {Minchev}, {Buck}, {Martig}, {Quillen}
  et~al.}]{hilmi2020}
{Hilmi} T, {Minchev} I, {Buck} T, {Martig} M, {Quillen} AC, et~al. 2020.
\textit{\mnras} 497:933--955

\bibitem[{{Ho} et~al.(2017){Ho}, {Seibert}, {Meidt}, {Kudritzki}, {Kobayashi}
  et~al.}]{ho2017}
{Ho} IT, {Seibert} M, {Meidt} SE, {Kudritzki} RP, {Kobayashi} C, et~al. 2017.
\textit{\apj} 846:39

\bibitem[{{Hollyhead} et~al.(2015){Hollyhead}, {Bastian}, {Adamo},
  {Silva-Villa}, {Dale} et~al.}]{HOLLYHEAD15CLUSTERS}
{Hollyhead} K, {Bastian} N, {Adamo} A, {Silva-Villa} E, {Dale} J, et~al. 2015.
\textit{\mnras} 449:1106--1117

\bibitem[{{Hopkins} et~al.(2012){Hopkins}, {Quataert} \&
  {Murray}}]{HOPKINS12GALAXIES}
{Hopkins} PF, {Quataert} E, {Murray} N. 2012.
\textit{\mnras} 421:3522--3537

\bibitem[{{Hoyle}(1953)}]{HOYLE53REVIEW}
{Hoyle} F. 1953.
\textit{\apj} 118:513

\bibitem[{{Hu} et~al.(2022){Hu}, {Schruba}, {Sternberg} \& {van
  Dishoeck}}]{HU22XCO}
{Hu} CY, {Schruba} A, {Sternberg} A, {van Dishoeck} EF. 2022.
\textit{\apj} 931:28

\bibitem[{{Hughes} et~al.(2013){Hughes}, {Meidt}, {Colombo}, {Schinnerer},
  {Pety} et~al.}]{HUGHES13GMCS}
{Hughes} A, {Meidt} SE, {Colombo} D, {Schinnerer} E, {Pety} J, et~al. 2013.
\textit{\apj} 779:46

\bibitem[{{Hummer} \& {Storey}(1987)}]{HUMMER87RECOMB}
{Hummer} DG, {Storey} PJ. 1987.
\textit{\mnras} 224:801--820

\bibitem[{{Hunt} et~al.(2020){Hunt}, {Tortora}, {Ginolfi} \&
  {Schneider}}]{HUNT20XCO}
{Hunt} LK, {Tortora} C, {Ginolfi} M, {Schneider} R. 2020.
\textit{\aap} 643:A180

\bibitem[{{Hunter} et~al.(2012){Hunter}, {Ficut-Vicas}, {Ashley}, {Brinks},
  {Cigan} et~al.}]{LITTLETHINGS12SURVEY}
{Hunter} DA, {Ficut-Vicas} D, {Ashley} T, {Brinks} E, {Cigan} P, et~al. 2012.
\textit{\aj} 144:134

\bibitem[{{Iffrig} \& {Hennebelle}(2015)}]{IFFRIG15SNE}
{Iffrig} O, {Hennebelle} P. 2015.
\textit{\aap} 576:A95

\bibitem[{{Imanishi} et~al.(2023){Imanishi}, {Baba}, {Nakanishi} \&
  {Izumi}}]{IMANISHI23DENSE}
{Imanishi} M, {Baba} S, {Nakanishi} K, {Izumi} T. 2023.
\textit{\apj} 950:75

\bibitem[{{Inoue} \& {Fukui}(2013)}]{INOUE13COLLIDE}
{Inoue} T, {Fukui} Y. 2013.
\textit{\apjl} 774:L31

\bibitem[{{Israel}(2020)}]{ISRAEL20XCO}
{Israel} FP. 2020.
\textit{\aap} 635:A131

\bibitem[{{Jameson} et~al.(2016){Jameson}, {Bolatto}, {Leroy}, {Meixner},
  {Roman-Duval} et~al.}]{JAMESON16XCO}
{Jameson} KE, {Bolatto} AD, {Leroy} AK, {Meixner} M, {Roman-Duval} J, et~al.
  2016.
\textit{\apj} 825:12

\bibitem[{{Jeffreson} et~al.(2022){Jeffreson}, {Sun} \&
  {Wilson}}]{JEFFRESON22GMCS}
{Jeffreson} SMR, {Sun} J, {Wilson} CD. 2022.
\textit{\mnras} 515:1663--1675

\bibitem[{{Jiang} et~al.(2010){Jiang}, {Chen}, {Wang}, {Su}, {Zhou}
  et~al.}]{JIANG10SNR}
{Jiang} B, {Chen} Y, {Wang} J, {Su} Y, {Zhou} X, et~al. 2010.
\textit{\apj} 712:1147--1156

\bibitem[{{Jim{\'e}nez-Donaire}
  et~al.(2017{\natexlab{a}}){Jim{\'e}nez-Donaire}, {Bigiel}, {Leroy},
  {Cormier}, {Gallagher} et~al.}]{JIMENEZ17DENSE}
{Jim{\'e}nez-Donaire} MJ, {Bigiel} F, {Leroy} AK, {Cormier} D, {Gallagher} M,
  et~al. 2017{\natexlab{a}}.
\textit{\mnras} 466:49--62

\bibitem[{{Jim{\'e}nez-Donaire} et~al.(2019){Jim{\'e}nez-Donaire}, {Bigiel},
  {Leroy}, {Usero}, {Cormier} et~al.}]{JIMENEZ19DENSE}
{Jim{\'e}nez-Donaire} MJ, {Bigiel} F, {Leroy} AK, {Usero} A, {Cormier} D,
  et~al. 2019.
\textit{\apj} 880:127

\bibitem[{{Jim{\'e}nez-Donaire}
  et~al.(2017{\natexlab{b}}){Jim{\'e}nez-Donaire}, {Cormier}, {Bigiel},
  {Leroy}, {Gallagher} et~al.}]{JIMENEZ17COLINES}
{Jim{\'e}nez-Donaire} MJ, {Cormier} D, {Bigiel} F, {Leroy} AK, {Gallagher} M,
  et~al. 2017{\natexlab{b}}.
\textit{\apjl} 836:L29

\bibitem[{{Jim{\'e}nez-Donaire} et~al.(2023){Jim{\'e}nez-Donaire}, {Usero},
  {Be{\v{s}}li{\'c}}, {Tafalla}, {Chac{\'o}n-Tanarro} et~al.}]{JIMENEZ23DENSE}
{Jim{\'e}nez-Donaire} MJ, {Usero} A, {Be{\v{s}}li{\'c}} I, {Tafalla} M,
  {Chac{\'o}n-Tanarro} A, et~al. 2023.
\textit{\aap} 676:L11

\bibitem[{{Jones} et~al.(2023){Jones}, {Clark}, {Glover} \&
  {Hacar}}]{JONES23DENSE}
{Jones} GH, {Clark} PC, {Glover} SCO, {Hacar} A. 2023.
\textit{\mnras} 520:1005--1021

\bibitem[{{Jung} et~al.(2023){Jung}, {Calzetti}, {Messa}, {Heyer}, {Sirressi}
  et~al.}]{JUNG23IMF}
{Jung} DE, {Calzetti} D, {Messa} M, {Heyer} M, {Sirressi} M, et~al. 2023.
\textit{\apj} 954:136

\bibitem[{{Kainulainen} et~al.(2009){Kainulainen}, {Beuther}, {Henning} \&
  {Plume}}]{KAINULAINEN09DENSE}
{Kainulainen} J, {Beuther} H, {Henning} T, {Plume} R. 2009.
\textit{\aap} 508:L35--L38

\bibitem[{{Kauffmann} et~al.(2017){Kauffmann}, {Goldsmith}, {Melnick}, {Tolls},
  {Guzman} \& {Menten}}]{KAUFFMANN17DENSE}
{Kauffmann} J, {Goldsmith} PF, {Melnick} G, {Tolls} V, {Guzman} A, {Menten} KM.
  2017.
\textit{\aap} 605:L5

\bibitem[{{Kawamura} et~al.(2009){Kawamura}, {Mizuno}, {Minamidani},
  {Filipovi{\'c}}, {Staveley-Smith} et~al.}]{KAWAMURA09TIMES}
{Kawamura} A, {Mizuno} Y, {Minamidani} T, {Filipovi{\'c}} MD, {Staveley-Smith}
  L, et~al. 2009.
\textit{\apjs} 184:1--17

\bibitem[{{Kenney} et~al.(1992){Kenney}, {Wilson}, {Scoville}, {Devereux} \&
  {Young}}]{kenney1992}
{Kenney} JDP, {Wilson} CD, {Scoville} NZ, {Devereux} NA, {Young} JS. 1992.
\textit{\apjl} 395:L79

\bibitem[{{Kennicutt}(1998)}]{KENNICUTT98SFGAS}
{Kennicutt} Robert~C. J. 1998.
\textit{\apj} 498:541--552

\bibitem[{{Kennicutt} \& {Evans}(2012)}]{KENNICUTT12REVIEW}
{Kennicutt} RC, {Evans} NJ. 2012.
\textit{\araa} 50:531--608

\bibitem[{{Keto} \& {Myers}(1986)}]{KETO86GMCS}
{Keto} ER, {Myers} PC. 1986.
\textit{\apj} 304:466

\bibitem[{{Kewley} et~al.(2019){Kewley}, {Nicholls} \&
  {Sutherland}}]{kewley2019}
{Kewley} LJ, {Nicholls} DC, {Sutherland} RS. 2019.
\textit{\araa} 57:511--570

\bibitem[{{Kim} \& {Ostriker}(2015)}]{KIM15SNE}
{Kim} CG, {Ostriker} EC. 2015.
\textit{\apj} 802:99

\bibitem[{{Kim} et~al.(2013){Kim}, {Ostriker} \& {Kim}}]{KIM13PRESS}
{Kim} CG, {Ostriker} EC, {Kim} WT. 2013.
\textit{\apj} 776:1

\bibitem[{{Kim} et~al.(2017){Kim}, {Ostriker} \& {Raileanu}}]{KIM17PRESS}
{Kim} CG, {Ostriker} EC, {Raileanu} R. 2017.
\textit{\apj} 834:25

\bibitem[{{Kim} et~al.(2023){Kim}, {Chevance}, {Kruijssen}, {Barnes}, {Bigiel}
  et~al.}]{KIM23TIMES}
{Kim} J, {Chevance} M, {Kruijssen} JMD, {Barnes} AT, {Bigiel} F, et~al. 2023.
\textit{\apjl} 944:L20

\bibitem[{{Kim} et~al.(2022){Kim}, {Chevance}, {Kruijssen}, {Leroy}, {Schruba}
  et~al.}]{KIM22TIMES}
{Kim} J, {Chevance} M, {Kruijssen} JMD, {Leroy} AK, {Schruba} A, et~al. 2022.
\textit{\mnras} 516:3006--3028

\bibitem[{{Kim} et~al.(2021{\natexlab{a}}){Kim}, {Chevance}, {Kruijssen},
  {Schruba}, {Sandstrom} et~al.}]{KIM21TIMES}
{Kim} J, {Chevance} M, {Kruijssen} JMD, {Schruba} A, {Sandstrom} K, et~al.
  2021{\natexlab{a}}.
\textit{\mnras} 504:487--509

\bibitem[{{Kim} et~al.(2016){Kim}, {Kim} \& {Ostriker}}]{kim2016}
{Kim} JG, {Kim} WT, {Ostriker} EC. 2016.
\textit{\apj} 819:137

\bibitem[{{Kim} et~al.(2018){Kim}, {Kim} \& {Ostriker}}]{kim2018}
{Kim} JG, {Kim} WT, {Ostriker} EC. 2018.
\textit{\apj} 859:68

\bibitem[{{Kim} et~al.(2019){Kim}, {Kim} \& {Ostriker}}]{kim2019}
{Kim} JG, {Kim} WT, {Ostriker} EC. 2019.
\textit{\apj} 883:102

\bibitem[{{Kim} et~al.(2021{\natexlab{b}}){Kim}, {Ostriker} \&
  {Filippova}}]{KIM21FEEDBACK}
{Kim} JG, {Ostriker} EC, {Filippova} N. 2021{\natexlab{b}}.
\textit{\apj} 911:128

\bibitem[{{Koda}(2021)}]{koda2021}
{Koda} J. 2021.
\textit{Research Notes of the American Astronomical Society} 5:222

\bibitem[{{Koda} et~al.(2023){Koda}, {Hirota}, {Egusa}, {Sakamoto}, {Sawada}
  et~al.}]{koda2023}
{Koda} J, {Hirota} A, {Egusa} F, {Sakamoto} K, {Sawada} T, et~al. 2023.
\textit{\apj} 949:108

\bibitem[{{Kormendy} \& {Kennicutt}(2004)}]{kormendy2004}
{Kormendy} J, {Kennicutt} Robert~C. J. 2004.
\textit{\araa} 42:603--683

\bibitem[{{Koyama} \& {Ostriker}(2009)}]{KOYAMA09PRESS}
{Koyama} H, {Ostriker} EC. 2009.
\textit{\apj} 693:1346--1359

\bibitem[{{Kreckel} et~al.(2018){Kreckel}, {Faesi}, {Kruijssen}, {Schruba},
  {Groves} et~al.}]{KRECKEL18SFR}
{Kreckel} K, {Faesi} C, {Kruijssen} JMD, {Schruba} A, {Groves} B, et~al. 2018.
\textit{\apjl} 863:L21

\bibitem[{{Kreckel} et~al.(2020){Kreckel}, {Ho}, {Blanc}, {Glover}, {Groves}
  et~al.}]{KRECKEL20METALS}
{Kreckel} K, {Ho} IT, {Blanc} GA, {Glover} SCO, {Groves} B, et~al. 2020.
\textit{\mnras} 499:193--209

\bibitem[{{Kreckel} et~al.(2019){Kreckel}, {Ho}, {Blanc}, {Groves}, {Santoro}
  et~al.}]{kreckel2019}
{Kreckel} K, {Ho} IT, {Blanc} GA, {Groves} B, {Santoro} F, et~al. 2019.
\textit{\apj} 887:80

\bibitem[{{Krieger} et~al.(2020){Krieger}, {Bolatto}, {Koch}, {Leroy},
  {Rosolowsky} et~al.}]{krieger2020}
{Krieger} N, {Bolatto} AD, {Koch} EW, {Leroy} AK, {Rosolowsky} E, et~al. 2020.
\textit{\apj} 899:158

\bibitem[{{Krieger} et~al.(2019){Krieger}, {Bolatto}, {Walter}, {Leroy},
  {Zschaechner} et~al.}]{krieger2019}
{Krieger} N, {Bolatto} AD, {Walter} F, {Leroy} AK, {Zschaechner} LK, et~al.
  2019.
\textit{\apj} 881:43

\bibitem[{{Kroupa}(2001)}]{KROUPA01IMF}
{Kroupa} P. 2001.
\textit{\mnras} 322:231--246

\bibitem[{{Kruijssen} et~al.(2015){Kruijssen}, {Dale} \&
  {Longmore}}]{KRUIJSSEN15CENTER}
{Kruijssen} JMD, {Dale} JE, {Longmore} SN. 2015.
\textit{\mnras} 447:1059--1079

\bibitem[{{Kruijssen} \& {Longmore}(2014)}]{KRUIJSSEN14TIMES}
{Kruijssen} JMD, {Longmore} SN. 2014.
\textit{\mnras} 439:3239--3252

\bibitem[{{Kruijssen} et~al.(2019){Kruijssen}, {Schruba}, {Chevance},
  {Longmore}, {Hygate} et~al.}]{KRUIJSSEN19TIMES}
{Kruijssen} JMD, {Schruba} A, {Chevance} M, {Longmore} SN, {Hygate} APS, et~al.
  2019.
\textit{\nat} 569:519--522

\bibitem[{{Krumholz}(2014)}]{KRUMHOLZ14REVIEW}
{Krumholz} MR. 2014.
\textit{\physrep} 539:49--134

\bibitem[{{Krumholz} et~al.(2015){Krumholz}, {Fumagalli}, {da Silva}, {Rendahl}
  \& {Parra}}]{KRUMHOLZ15SFR}
{Krumholz} MR, {Fumagalli} M, {da Silva} RL, {Rendahl} T, {Parra} J. 2015.
\textit{\mnras} 452:1447--1467

\bibitem[{{Krumholz} \& {Matzner}(2009)}]{krumholz2009}
{Krumholz} MR, {Matzner} CD. 2009.
\textit{\apj} 703:1352--1362

\bibitem[{{Krumholz} \& {McKee}(2005)}]{KRUMHOLZ05EFF}
{Krumholz} MR, {McKee} CF. 2005.
\textit{\apj} 630:250--268

\bibitem[{{Krumholz} et~al.(2019){Krumholz}, {McKee} \&
  {Bland-Hawthorn}}]{KRUMHOLZ19REVIEW}
{Krumholz} MR, {McKee} CF, {Bland-Hawthorn} J. 2019.
\textit{\araa} 57:227--303

\bibitem[{{Krumholz} \& {Tan}(2007)}]{KRUMHOLZ07EFF}
{Krumholz} MR, {Tan} JC. 2007.
\textit{\apj} 654:304--315

\bibitem[{{Krumholz} \& {Thompson}(2007)}]{KRUMHOLZ07DENSE}
{Krumholz} MR, {Thompson} TA. 2007.
\textit{\apj} 669:289--298

\bibitem[{{Krumholz} \& {Ting}(2018)}]{krumholz2018}
{Krumholz} MR, {Ting} YS. 2018.
\textit{\mnras} 475:2236--2252

\bibitem[{{Lada} et~al.(2012){Lada}, {Forbrich}, {Lombardi} \&
  {Alves}}]{LADA12DENSE}
{Lada} CJ, {Forbrich} J, {Lombardi} M, {Alves} JF. 2012.
\textit{\apj} 745:190

\bibitem[{{Lada} et~al.(2010){Lada}, {Lombardi} \& {Alves}}]{LADA10DENSE}
{Lada} CJ, {Lombardi} M, {Alves} JF. 2010.
\textit{\apj} 724:687--693

\bibitem[{{Lamperti} et~al.(2020){Lamperti}, {Saintonge}, {Koss}, {Viti},
  {Wilson} et~al.}]{LAMPERTI20LINES}
{Lamperti} I, {Saintonge} A, {Koss} M, {Viti} S, {Wilson} CD, et~al. 2020.
\textit{\apj} 889:103

\bibitem[{{Lancaster} et~al.(2021{\natexlab{a}}){Lancaster}, {Ostriker}, {Kim}
  \& {Kim}}]{lancaster2021a}
{Lancaster} L, {Ostriker} EC, {Kim} JG, {Kim} CG. 2021{\natexlab{a}}.
\textit{\apj} 914:89

\bibitem[{{Lancaster} et~al.(2021{\natexlab{b}}){Lancaster}, {Ostriker}, {Kim}
  \& {Kim}}]{lancaster2021b}
{Lancaster} L, {Ostriker} EC, {Kim} JG, {Kim} CG. 2021{\natexlab{b}}.
\textit{\apj} 914:90

\bibitem[{{Lancaster} et~al.(2021{\natexlab{c}}){Lancaster}, {Ostriker}, {Kim}
  \& {Kim}}]{lancaster2021c}
{Lancaster} L, {Ostriker} EC, {Kim} JG, {Kim} CG. 2021{\natexlab{c}}.
\textit{\apjl} 922:L3

\bibitem[{{Lang} et~al.(2020){Lang}, {Meidt}, {Rosolowsky}, {Nofech},
  {Schinnerer} et~al.}]{LANG20KINEMATICS}
{Lang} P, {Meidt} SE, {Rosolowsky} E, {Nofech} J, {Schinnerer} E, et~al. 2020.
\textit{\apj} 897:122

\bibitem[{{Larson} et~al.(2023){Larson}, {Lee}, {Thilker}, {Whitmore}, {Deger}
  et~al.}]{larson2023}
{Larson} KL, {Lee} JC, {Thilker} DA, {Whitmore} BC, {Deger} S, et~al. 2023.
\textit{\mnras} 523:6061--6081

\bibitem[{{Larson}(1981)}]{LARSON81GMCS}
{Larson} RB. 1981.
\textit{\mnras} 194:809--826

\bibitem[{{Lee} et~al.(2016){Lee}, {Miville-Desch{\^e}nes} \&
  {Murray}}]{LEE16LIFETIMES}
{Lee} EJ, {Miville-Desch{\^e}nes} MA, {Murray} NW. 2016.
\textit{\apj} 833:229

\bibitem[{{Lee} et~al.(2023){Lee}, {Sandstrom}, {Leroy}, {Thilker},
  {Schinnerer} et~al.}]{PHANGSJWST23SURVEY}
{Lee} JC, {Sandstrom} KM, {Leroy} AK, {Thilker} DA, {Schinnerer} E, et~al.
  2023.
\textit{\apjl} 944:L17

\bibitem[{{Lee} et~al.(2022){Lee}, {Whitmore}, {Thilker}, {Deger}, {Larson}
  et~al.}]{PHANGSHST22SURVEY}
{Lee} JC, {Whitmore} BC, {Thilker} DA, {Deger} S, {Larson} KL, et~al. 2022.
\textit{\apjs} 258:10

\bibitem[{{Leitherer} et~al.(2014){Leitherer}, {Ekstr{\"o}m}, {Meynet},
  {Schaerer}, {Agienko} \& {Levesque}}]{LEITHERER14STARBURST}
{Leitherer} C, {Ekstr{\"o}m} S, {Meynet} G, {Schaerer} D, {Agienko} KB,
  {Levesque} EM. 2014.
\textit{\apjs} 212:14

\bibitem[{{Leitherer} et~al.(1999){Leitherer}, {Schaerer}, {Goldader},
  {Delgado}, {Robert} et~al.}]{STARBURST99}
{Leitherer} C, {Schaerer} D, {Goldader} JD, {Delgado} RMG, {Robert} C, et~al.
  1999.
\textit{\apjs} 123:3--40

\bibitem[{{Lenki{\'c}} et~al.(2023){Lenki{\'c}}, {Nally}, {Jones}, {Boyer},
  {Kavanagh} et~al.}]{LENKIC23YSOS}
{Lenki{\'c}} L, {Nally} C, {Jones} OC, {Boyer} ML, {Kavanagh} PJ, et~al. 2023.
\textit{arXiv e-prints} :arXiv:2307.15704

\bibitem[{{Leroy} et~al.(2011){Leroy}, {Bolatto}, {Gordon}, {Sandstrom},
  {Gratier} et~al.}]{LEROY11XCO}
{Leroy} AK, {Bolatto} A, {Gordon} K, {Sandstrom} K, {Gratier} P, et~al. 2011.
\textit{\apj} 737:12

\bibitem[{{Leroy} et~al.(2015){Leroy}, {Bolatto}, {Ostriker}, {Rosolowsky},
  {Walter} et~al.}]{LEROY15GMCS}
{Leroy} AK, {Bolatto} AD, {Ostriker} EC, {Rosolowsky} E, {Walter} F, et~al.
  2015.
\textit{\apj} 801:25

\bibitem[{{Leroy} et~al.(2018){Leroy}, {Bolatto}, {Ostriker}, {Walter},
  {Gorski} et~al.}]{leroy2018}
{Leroy} AK, {Bolatto} AD, {Ostriker} EC, {Walter} F, {Gorski} M, et~al. 2018.
\textit{\apj} 869:126

\bibitem[{{Leroy} et~al.(2016){Leroy}, {Hughes}, {Schruba}, {Rosolowsky},
  {Blanc} et~al.}]{LEROY16GMCS}
{Leroy} AK, {Hughes} A, {Schruba} A, {Rosolowsky} E, {Blanc} GA, et~al. 2016.
\textit{\apj} 831:16

\bibitem[{{Leroy} et~al.(2022){Leroy}, {Rosolowsky}, {Usero}, {Sandstrom},
  {Schinnerer} et~al.}]{LEROY22LINES}
{Leroy} AK, {Rosolowsky} E, {Usero} A, {Sandstrom} K, {Schinnerer} E, et~al.
  2022.
\textit{\apj} 927:149

\bibitem[{{Leroy} et~al.(2023){Leroy}, {Sandstrom}, {Rosolowsky}, {Belfiore},
  {Bolatto} et~al.}]{LEROY23JWST}
{Leroy} AK, {Sandstrom} K, {Rosolowsky} E, {Belfiore} F, {Bolatto} AD, et~al.
  2023.
\textit{\apjl} 944:L9

\bibitem[{{Leroy} et~al.(2017{\natexlab{a}}){Leroy}, {Schinnerer}, {Hughes},
  {Kruijssen}, {Meidt} et~al.}]{LEROY17SFGAS}
{Leroy} AK, {Schinnerer} E, {Hughes} A, {Kruijssen} JMD, {Meidt} S, et~al.
  2017{\natexlab{a}}.
\textit{\apj} 846:71

\bibitem[{{Leroy} et~al.(2021){Leroy}, {Schinnerer}, {Hughes}, {Rosolowsky},
  {Pety} et~al.}]{PHANGSALMA21SURVEY}
{Leroy} AK, {Schinnerer} E, {Hughes} A, {Rosolowsky} E, {Pety} J, et~al. 2021.
\textit{\apjs} 257:43

\bibitem[{{Leroy} et~al.(2017{\natexlab{b}}){Leroy}, {Usero}, {Schruba},
  {Bigiel}, {Kruijssen} et~al.}]{LEROY17DENSE}
{Leroy} AK, {Usero} A, {Schruba} A, {Bigiel} F, {Kruijssen} JMD, et~al.
  2017{\natexlab{b}}.
\textit{\apj} 835:217

\bibitem[{{Leroy} et~al.(2008){Leroy}, {Walter}, {Brinks}, {Bigiel}, {de Blok}
  et~al.}]{LEROY08SFGAS}
{Leroy} AK, {Walter} F, {Brinks} E, {Bigiel} F, {de Blok} WJG, et~al. 2008.
\textit{\aj} 136:2782--2845

\bibitem[{{Leroy} et~al.(2013){Leroy}, {Walter}, {Sandstrom}, {Schruba},
  {Munoz-Mateos} et~al.}]{LEROY13SFGAS}
{Leroy} AK, {Walter} F, {Sandstrom} K, {Schruba} A, {Munoz-Mateos} JC, et~al.
  2013.
\textit{\aj} 146:19

\bibitem[{{Levy} et~al.(2021){Levy}, {Bolatto}, {Leroy}, {Emig}, {Gorski}
  et~al.}]{LEVY21CLUSTERS}
{Levy} RC, {Bolatto} AD, {Leroy} AK, {Emig} KL, {Gorski} M, et~al. 2021.
\textit{\apj} 912:4

\bibitem[{{Levy} et~al.(2022){Levy}, {Bolatto}, {Leroy}, {Sormani}, {Emig}
  et~al.}]{levy2022}
{Levy} RC, {Bolatto} AD, {Leroy} AK, {Sormani} MC, {Emig} KL, et~al. 2022.
\textit{\apj} 935:19

\bibitem[{{Li}(2020)}]{LI20DUST}
{Li} A. 2020.
\textit{Nature Astronomy} 4:339--351

\bibitem[{{Li} et~al.(2023){Li}, {Wisnioski}, {Mendel}, {Krumholz}, {Kewley}
  et~al.}]{li2023}
{Li} Z, {Wisnioski} E, {Mendel} JT, {Krumholz} MR, {Kewley} LJ, et~al. 2023.
\textit{\mnras} 518:286--304

\bibitem[{{Linden} et~al.(2023){Linden}, {Evans}, {Armus}, {Rich}, {Larson}
  et~al.}]{LINDEN23CLUSTERS}
{Linden} ST, {Evans} AS, {Armus} L, {Rich} JA, {Larson} KL, et~al. 2023.
\textit{\apjl} 944:L55

\bibitem[{{Lique} et~al.(2010){Lique}, {Spielfiedel}, {Feautrier}, {Schneider},
  {K{\l}os} \& {Alexander}}]{LIQUE10COLLIDE}
{Lique} F, {Spielfiedel} A, {Feautrier} N, {Schneider} IF, {K{\l}os} J,
  {Alexander} MH. 2010.
\textit{\jcp} 132:024303--024303

\bibitem[{{Liu} et~al.(2023){Liu}, {Schinnerer}, {Cao}, {Leroy}, {Usero}
  et~al.}]{liu2023}
{Liu} D, {Schinnerer} E, {Cao} Y, {Leroy} A, {Usero} A, et~al. 2023.
\textit{\apjl} 944:L19

\bibitem[{{Liu} et~al.(2021){Liu}, {Bureau}, {Blitz}, {Davis}, {Onishi}
  et~al.}]{LIU21GMCS}
{Liu} L, {Bureau} M, {Blitz} L, {Davis} TA, {Onishi} K, et~al. 2021.
\textit{\mnras} 505:4048--4085

\bibitem[{{Long} et~al.(2022){Long}, {Blair}, {Winkler}, {Della Bruna}, {Adamo}
  et~al.}]{LONG22SNR}
{Long} KS, {Blair} WP, {Winkler} PF, {Della Bruna} L, {Adamo} A, et~al. 2022.
\textit{\apj} 929:144

\bibitem[{{Longmore} et~al.(2013){Longmore}, {Bally}, {Testi}, {Purcell},
  {Walsh} et~al.}]{LONGMORE13DENSE}
{Longmore} SN, {Bally} J, {Testi} L, {Purcell} CR, {Walsh} AJ, et~al. 2013.
\textit{\mnras} 429:987--1000

\bibitem[{{Lopez} et~al.(2011){Lopez}, {Krumholz}, {Bolatto}, {Prochaska} \&
  {Ramirez-Ruiz}}]{LOPEZ11FEEDBACK}
{Lopez} LA, {Krumholz} MR, {Bolatto} AD, {Prochaska} JX, {Ramirez-Ruiz} E.
  2011.
\textit{\apj} 731:91

\bibitem[{{Lopez} et~al.(2014){Lopez}, {Krumholz}, {Bolatto}, {Prochaska},
  {Ramirez-Ruiz} \& {Castro}}]{LOPEZ14FEEDBACK}
{Lopez} LA, {Krumholz} MR, {Bolatto} AD, {Prochaska} JX, {Ramirez-Ruiz} E,
  {Castro} D. 2014.
\textit{\apj} 795:121

\bibitem[{{Mac Low} \& {Klessen}(2004)}]{MACLOW04REVIEW}
{Mac Low} MM, {Klessen} RS. 2004.
\textit{Reviews of Modern Physics} 76:125--194

\bibitem[{{Maiolino} \& {Mannucci}(2019)}]{MAIOLINO19REVIEW}
{Maiolino} R, {Mannucci} F. 2019.
\textit{\aapr} 27:3

\bibitem[{{Maloney} \& {Black}(1988)}]{MALONEY88XCO}
{Maloney} P, {Black} JH. 1988.
\textit{\apj} 325:389

\bibitem[{{Mangum} \& {Shirley}(2015)}]{MANGUM15LINES}
{Mangum} JG, {Shirley} YL. 2015.
\textit{\pasp} 127:266

\bibitem[{{Martig} et~al.(2009){Martig}, {Bournaud}, {Teyssier} \&
  {Dekel}}]{martig2009}
{Martig} M, {Bournaud} F, {Teyssier} R, {Dekel} A. 2009.
\textit{\apj} 707:250--267

\bibitem[{{Mart{\'\i}n} et~al.(2021){Mart{\'\i}n}, {Mangum}, {Harada},
  {Costagliola}, {Sakamoto} et~al.}]{ALCHEMI21SURVEY}
{Mart{\'\i}n} S, {Mangum} JG, {Harada} N, {Costagliola} F, {Sakamoto} K, et~al.
  2021.
\textit{\aap} 656:A46

\bibitem[{{Martizzi} et~al.(2015){Martizzi}, {Faucher-Gigu{\`e}re} \&
  {Quataert}}]{MARTIZZI15FEEDBACK}
{Martizzi} D, {Faucher-Gigu{\`e}re} CA, {Quataert} E. 2015.
\textit{\mnras} 450:504--522

\bibitem[{{Mayker Chen} et~al.(2023){Mayker Chen}, {Leroy}, {Lopez},
  {Benincasa}, {Chevance} et~al.}]{mayker2023}
{Mayker Chen} N, {Leroy} AK, {Lopez} LA, {Benincasa} S, {Chevance} M, et~al.
  2023.
\textit{\apj} 944:110

\bibitem[{{McKee} \& {Ostriker}(2007)}]{MCKEE07REVIEW}
{McKee} CF, {Ostriker} EC. 2007.
\textit{\araa} 45:565--687

\bibitem[{{McKee} \& {Zweibel}(1992)}]{MCKEE92GMCS}
{McKee} CF, {Zweibel} EG. 1992.
\textit{\apj} 399:551

\bibitem[{{McLeod} et~al.(2021){McLeod}, {Ali}, {Chevance}, {Della Bruna},
  {Schruba} et~al.}]{MCLEOD21HII}
{McLeod} AF, {Ali} AA, {Chevance} M, {Della Bruna} L, {Schruba} A, et~al. 2021.
\textit{\mnras} 508:5425--5448

\bibitem[{{McLeod} et~al.(2019){McLeod}, {Dale}, {Evans}, {Ginsburg},
  {Kruijssen} et~al.}]{mcleod2019}
{McLeod} AF, {Dale} JE, {Evans} CJ, {Ginsburg} A, {Kruijssen} JMD, et~al. 2019.
\textit{\mnras} 486:5263--5288

\bibitem[{{Meidt} et~al.(2015){Meidt}, {Hughes}, {Dobbs}, {Pety}, {Thompson}
  et~al.}]{meidt2015}
{Meidt} SE, {Hughes} A, {Dobbs} CL, {Pety} J, {Thompson} TA, et~al. 2015.
\textit{\apj} 806:72

\bibitem[{{Meidt} et~al.(2021){Meidt}, {Leroy}, {Querejeta}, {Schinnerer},
  {Sun} et~al.}]{MEIDT21ARMS}
{Meidt} SE, {Leroy} AK, {Querejeta} M, {Schinnerer} E, {Sun} J, et~al. 2021.
\textit{\apj} 913:113

\bibitem[{{Meidt} et~al.(2018){Meidt}, {Leroy}, {Rosolowsky}, {Kruijssen},
  {Schinnerer} et~al.}]{MEIDT18GMCS}
{Meidt} SE, {Leroy} AK, {Rosolowsky} E, {Kruijssen} JMD, {Schinnerer} E, et~al.
  2018.
\textit{\apj} 854:100

\bibitem[{{Meier} \& {Turner}(2004)}]{meier2004}
{Meier} DS, {Turner} JL. 2004.
\textit{\aj} 127:2069--2084

\bibitem[{{Meier} \& {Turner}(2005)}]{MEIER05LINES}
{Meier} DS, {Turner} JL. 2005.
\textit{\apj} 618:259--280

\bibitem[{{Metha} et~al.(2021){Metha}, {Trenti} \& {Chu}}]{metha2021}
{Metha} B, {Trenti} M, {Chu} T. 2021.
\textit{\mnras} 508:489--507

\bibitem[{{Miura} et~al.(2012){Miura}, {Kohno}, {Tosaki}, {Espada}, {Hwang}
  et~al.}]{MIURA12TIMES}
{Miura} RE, {Kohno} K, {Tosaki} T, {Espada} D, {Hwang} N, et~al. 2012.
\textit{\apj} 761:37

\bibitem[{{Miville-Desch{\^e}nes} et~al.(2017){Miville-Desch{\^e}nes}, {Murray}
  \& {Lee}}]{MIVILLE17GMCS}
{Miville-Desch{\^e}nes} MA, {Murray} N, {Lee} EJ. 2017.
\textit{\apj} 834:57

\bibitem[{{Moon} et~al.(2022){Moon}, {Kim}, {Kim} \& {Ostriker}}]{moon2022}
{Moon} S, {Kim} WT, {Kim} CG, {Ostriker} EC. 2022.
\textit{\apj} 925:99

\bibitem[{{Moon} et~al.(2023){Moon}, {Kim}, {Kim} \& {Ostriker}}]{moon2023}
{Moon} S, {Kim} WT, {Kim} CG, {Ostriker} EC. 2023.
\textit{\apj} 946:114

\bibitem[{{Morokuma-Matsui} et~al.(2020){Morokuma-Matsui}, {Sorai}, {Sato},
  {Kuno}, {Takeuchi} et~al.}]{morokuma-matsui2020}
{Morokuma-Matsui} K, {Sorai} K, {Sato} Y, {Kuno} N, {Takeuchi} TT, et~al. 2020.
\textit{\pasj} 72:90

\bibitem[{{Murphy} et~al.(2011){Murphy}, {Condon}, {Schinnerer}, {Kennicutt},
  {Calzetti} et~al.}]{MURPHY11SFR}
{Murphy} EJ, {Condon} JJ, {Schinnerer} E, {Kennicutt} RC, {Calzetti} D, et~al.
  2011.
\textit{\apj} 737:67

\bibitem[{{Murray}(2011)}]{MURRAY11LIFETIMES}
{Murray} N. 2011.
\textit{\apj} 729:133

\bibitem[{{Murray} et~al.(2010){Murray}, {Quataert} \&
  {Thompson}}]{MURRAY10FEEDBACK}
{Murray} N, {Quataert} E, {Thompson} TA. 2010.
\textit{\apj} 709:191--209

\bibitem[{{Narayanan} et~al.(2012){Narayanan}, {Krumholz}, {Ostriker} \&
  {Hernquist}}]{NARAYANAN12XCO}
{Narayanan} D, {Krumholz} MR, {Ostriker} EC, {Hernquist} L. 2012.
\textit{\mnras} 421:3127--3146

\bibitem[{{Nath} et~al.(2020){Nath}, {Das} \& {Oey}}]{nath2020}
{Nath} BB, {Das} P, {Oey} MS. 2020.
\textit{\mnras} 493:1034--1043

\bibitem[{{Neumann} et~al.(2023){Neumann}, {Gallagher}, {Bigiel}, {Leroy},
  {Barnes} et~al.}]{NEUMANN23DENSE}
{Neumann} L, {Gallagher} MJ, {Bigiel} F, {Leroy} AK, {Barnes} AT, et~al. 2023.
\textit{\mnras} 521:3348--3383

\bibitem[{{Oey} et~al.(2007){Oey}, {Meurer}, {Yelda}, {Furst},
  {Caballero-Nieves} et~al.}]{OEY07DIG}
{Oey} MS, {Meurer} GR, {Yelda} S, {Furst} EJ, {Caballero-Nieves} SM, et~al.
  2007.
\textit{\apj} 661:801--814

\bibitem[{{Oey} et~al.(2003){Oey}, {Parker}, {Mikles} \&
  {Zhang}}]{OEY03HIIREGIONS}
{Oey} MS, {Parker} JS, {Mikles} VJ, {Zhang} X. 2003.
\textit{\aj} 126:2317--2329

\bibitem[{{Olivier} et~al.(2021){Olivier}, {Lopez}, {Rosen}, {Nayak}, {Reiter}
  et~al.}]{olivier2021}
{Olivier} GM, {Lopez} LA, {Rosen} AL, {Nayak} O, {Reiter} M, et~al. 2021.
\textit{\apj} 908:68

\bibitem[{{Onodera} et~al.(2010){Onodera}, {Kuno}, {Tosaki}, {Kohno},
  {Nakanishi} et~al.}]{ONODERA10SFGAS}
{Onodera} S, {Kuno} N, {Tosaki} T, {Kohno} K, {Nakanishi} K, et~al. 2010.
\textit{\apjl} 722:L127--L131

\bibitem[{{Onus} et~al.(2018){Onus}, {Krumholz} \& {Federrath}}]{ONUS18DENSE}
{Onus} A, {Krumholz} MR, {Federrath} C. 2018.
\textit{\mnras} 479:1702--1710

\bibitem[{{Osterbrock}(1989)}]{OSTERBROCK89BOOK}
{Osterbrock} DE. 1989.
\textit{{Astrophysics of gaseous nebulae and active galactic nuclei}}

\bibitem[{{Ostriker} \& {Kim}(2022)}]{OSTRIKER22PRESS}
{Ostriker} EC, {Kim} CG. 2022.
\textit{\apj} 936:137

\bibitem[{{Ostriker} et~al.(2010){Ostriker}, {McKee} \&
  {Leroy}}]{OSTRIKER10PRESS}
{Ostriker} EC, {McKee} CF, {Leroy} AK. 2010.
\textit{\apj} 721:975--994

\bibitem[{{Padoan} et~al.(2012){Padoan}, {Haugb{\o}lle} \&
  {Nordlund}}]{PADOAN12EFF}
{Padoan} P, {Haugb{\o}lle} T, {Nordlund} {\r{A}}. 2012.
\textit{\apjl} 759:L27

\bibitem[{{Padoan} \& {Nordlund}(2002)}]{PADOAN02TURB}
{Padoan} P, {Nordlund} {\r{A}}. 2002.
\textit{\apj} 576:870--879

\bibitem[{{Padoan} \& {Nordlund}(2011)}]{PADOAN11EFF}
{Padoan} P, {Nordlund} {\r{A}}. 2011.
\textit{\apj} 730:40

\bibitem[{{Pan} et~al.(2022){Pan}, {Schinnerer}, {Hughes}, {Leroy}, {Groves}
  et~al.}]{PAN22TIMES}
{Pan} HA, {Schinnerer} E, {Hughes} A, {Leroy} A, {Groves} B, et~al. 2022.
\textit{\apj} 927:9

\bibitem[{{Paradis} et~al.(2012){Paradis}, {Dobashi}, {Shimoikura}, {Kawamura},
  {Onishi} et~al.}]{PARADIS12XCO}
{Paradis} D, {Dobashi} K, {Shimoikura} T, {Kawamura} A, {Onishi} T, et~al.
  2012.
\textit{\aap} 543:A103

\bibitem[{{Pathak} et~al.(2024){Pathak}, {Leroy}, {Thompson}, {Lopez},
  {Belfiore} et~al.}]{PATHAK24MIDIR}
{Pathak} D, {Leroy} AK, {Thompson} TA, {Lopez} LA, {Belfiore} F, et~al. 2024.
\textit{\aj} 167:39

\bibitem[{{Pejcha} \& {Prieto}(2015)}]{PEJCHA15SNR}
{Pejcha} O, {Prieto} JL. 2015.
\textit{\apj} 799:215

\bibitem[{{Peltonen} et~al.(2023){Peltonen}, {Rosolowsky}, {Johnson}, {Seth},
  {Dalcanton} et~al.}]{peltonen2023}
{Peltonen} J, {Rosolowsky} E, {Johnson} LC, {Seth} AC, {Dalcanton} J, et~al.
  2023.
\textit{\mnras} 522:6137--6149

\bibitem[{{Pereira-Santaella} et~al.(2021){Pereira-Santaella}, {Colina},
  {Garc{\'\i}a-Burillo}, {Lamperti}, {Gonz{\'a}lez-Alfonso}
  et~al.}]{PUMA21SURVEYTWO}
{Pereira-Santaella} M, {Colina} L, {Garc{\'\i}a-Burillo} S, {Lamperti} I,
  {Gonz{\'a}lez-Alfonso} E, et~al. 2021.
\textit{\aap} 651:A42

\bibitem[{{Pessa} et~al.(2023){Pessa}, {Schinnerer}, {Sanchez-Blazquez},
  {Belfiore}, {Groves} et~al.}]{pessa2023}
{Pessa} I, {Schinnerer} E, {Sanchez-Blazquez} P, {Belfiore} F, {Groves} B,
  et~al. 2023.
\textit{\aap} 673:A147

\bibitem[{{Pety} et~al.(2017){Pety}, {Guzm{\'a}n}, {Orkisz}, {Liszt}, {Gerin}
  et~al.}]{PETY17DENSE}
{Pety} J, {Guzm{\'a}n} VV, {Orkisz} JH, {Liszt} HS, {Gerin} M, et~al. 2017.
\textit{\aap} 599:A98

\bibitem[{{Pineda} et~al.(2009){Pineda}, {Rosolowsky} \&
  {Goodman}}]{PINEDA09GMCS}
{Pineda} JE, {Rosolowsky} EW, {Goodman} AA. 2009.
\textit{\apjl} 699:L134--L138

\bibitem[{{Pineda} et~al.(2013){Pineda}, {Langer}, {Velusamy} \&
  {Goldsmith}}]{PINEDA13XCO}
{Pineda} JL, {Langer} WD, {Velusamy} T, {Goldsmith} PF. 2013.
\textit{\aap} 554:A103

\bibitem[{{Pokhrel} et~al.(2020){Pokhrel}, {Simpson} \&
  {Bagetakos}}]{POKHREL20HOLES}
{Pokhrel} NR, {Simpson} CE, {Bagetakos} I. 2020.
\textit{\aj} 160:66

\bibitem[{{Popescu} et~al.(2012){Popescu}, {Hanson} \&
  {Elmegreen}}]{POPESCU12SFR}
{Popescu} B, {Hanson} MM, {Elmegreen} BG. 2012.
\textit{\apj} 751:122

\bibitem[{{Privon} et~al.(2015){Privon}, {Herrero-Illana}, {Evans}, {Iwasawa},
  {Perez-Torres} et~al.}]{PRIVON15DENSE}
{Privon} GC, {Herrero-Illana} R, {Evans} AS, {Iwasawa} K, {Perez-Torres} MA,
  et~al. 2015.
\textit{\apj} 814:39

\bibitem[{{Querejeta} et~al.(2016){Querejeta}, {Schinnerer},
  {Garc{\'\i}a-Burillo}, {Bigiel}, {Blanc} et~al.}]{querejeta2016}
{Querejeta} M, {Schinnerer} E, {Garc{\'\i}a-Burillo} S, {Bigiel} F, {Blanc} GA,
  et~al. 2016.
\textit{\aap} 593:A118

\bibitem[{{Querejeta} et~al.(2021){Querejeta}, {Schinnerer}, {Meidt}, {Sun},
  {Leroy} et~al.}]{querejeta2021}
{Querejeta} M, {Schinnerer} E, {Meidt} S, {Sun} J, {Leroy} AK, et~al. 2021.
\textit{\aap} 656:A133

\bibitem[{{Querejeta} et~al.(2019){Querejeta}, {Schinnerer}, {Schruba},
  {Murphy}, {Meidt} et~al.}]{querejeta2019}
{Querejeta} M, {Schinnerer} E, {Schruba} A, {Murphy} E, {Meidt} S, et~al. 2019.
\textit{\aap} 625:A19

\bibitem[{{Regan} et~al.(2001){Regan}, {Thornley}, {Helfer}, {Sheth}, {Wong}
  et~al.}]{REGAN01SFGAS}
{Regan} MW, {Thornley} MD, {Helfer} TT, {Sheth} K, {Wong} T, et~al. 2001.
\textit{\apj} 561:218--237

\bibitem[{{Rosolowsky} et~al.(2021){Rosolowsky}, {Hughes}, {Leroy}, {Sun},
  {Querejeta} et~al.}]{ROSOLOWSKY21GMCS}
{Rosolowsky} E, {Hughes} A, {Leroy} AK, {Sun} J, {Querejeta} M, et~al. 2021.
\textit{\mnras} 502:1218--1245

\bibitem[{{Rubin}(1968)}]{RUBIN68SFR}
{Rubin} RH. 1968.
\textit{\apj} 153:761

\bibitem[{{Saintonge} \& {Catinella}(2022)}]{SAINTONGE22REVIEW}
{Saintonge} A, {Catinella} B. 2022.
\textit{\araa} 60:319--361

\bibitem[{{Saintonge} et~al.(2017){Saintonge}, {Catinella}, {Tacconi},
  {Kauffmann}, {Genzel} et~al.}]{SAINTONGE17SFGAS}
{Saintonge} A, {Catinella} B, {Tacconi} LJ, {Kauffmann} G, {Genzel} R, et~al.
  2017.
\textit{\apjs} 233:22

\bibitem[{{Saito} et~al.(2022){Saito}, {Takano}, {Harada}, {Nakajima},
  {Schinnerer} et~al.}]{saito2022}
{Saito} T, {Takano} S, {Harada} N, {Nakajima} T, {Schinnerer} E, et~al. 2022.
\textit{\apj} 935:155

\bibitem[{{Salak} et~al.(2017){Salak}, {Tomiyasu}, {Nakai}, {Kuno}, {Miyamoto}
  \& {Kaneko}}]{salak2017}
{Salak} D, {Tomiyasu} Y, {Nakai} N, {Kuno} N, {Miyamoto} Y, {Kaneko} H. 2017.
\textit{\apj} 849:90

\bibitem[{{S{\'a}nchez}(2020)}]{SANCHEZ20REVIEW}
{S{\'a}nchez} SF. 2020.
\textit{\araa} 58:99--155

\bibitem[{{S{\'a}nchez-Menguiano} et~al.(2020){S{\'a}nchez-Menguiano},
  {S{\'a}nchez}, {P{\'e}rez}, {Ruiz-Lara}, {Galbany}
  et~al.}]{sanchez-menguiano2020}
{S{\'a}nchez-Menguiano} L, {S{\'a}nchez} SF, {P{\'e}rez} I, {Ruiz-Lara} T,
  {Galbany} L, et~al. 2020.
\textit{\mnras} 492:4149--4163

\bibitem[{{Sandstrom} et~al.(2023){Sandstrom}, {Koch}, {Leroy}, {Rosolowsky},
  {Emsellem} et~al.}]{SANDSTROM23JWST}
{Sandstrom} KM, {Koch} EW, {Leroy} AK, {Rosolowsky} E, {Emsellem} E, et~al.
  2023.
\textit{\apjl} 944:L8

\bibitem[{{Sandstrom} et~al.(2013){Sandstrom}, {Leroy}, {Walter}, {Bolatto},
  {Croxall} et~al.}]{SANDSTROM13XCO}
{Sandstrom} KM, {Leroy} AK, {Walter} F, {Bolatto} AD, {Croxall} KV, et~al.
  2013.
\textit{\apj} 777:5

\bibitem[{{Santoro} et~al.(2022){Santoro}, {Kreckel}, {Belfiore}, {Groves},
  {Congiu} et~al.}]{SANTORO22SFR}
{Santoro} F, {Kreckel} K, {Belfiore} F, {Groves} B, {Congiu} E, et~al. 2022.
\textit{\aap} 658:A188

\bibitem[{{Sarbadhicary} et~al.(2023){Sarbadhicary}, {Wagner}, {Koch}, {Mayker
  Chen}, {Leroy} et~al.}]{SARBADHICARY23SNR}
{Sarbadhicary} SK, {Wagner} J, {Koch} EW, {Mayker Chen} N, {Leroy} AK, et~al.
  2023.
\textit{arXiv e-prints} :arXiv:2310.17694

\bibitem[{{Schinnerer} et~al.(2023){Schinnerer}, {Emsellem}, {Henshaw}, {Liu},
  {Meidt} et~al.}]{schinnerer2023}
{Schinnerer} E, {Emsellem} E, {Henshaw} JD, {Liu} D, {Meidt} SE, et~al. 2023.
\textit{\apjl} 944:L15

\bibitem[{{Schinnerer} et~al.(2019){Schinnerer}, {Hughes}, {Leroy}, {Groves},
  {Blanc} et~al.}]{SCHINNERER19TIMES}
{Schinnerer} E, {Hughes} A, {Leroy} A, {Groves} B, {Blanc} GA, et~al. 2019.
\textit{\apj} 887:49

\bibitem[{{Schinnerer} et~al.(2013){Schinnerer}, {Meidt}, {Pety}, {Hughes},
  {Colombo} et~al.}]{PAWS13SURVEY}
{Schinnerer} E, {Meidt} SE, {Pety} J, {Hughes} A, {Colombo} D, et~al. 2013.
\textit{\apj} 779:42

\bibitem[{{Schruba} et~al.(2019){Schruba}, {Kruijssen} \&
  {Leroy}}]{SCHRUBA19GMCS}
{Schruba} A, {Kruijssen} JMD, {Leroy} AK. 2019.
\textit{\apj} 883:2

\bibitem[{{Schruba} et~al.(2010){Schruba}, {Leroy}, {Walter}, {Sandstrom} \&
  {Rosolowsky}}]{SCHRUBA10SFGAS}
{Schruba} A, {Leroy} AK, {Walter} F, {Sandstrom} K, {Rosolowsky} E. 2010.
\textit{\apj} 722:1699--1706

\bibitem[{{Scoville} et~al.(2014){Scoville}, {Aussel}, {Sheth}, {Scott},
  {Sanders} et~al.}]{SCOVILLE14DUST}
{Scoville} N, {Aussel} H, {Sheth} K, {Scott} KS, {Sanders} D, et~al. 2014.
\textit{\apj} 783:84

\bibitem[{{Scoville} et~al.(1987){Scoville}, {Yun}, {Clemens}, {Sanders} \&
  {Waller}}]{SCOVILLE87GMCS}
{Scoville} NZ, {Yun} MS, {Clemens} DP, {Sanders} DB, {Waller} WH. 1987.
\textit{\apjs} 63:821

\bibitem[{{Seifried} et~al.(2020){Seifried}, {Haid}, {Walch}, {Borchert} \&
  {Bisbas}}]{SEIFRIED20XCO}
{Seifried} D, {Haid} S, {Walch} S, {Borchert} EMA, {Bisbas} TG. 2020.
\textit{\mnras} 492:1465--1483

\bibitem[{{Seo} \& {Kim}(2013)}]{seo2013}
{Seo} WY, {Kim} WT. 2013.
\textit{\apj} 769:100

\bibitem[{{Serabyn} \& {Morris}(1996)}]{serabyn1996}
{Serabyn} E, {Morris} M. 1996.
\textit{\nat} 382:602--604

\bibitem[{{Shi} et~al.(2011){Shi}, {Helou}, {Yan}, {Armus}, {Wu}
  et~al.}]{SHI11SFGAS}
{Shi} Y, {Helou} G, {Yan} L, {Armus} L, {Wu} Y, et~al. 2011.
\textit{\apj} 733:87

\bibitem[{{Shirley}(2015)}]{SHIRLEY15LINES}
{Shirley} YL. 2015.
\textit{\pasp} 127:299

\bibitem[{{Smith} et~al.(2014){Smith}, {Glover}, {Clark}, {Klessen} \&
  {Springel}}]{SMITH14XCO}
{Smith} RJ, {Glover} SCO, {Clark} PC, {Klessen} RS, {Springel} V. 2014.
\textit{\mnras} 441:1628--1645

\bibitem[{{Solomon} et~al.(1987){Solomon}, {Rivolo}, {Barrett} \&
  {Yahil}}]{SOLOMON87GMCS}
{Solomon} PM, {Rivolo} AR, {Barrett} J, {Yahil} A. 1987.
\textit{\apj} 319:730

\bibitem[{{Sormani} et~al.(2023){Sormani}, {Sobacchi} \&
  {Sanders}}]{SORMANI23BAR}
{Sormani} MC, {Sobacchi} E, {Sanders} JL. 2023.
\textit{arXiv e-prints} :arXiv:2309.14093

\bibitem[{{Sormani} et~al.(2020){Sormani}, {Tress}, {Glover}, {Klessen},
  {Battersby} et~al.}]{sormani2020}
{Sormani} MC, {Tress} RG, {Glover} SCO, {Klessen} RS, {Battersby} CD, et~al.
  2020.
\textit{\mnras} 497:5024--5040

\bibitem[{{Spitzer}(1968)}]{SPITZER68BOOK}
{Spitzer} Lyman J. 1968.
\textit{{Dynamics of Interstellar Matter and the Formation of Stars}}. In
  \textit{Nebulae and Interstellar Matter}, eds. BM~{Middlehurst}, LH~{Aller}.
  ~1

\bibitem[{{Spitzer}(1978)}]{spitzer1978}
{Spitzer} L. 1978.
\textit{{Physical processes in the interstellar medium}}

\bibitem[{{Stuber} et~al.(2023{\natexlab{a}}){Stuber}, {Pety}, {Schinnerer},
  {Bigiel}, {Usero} et~al.}]{STUBER23DENSE}
{Stuber} SK, {Pety} J, {Schinnerer} E, {Bigiel} F, {Usero} A, et~al.
  2023{\natexlab{a}}.
\textit{\aap} 680:L20

\bibitem[{{Stuber} et~al.(2023{\natexlab{b}}){Stuber}, {Schinnerer},
  {Williams}, {Querejeta}, {Meidt} et~al.}]{stuber2023}
{Stuber} SK, {Schinnerer} E, {Williams} TG, {Querejeta} M, {Meidt} S, et~al.
  2023{\natexlab{b}}.
\textit{\aap} 676:A113

\bibitem[{{Sukhbold} et~al.(2016){Sukhbold}, {Ertl}, {Woosley}, {Brown} \&
  {Janka}}]{SUKHBOLD16SNE}
{Sukhbold} T, {Ertl} T, {Woosley} SE, {Brown} JM, {Janka} HT. 2016.
\textit{\apj} 821:38

\bibitem[{{Sun} et~al.(2020{\natexlab{a}}){Sun}, {Leroy}, {Ostriker}, {Hughes},
  {Rosolowsky} et~al.}]{SUN20PRESS}
{Sun} J, {Leroy} AK, {Ostriker} EC, {Hughes} A, {Rosolowsky} E, et~al.
  2020{\natexlab{a}}.
\textit{\apj} 892:148

\bibitem[{{Sun} et~al.(2023){Sun}, {Leroy}, {Ostriker}, {Meidt}, {Rosolowsky}
  et~al.}]{SUN23SFGAS}
{Sun} J, {Leroy} AK, {Ostriker} EC, {Meidt} S, {Rosolowsky} E, et~al. 2023.
\textit{\apjl} 945:L19

\bibitem[{{Sun} et~al.(2022){Sun}, {Leroy}, {Rosolowsky}, {Hughes},
  {Schinnerer} et~al.}]{SUN22CLOUDS}
{Sun} J, {Leroy} AK, {Rosolowsky} E, {Hughes} A, {Schinnerer} E, et~al. 2022.
\textit{\aj} 164:43

\bibitem[{{Sun} et~al.(2020{\natexlab{b}}){Sun}, {Leroy}, {Schinnerer},
  {Hughes}, {Rosolowsky} et~al.}]{SUN20GMCS}
{Sun} J, {Leroy} AK, {Schinnerer} E, {Hughes} A, {Rosolowsky} E, et~al.
  2020{\natexlab{b}}.
\textit{\apjl} 901:L8

\bibitem[{{Sun} et~al.(2018){Sun}, {Leroy}, {Schruba}, {Rosolowsky}, {Hughes}
  et~al.}]{SUN18CLOUDS}
{Sun} J, {Leroy} AK, {Schruba} A, {Rosolowsky} E, {Hughes} A, et~al. 2018.
\textit{\apj} 860:172

\bibitem[{{Tacchella} et~al.(2022){Tacchella}, {Smith}, {Kannan}, {Marinacci},
  {Hernquist} et~al.}]{TACHELLA22SFR}
{Tacchella} S, {Smith} A, {Kannan} R, {Marinacci} F, {Hernquist} L, et~al.
  2022.
\textit{\mnras} 513:2904--2929

\bibitem[{{Tacconi} et~al.(2020){Tacconi}, {Genzel} \&
  {Sternberg}}]{TACCONI20REVIEW}
{Tacconi} LJ, {Genzel} R, {Sternberg} A. 2020.
\textit{\araa} 58:157--203

\bibitem[{{Tafalla} et~al.(2021){Tafalla}, {Usero} \& {Hacar}}]{TAFALLA21LINES}
{Tafalla} M, {Usero} A, {Hacar} A. 2021.
\textit{\aap} 646:A97

\bibitem[{{Tafalla} et~al.(2023){Tafalla}, {Usero} \& {Hacar}}]{TAFALLA23LINES}
{Tafalla} M, {Usero} A, {Hacar} A. 2023.
\textit{\aap} 679:A112

\bibitem[{{Tan}(2000)}]{TAN00COLLIDE}
{Tan} JC. 2000.
\textit{\apj} 536:173--184

\bibitem[{{Teng} et~al.(2023){Teng}, {Sandstrom}, {Sun}, {Gong}, {Bolatto}
  et~al.}]{TENG23XCO}
{Teng} YH, {Sandstrom} KM, {Sun} J, {Gong} M, {Bolatto} AD, et~al. 2023.
\textit{\apj} 950:119

\bibitem[{{Teng} et~al.(2022){Teng}, {Sandstrom}, {Sun}, {Leroy}, {Johnson}
  et~al.}]{TENG22XCO}
{Teng} YH, {Sandstrom} KM, {Sun} J, {Leroy} AK, {Johnson} LC, et~al. 2022.
\textit{\apj} 925:72

\bibitem[{{Tenorio-Tagle}(1979)}]{tenorio-tagle1979}
{Tenorio-Tagle} G. 1979.
\textit{\aap} 71:59--65

\bibitem[{{Thilker} et~al.(2002){Thilker}, {Walterbos}, {Braun} \&
  {Hoopes}}]{THILKER02DIG}
{Thilker} DA, {Walterbos} RAM, {Braun} R, {Hoopes} CG. 2002.
\textit{\aj} 124:3118--3134

\bibitem[{{Thompson} et~al.(2005){Thompson}, {Quataert} \&
  {Murray}}]{THOMPSON05PRESS}
{Thompson} TA, {Quataert} E, {Murray} N. 2005.
\textit{\apj} 630:167--185

\bibitem[{{Toomre}(1964)}]{TOOMRE64DISKS}
{Toomre} A. 1964.
\textit{\apj} 139:1217--1238

\bibitem[{{Tress} et~al.(2020){Tress}, {Sormani}, {Glover}, {Klessen},
  {Battersby} et~al.}]{tress2020}
{Tress} RG, {Sormani} MC, {Glover} SCO, {Klessen} RS, {Battersby} CD, et~al.
  2020.
\textit{\mnras} 499:4455--4478

\bibitem[{{Turner} et~al.(2022){Turner}, {Dale}, {Lilly}, {Boquien}, {Deger}
  et~al.}]{TURNER22TIMES}
{Turner} JA, {Dale} DA, {Lilly} J, {Boquien} M, {Deger} S, et~al. 2022.
\textit{\mnras} 516:4612--4626

\bibitem[{{Ueda} et~al.(2021){Ueda}, {Iono}, {Yun}, {Michiyama}, {Watanabe}
  et~al.}]{UEDA21DENSE}
{Ueda} J, {Iono} D, {Yun} MS, {Michiyama} T, {Watanabe} Y, et~al. 2021.
\textit{\apjs} 257:57

\bibitem[{{Usero} et~al.(2015){Usero}, {Leroy}, {Walter}, {Schruba},
  {Garc{\'\i}a-Burillo} et~al.}]{USERO15DENSE}
{Usero} A, {Leroy} AK, {Walter} F, {Schruba} A, {Garc{\'\i}a-Burillo} S, et~al.
  2015.
\textit{\aj} 150:115

\bibitem[{{Utomo} et~al.(2018){Utomo}, {Sun}, {Leroy}, {Kruijssen},
  {Schinnerer} et~al.}]{UTOMO18EFF}
{Utomo} D, {Sun} J, {Leroy} AK, {Kruijssen} JMD, {Schinnerer} E, et~al. 2018.
\textit{\apjl} 861:L18

\bibitem[{{Veilleux} et~al.(2020){Veilleux}, {Maiolino}, {Bolatto} \&
  {Aalto}}]{veilleux2020}
{Veilleux} S, {Maiolino} R, {Bolatto} AD, {Aalto} S. 2020.
\textit{\aapr} 28:2

\bibitem[{{Walch} et~al.(2015){Walch}, {Girichidis}, {Naab}, {Gatto}, {Glover}
  et~al.}]{WALCH15FEEDBACK}
{Walch} S, {Girichidis} P, {Naab} T, {Gatto} A, {Glover} SCO, et~al. 2015.
\textit{\mnras} 454:238--268

\bibitem[{{Walch} \& {Naab}(2015)}]{WALCH15SN}
{Walch} S, {Naab} T. 2015.
\textit{\mnras} 451:2757--2771

\bibitem[{{Walcher} et~al.(2011){Walcher}, {Groves}, {Budav{\'a}ri} \&
  {Dale}}]{walcher2011}
{Walcher} J, {Groves} B, {Budav{\'a}ri} T, {Dale} D. 2011.
\textit{\apss} 331:1--52

\bibitem[{{Walter} et~al.(2017){Walter}, {Bolatto}, {Leroy}, {Veilleux},
  {Warren} et~al.}]{WALTER17DENSE}
{Walter} F, {Bolatto} AD, {Leroy} AK, {Veilleux} S, {Warren} SR, et~al. 2017.
\textit{\apj} 835:265

\bibitem[{{Ward} et~al.(2022){Ward}, {Kruijssen}, {Chevance}, {Kim} \&
  {Longmore}}]{ward2022}
{Ward} JL, {Kruijssen} JMD, {Chevance} M, {Kim} J, {Longmore} SN. 2022.
\textit{\mnras} 516:4025--4042

\bibitem[{{Watkins} et~al.(2023{\natexlab{a}}){Watkins}, {Barnes}, {Henny},
  {Kim}, {Kreckel} et~al.}]{watkins2023a}
{Watkins} EJ, {Barnes} AT, {Henny} K, {Kim} H, {Kreckel} K, et~al.
  2023{\natexlab{a}}.
\textit{\apjl} 944:L24

\bibitem[{{Watkins} et~al.(2023{\natexlab{b}}){Watkins}, {Kreckel}, {Groves},
  {Glover}, {Whitmore} et~al.}]{watkins2023b}
{Watkins} EJ, {Kreckel} K, {Groves} B, {Glover} SCO, {Whitmore} BC, et~al.
  2023{\natexlab{b}}.
\textit{\aap} 676:A67

\bibitem[{{Wei{\ss}} et~al.(2005){Wei{\ss}}, {Walter} \&
  {Scoville}}]{WEISS05LINES}
{Wei{\ss}} A, {Walter} F, {Scoville} NZ. 2005.
\textit{\aap} 438:533--544

\bibitem[{{Weisz} et~al.(2009){Weisz}, {Skillman}, {Cannon}, {Dolphin},
  {Kennicutt} et~al.}]{WEISZ09HOLES}
{Weisz} DR, {Skillman} ED, {Cannon} JM, {Dolphin} AE, {Kennicutt} Robert~C. J,
  et~al. 2009.
\textit{\apj} 704:1538--1569

\bibitem[{{Whitcomb} et~al.(2023){Whitcomb}, {Sandstrom}, {Leroy} \&
  {Smith}}]{WHITCOMB23SFR}
{Whitcomb} CM, {Sandstrom} K, {Leroy} A, {Smith} JDT. 2023.
\textit{\apj} 948:88

\bibitem[{{Whitmore} et~al.(2023{\natexlab{a}}){Whitmore}, {Chandar}, {Lee},
  {Floyd}, {Deger} et~al.}]{whitmore2023b}
{Whitmore} BC, {Chandar} R, {Lee} JC, {Floyd} M, {Deger} S, et~al.
  2023{\natexlab{a}}.
\textit{\mnras} 520:63--88

\bibitem[{{Whitmore} et~al.(2023{\natexlab{b}}){Whitmore}, {Chandar},
  {Rodr{\'\i}guez}, {Lee}, {Emsellem} et~al.}]{whitmore2023}
{Whitmore} BC, {Chandar} R, {Rodr{\'\i}guez} MJ, {Lee} JC, {Emsellem} E, et~al.
  2023{\natexlab{b}}.
\textit{\apjl} 944:L14

\bibitem[{{Whitworth}(1979)}]{whitworth1979}
{Whitworth} A. 1979.
\textit{\mnras} 186:59--67

\bibitem[{{Williams} et~al.(2023){Williams}, {Bureau}, {Davis}, {Cappellari},
  {Choi} et~al.}]{WILLIAMS23GMCS}
{Williams} TG, {Bureau} M, {Davis} TA, {Cappellari} M, {Choi} W, et~al. 2023.
\textit{\mnras}

\bibitem[{{Williams} et~al.(2022){Williams}, {Kreckel}, {Belfiore}, {Groves},
  {Sandstrom} et~al.}]{williams2022}
{Williams} TG, {Kreckel} K, {Belfiore} F, {Groves} B, {Sandstrom} K, et~al.
  2022.
\textit{\mnras} 509:1303--1322

\bibitem[{{Wilson}(2018)}]{WILSON18LINES}
{Wilson} CD. 2018.
\textit{\mnras} 477:2926--2942

\bibitem[{{Wilson} et~al.(2023){Wilson}, {Bemis}, {Ledger} \&
  {Klimi}}]{WILSON23LINES}
{Wilson} CD, {Bemis} A, {Ledger} B, {Klimi} O. 2023.
\textit{\mnras} 521:717--736

\bibitem[{{Wu} et~al.(2010){Wu}, {Evans}, {Shirley} \& {Knez}}]{WU10DENSE}
{Wu} J, {Evans} Neal~J. I, {Shirley} YL, {Knez} C. 2010.
\textit{\apjs} 188:313--357

\bibitem[{{Yajima} et~al.(2021){Yajima}, {Sorai}, {Miyamoto}, {Muraoka}, {Kuno}
  et~al.}]{YAJIMA21LINES}
{Yajima} Y, {Sorai} K, {Miyamoto} Y, {Muraoka} K, {Kuno} N, et~al. 2021.
\textit{\pasj} 73:257--285

\bibitem[{{Yang} et~al.(2010){Yang}, {Stancil}, {Balakrishnan} \&
  {Forrey}}]{YANG10COLLIDE}
{Yang} B, {Stancil} PC, {Balakrishnan} N, {Forrey} RC. 2010.
\textit{\apj} 718:1062--1069

\bibitem[{{Yim} et~al.(2014){Yim}, {Wong}, {Xue}, {Rand}, {Rosolowsky}
  et~al.}]{YIM14EDGEON}
{Yim} K, {Wong} T, {Xue} R, {Rand} RJ, {Rosolowsky} E, et~al. 2014.
\textit{\aj} 148:127

\bibitem[{{Yoda} et~al.(2010){Yoda}, {Handa}, {Kohno}, {Nakajima}, {Kaiden}
  et~al.}]{YODA10LINES}
{Yoda} T, {Handa} T, {Kohno} K, {Nakajima} T, {Kaiden} M, et~al. 2010.
\textit{\pasj} 62:1277--1289

\bibitem[{{Young} et~al.(1996){Young}, {Allen}, {Kenney}, {Lesser} \&
  {Rownd}}]{YOUNG96SFGAS}
{Young} JS, {Allen} L, {Kenney} JDP, {Lesser} A, {Rownd} B. 1996.
\textit{\aj} 112:1903

\bibitem[{{Young} et~al.(1995){Young}, {Xie}, {Tacconi}, {Knezek}, {Viscuso}
  et~al.}]{YOUNG95SFGAS}
{Young} JS, {Xie} S, {Tacconi} L, {Knezek} P, {Viscuso} P, et~al. 1995.
\textit{\apjs} 98:219

\end{thebibliography}
\bibliographystyle{ar-style2}

\end{document}